%% file: New_joint_source-channel_coding_scheme_for_MARC.tex
\documentclass[onecolumn,10pt,draftcls]{IEEEtran}

\usepackage{times}
\usepackage{amsmath,dsfont}
\usepackage{amssymb,amsthm}
\usepackage{epsfig,verbatim}
\usepackage{mathrsfs}
\usepackage{verbatim}
\usepackage{syntonly}
\usepackage{graphicx}
\usepackage{epstopdf,subfig}
\usepackage{multirow}
\usepackage{diagbox}
\usepackage{scalefnt}
\usepackage{mathtools}
\usepackage{ifthen}

\newboolean{SquizFlag}
\setboolean{SquizFlag}{true}

\input{newCommands.tex}

 \IEEEoverridecommandlockouts
\title{On Joint Source-Channel Coding for Correlated Sources Over Multiple-Access Relay Channels
\thanks{
\noindent
This work was partially supported by the European Commission's Marie Curie IRG Fellowship
PIRG05-GA-2009-246657 under the Seventh Framework Programme, and by the Israel Science Foundation under grant 396/11. 
Parts of this work were presented at the IEEE International Symposium on Wireless Communication Systems (ISWCS), August 2012, Paris, France, and at the IEEE International Symposium on Information Theory (ISIT), July 2013, Istanbul, Turkey.
}
}

\author{
\IEEEauthorblockN{Yonathan Murin$^{\dagger}$, Ron Dabora$^{\dagger}$, and Deniz G\"und\"uz$^{\ast}$} \\
\authorblockA{$^{\dagger}$Dept. of Electrical and Computer Engineering, Ben-Gurion University, Israel \\
$^{\ast}$Dept. of Electrical and Electronic Engineering, Imperial College London, United Kingdom \\
Email: moriny@ee.bgu.ac.il, ron@ee.bgu.ac.il, d.gunduz@imperial.ac.uk}

\vspace{-1.1cm}

}

\vspace{-0.7cm}

\begin{document}
\maketitle
\thispagestyle{empty} 

\begin{abstract}
	We study the transmission of correlated sources over discrete memoryless (DM) multiple-access-relay channels (MARCs), in which both the relay and the destination have access to side information arbitrarily correlated with the sources. As the optimal transmission scheme is an open problem, in this work we propose a new joint source-channel coding scheme based on a novel combination of the correlation preserving mapping (CPM) technique with Slepian-Wolf (SW) source coding, and obtain the corresponding sufficient conditions. The proposed coding scheme is based on the decode-and-forward strategy, and utilizes CPM for encoding information simultaneously to the relay and the destination, whereas the cooperation information from the relay is encoded via SW source coding. 
		It is shown that there are cases in which the new scheme strictly outperforms the schemes available in the literature. 
		This is the first instance of a source-channel code that uses CPM for encoding information to two different nodes (relay and destination).
		In addition to sufficient conditions, we present three different sets of single-letter necessary conditions for reliable transmission of correlated sources over DM MARCs. The newly derived conditions
		are shown to be at least as tight as the previously known necessary conditions.	   
\end{abstract}




\vspace{-0.3cm}
\section{Introduction}
The multiple-access relay channel (MARC) is a multiuser network in which several sources communicate with a single destination with the help of a relay \cite{Kramer:2005}, \cite{Sankar:07}. 
This model represents cooperative uplink communication in wireless networks.
In this work, we study the lossless transmission of arbitrarily correlated sources over MARCs, in which both the relay and the destination have access to side information correlated with the sources. 

It is well known \cite{Shannon:48} that a source can be reliably transmitted over a memoryless point-to-point (PtP) channel, if its entropy is less than the channel capacity. Conversely, if the source entropy is larger than the channel capacity, then reliable transmission is not possible. Therefore, for memoryless PtP channels,  a separate design of the source and channel codes achieves the optimal end-to-end performance. 
However, the optimality of separate designs does not generalize to multiuser networks \cite{Cover:80}, \cite{GunduzErkip:09}, \cite{Murin:ISIT12}.

Since the MARC combines both the multiple access channel (MAC) and the relay channel models,
and since separate source-channel coding is not optimal for MAC with correlated sources \cite{Cover:80}, we conclude that separate designs are not optimal for MARCs. 
Therefore, it is important to develop methods for joint source-channel coding (JSCC) for this network.
In this work we derive separate sets of sufficient and necessary conditions, which are not necessarily tight.
In deriving our sufficiency conditions we focus on cooperation schemes based on the decode-and-forward (DF) protocol, such that the sequences of both sources are decoded at the relay.
Accordingly, transmission to both the relay and the destination can benefit from joint design of the source and channel codes. 

\vspace{-0.25cm}
\subsection{Prior Work}

\vspace{-0.1cm}
The MARC has received a lot of attention in recent years, especially from a channel coding perspective. 
In \cite{Kramer:2005}, Kramer et al. derived an achievable rate region for the MARC with independent messages, using a coding scheme based on DF relaying, regular encoding, successive decoding at the relay, and backward decoding at the destination. 
In \cite{Sankar:07} it was shown that for the MARC, in contrast to the relay channel, DF schemes with different decoding techniques at the destination yield different rate regions. Specifically, backward decoding can support a larger rate region than sliding window decoding. Another DF-based coding scheme, which uses offset encoding, successive decoding at the relay and sliding window decoding at the destination, was presented in \cite{Sankar:07}. This scheme was shown to be at least as good as sliding window decoding. Moreover, this scheme achieves the corner points of the backward decoding rate region, but with a smaller delay.
While the focus of \cite{Kramer:2005} and \cite{Sankar:07} was mainly on achievable rate regions, outer bounds on the capacity region of MARCs were derived in \cite{KramerMandayam:04}.
More recently, in \cite{Tandon:CISS:11}, Tandon and Poor derived the capacity region of two classes of MARCs, which include a primitive relay assisting the transmitters through an orthogonal finite-capacity link to the destination.

While the works \cite{Kramer:2005}, \cite{Sankar:07}, \cite{KramerMandayam:04} and \cite{Tandon:CISS:11} considered channel coding for MARCs, in \cite{Murin:IT11} we studied source-channel coding for MARCs with correlated sources. 
In \cite{Murin:ISIT12} we presented an explicit example in which separate source and channel code design is suboptimal for this model.
The suboptimality of separate source and channel coding for multiuser scenario was first shown by Shannon in \cite{Shannon:61} by considering the transmission of correlated sources over a two-way channel.

Lossless transmission of correlated sources over relay channels with correlated side information was studied in \cite{ErkipGunduz:07}, \cite{Gunduz:12}, \cite{ElGamalCioffi:07} and \cite{Sefidgaran:ISIT:09}. Specifically, in \cite{ErkipGunduz:07} G\"und\"uz and Erkip proposed a DF based achievability scheme and showed that separation is optimal for physically degraded relay channels as well as for cooperative relay-broadcast channels. This work was later extended to multiple relay networks in \cite{Gunduz:12}. 
	The relay channel with arbitrarily correlated sources, in which one of the sources is available at the transmitter while the other is known at the relay, and the destination is interested in a lossless reconstruction of both sources, was considered in \cite{SmithVishwanath:04}, \cite{SalehkalaibarAref:ISIT11} and \cite{SalehkalaibarAref:ISIT12}. 
The work \cite{SmithVishwanath:04} used block Markov irregular encoding with list decoding (based on \cite{CoverG:79}), at both the relay and the destination, to characterize sufficient conditions for reliable transmission using a separation-based source-channel code. The works \cite{SalehkalaibarAref:ISIT11} and \cite{SalehkalaibarAref:ISIT12} used block Markov regular encoding with backward decoding, in which the relay partially decodes the sequence transmitted from the transmitter prior to sending both its own source sequence and the cooperation information to the destination.
%

As shown in \cite{Murin:ISIT12}, source-channel separation is suboptimal for general MARCS. Therefore, optimal performance require employing a joint source-channel code.
An important technique for JSCC is the correlation preserving mapping (CPM) technique in which the channel codewords are correlated with the source sequences. CPM was introduced in \cite{Cover:80} in which it was used to obtain single-letter sufficiency conditions for reliable transmission of discrete, memoryless (DM) arbitrarily correlated sources over a MAC. CPM typically enlarges the set of feasible input distribution, thereby enlarging the set of sources which can be reliably transmitted compared to separate source and channel coding.

The CPM technique of \cite{Cover:80} was extended to source coding with side information for MACs in \cite{Ahlswede:83}, to broadcast channels with correlated sources in \cite{HanCosta:87} (with a correction in \cite{KramerNair:09}), and to the transmission of correlated sources over interference channels (ICs) in \cite{SalehiKurtas:93}. However, when the sources are independent, the region obtained from \cite{SalehiKurtas:93} does not specialize to the Han and Kobayashi (HK) region of \cite{HK:81}. Sufficient conditions for reliable transmission, based on the CPM technique, which specialize to the HK region were derived in \cite{LiuChen:2011}. 
The transmission of independent sources over ICs with correlated receiver side information was studied in \cite{GunduzLiu:Rev}, where it was shown that separation is optimal when each receiver has access to side information correlated only with its own desired source. When each receiver has access to side information correlated only with the interfering transmitter's source, \cite{GunduzLiu:Rev} provided sufficient conditions for reliable transmission based on the CPM technique together with the HK superposition encoding and partial interference cancellation.

Although CPM implements JSCC, in \cite{Dueck:81} Dueck observed that the sufficiency conditions derived in \cite{Cover:80} are not necessary. Therefore, in this work, in addition to sufficient conditions, necessary conditions are considered as well.
Observe that the feasible joint distributions of the sources and the respective channel inputs for the MAC (and for the MARC), must satisfy a Markov relationship which reflects the fact that the channel inputs at the transmitters are correlated {\em only via the correlation of the sources}.
In \cite{Cover:80}, in addition to the single-letter sufficient conditions, multi-letter necessary and sufficient conditions, which account for the above constraint, were also presented. 
However, as noted in \cite{Cover:80}, these conditions are based on $n$-letter mutual information expressions, and thereby not computable.
The work \cite{Mitran:11} followed the lines of \cite{Cover:80}, and established necessary conditions for reliable transmission of correlated sources over DM MARCs, which are based on $n$-letter expressions. Furthermore, \cite{Mitran:11} showed that in some cases source-channel separation is optimal and the $n$-letter expressions specialize to single-letter expressions.
In contrast to \cite{Cover:80}, in \cite{Kang:2011} Kang and Ulukus used the above constraint to derive a new set of {\em single-letter} necessary conditions for reliable transmission of correlated sources over a MAC.

\vspace{-0.3cm}
\subsection{Main Contributions}

\vspace{-0.1cm}
This work has a number of important contributions:
\begin{enumerate}
	\item 
		We derive a novel JSCC achievable scheme for MARCs. The scheme uses CPM for encoding information from the sources to {\em both} the relay and the destination. The relay, on the other hand, uses SW source coding\footnote{Throughout this work we refer to separate source-channel coding (i.e., a source code followed by a channel code) as encoding using SW source coding.} for forwarding its cooperation information. Therefore, the sources and the relay send {\em different} types of information to the destination: the sources send source-channel codewords, while the relay sends binning information (SW bin indices). This is in contrast to the schemes of \cite[Thm. 1, Thm. 2]{Murin:ISIT12}, and to \cite{SalehkalaibarAref:ISIT11}, in which the same type of information is sent to the destination from the sources as well as from the relay (either SW bin indices or source-channel codewords). 
		The new scheme uses the DF strategy with successive decoding at the relay and simultaneous backward decoding of both cooperation information and source sequences at the destination.
		This scheme achieves {\em the best known results for all previously characterized} special cases. 
	\item
		We show that, similarly to the capacity analysis for MARCs, also for JSCC simultaneous backward decoding of the cooperation information and source sequences at the destination, outperforms sequential backward decoding at the destination. We also show that simultaneous backward decoding at the destination outperforms the scheme derived in \cite[Thm. 1]{Murin:ISIT12}.
		Additionally, we show that there are cases in which simultaneous backward decoding at the destination strictly outperform the schemes derived in \cite{Murin:ISIT12}. This is proved through an explicit analysis of the error probability for a specific MARC model.
		

	\item
		We derive three new sets of single-letter necessary conditions for reliable transmission of correlated sources over DM MARCs. The first set of conditions is a ``MAC-type" bound, considering the cut around the sources and the relay, while the other two sets are ``broadcast-type" bounds, derived using the cut around the destination and the relay. 
		The new sets of necessary conditions are shown to be at least as tight as previously known conditions, and in some scenarios, the new sets are strictly tighter than known conditions.
\end{enumerate}

The rest of this paper is organized as follows: in Section \ref{sec:Preliminaries} we introduce the notations and the channel model. In Section \ref{sec:prevSchemes} we briefly review the existing schemes and give motivation for a new JSCC scheme. In Section \ref{sec:MixedJointAchiev} we present the new achievability scheme and derive it's corresponding set of sufficiency conditions. In Section \ref{sec:Comparison} a comparison between the existing schemes and the new scheme is presented. Necessary conditions are presented in Section \ref{sec:NecessaryConditions}, and concluding remarks are provided in Section \ref{sec:conclusions}.

\vspace{-0.15cm}
\section{Preliminaries} \label{sec:Preliminaries}

\vspace{-0.2cm}
\subsection{Notations}

\vspace{-0.15cm}
\input{Notations.tex}

\vspace{-0.3cm}
\subsection{System Model} \label{subsec:model}

\begin{figure}[ht]
    \vspace{-0.3cm}
		\centering
		\captionsetup{font=small}
    \scalebox{0.5}{\includegraphics{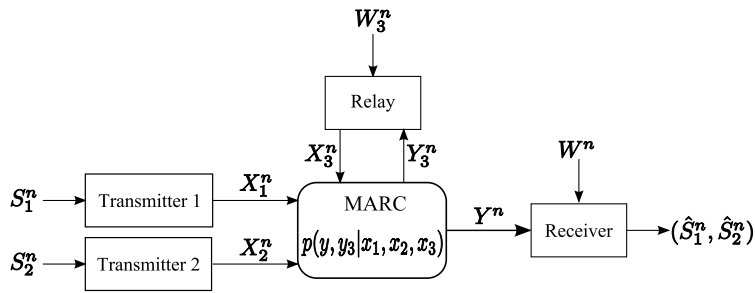}}
    \caption{The multiple-access relay channel with correlated side information.
    $(\hat{S}^n_{1}, \hat{S}^n_{2})$ are the reconstructions at the destination.}
    \label{fig:MARCsideInfo}
		\vspace{-0.65cm}
\end{figure}

\vspace{-0.1cm}
The MARC consists of two transmitters (sources), a receiver (destination) and a relay.
Transmitter $i$ observes the source sequence $S_i^n$, for $i=1,2$.
The receiver is interested in a lossless reconstruction of the source sequences observed by the two transmitters, and the objective of the relay is to help the transmitters and the receiver in reconstructing the source sequences.
The relay and the receiver each observes its own side information, denoted by $W_3^n$ and $W^n$, respectively, correlated with the source sequences.
Figure \ref{fig:MARCsideInfo} depicts the MARC with side information scenario.

The sources and the side information sequences, $\{ S_{1,k},S_{2,k},W_{k},W_{3,k} \}_{k=1}^{n}$, are
arbitrarily correlated at each sample index $k$, according to the joint distribution $p(s_1,s_2,w,w_3)$ defined over a
finite alphabet $\mS_1 \times \mS_2 \times \mW \times \mW_3$, and independent across different sample indices $k$.
This joint distribution is known at all nodes.
For transmission, a DM MARC with inputs $X_i \in \mX_i, i=1,2,3$,
and outputs $Y, Y_3$ over finite output alphabets $\mY,\mY_3$, respectively, is available.
The MARC is causal and memoryless in the sense of \cite{Massey:90}:
\vspace{-0.2cm}
\begin{equation}
    p(y_{k},y_{3,k}|y^{k-1},y_3^{k-1},x_1^k,x_2^k,x_3^k, s_1^n, s_2^n, w_3^n, w^n) = p(y_{k},y_{3,k}|x_{1,k},x_{2,k},x_{3,k}), \quad k=1,2,\dots,n.
\label{eq:MARCchanDist}
\end{equation}

\vspace{-0.3cm}
\begin{MyDefinition}
    \label{def:MABRCcodeDef}
    \textnormal{ A {\em source-channel code} for the MARC with correlated side information consists of two encoding functions at the transmitters,
    \vspace{-0.2cm}
		\begin{equation}
			f_i^{(n)} : \mS_i^n \mapsto \mX_i^n, \quad i=1,2,
		\label{eq:MARC_encFunc}
		\vspace{-0.2cm}
		\end{equation}
a set of causal encoding functions at the relay, $\{ f_{3,k}^{(n)} \}_{k=1}^n$, such that
    \vspace{-0.1cm}
		\begin{equation}
        x_{3,k} = f_{3,k}^{(n)}(y_{3,1}^{k-1},w_{3,1}^n), \quad k=1,2,\dots,n,
    \label{eq:MARC_relayEncFunc}
		\vspace{-0.3cm}
    \end{equation}
and a decoding function at the destination
    \vspace{-0.25cm}
		\begin{eqnarray}
        g^{(n)} &:& \mY^n \times \mW^n \mapsto \mS_1^n \times \mS_2^n.
    \label{eq:MABRC_decFunc}
    \end{eqnarray}
    }    
\end{MyDefinition}

\vspace{-0.15cm}
\begin{MyDefinition}
    \label{def:MARCpErr}
    \textnormal{Let $\hat{S}_i^n, i=1,2$, denote the reconstruction of $S_i^n, i=1,2,$ respectively, at the receiver, i.e., $(\hat{S}_1^n,\hat{S}_2^n) = g^{(n)}(Y^n, W^n)$. The {\em average probability of error}, $P_e^{(n)}$, of a source-channel code for the MARC is defined as:
    \vspace{-0.15cm}
		\begin{eqnarray}
        P_e^{(n)} & \triangleq & \Pr \Big((\hat{S}_1^n,\hat{S}_2^n) \neq (S_1^n,S_2^n) \Big)  .
    \label{eq:MARC_pErr}
		\vspace{-0.15cm}
    \end{eqnarray}}
\end{MyDefinition}

\vspace{-0.15cm}
\begin{MyDefinition}
	\textnormal{
		The sources $S_1$ and $S_2$ can be \emph{reliably transmitted} over the MARC with side information if there exists a sequence of source-channel codes such that $P_e^{(n)} \rightarrow 0$ as $n \rightarrow \infty$.
	}
\end{MyDefinition}

\vspace{-0.4cm}
\subsection{The Primitive Semi-Orthogonal MARC} \label{subsec:PrimitiveSOMARC}

\vspace{-0.1cm}
The DM semi-orthogonal MARC (SOMARC) is a MARC in which the relay-destination link is orthogonal to the channels from the sources to the relay and the destination. Let $Y_R$ denote the signal received at the destination due to the relay channel input $X_3$, and $Y_S$ denote the signal received at the destination due to the transmission of $X_1$ and $X_2$. The conditional distribution function of the SOMARC is:
\vspace{-0.25cm}
\begin{equation}
	p(y_R,y_S,y_3|x_1,x_2,x_3)=p(y_R|x_3)p(y_S,y_3|x_1,x_2).
\label{eq:SOMARC_chain}
\end{equation}

\begin{figure}[ht]
    \vspace{-0.1cm}
    \centering
		\captionsetup{font=small}
    \scalebox{0.50}{\includegraphics{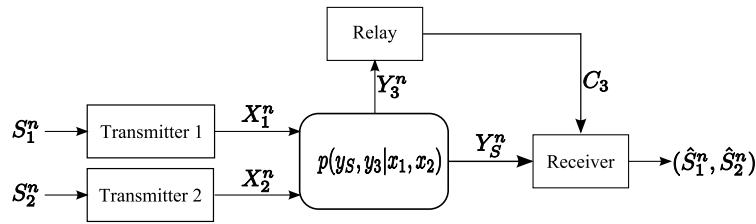}}
    \caption{Primitive semi-orthogonal multiple-access relay channel (PSOMARC).}
    \label{fig:PrimitiveSOMARC}
    \vspace{-0.5cm}
\end{figure}

\vspace{-0.15cm}
A special case of the SOMARC, called the primitive SOMARC (PSOMARC), was considered by Tandon and Poor in \cite{Tandon:CISS:11}. In this channel the relay-destination link $X_3-Y_R$ is replaced with a finite-capacity link whose capacity is $C_3$. This model is depicted in Figure \ref{fig:PrimitiveSOMARC}.
Observe that in the PSOMARC setup there is no side-information at either the relay or destination.

\vspace{-0.4cm}
\subsection{Implementing JSCC via CPM} \label{subsec:CPMTechnique}

\vspace{-0.1cm}
%
	JSCC is implemented via CPM by generating the channel inputs (codewords) statistically dependent with the source sequences, thus, the channel codewords ``preserve" some of the correlation exhibited among the sources. 
	For example, if two sources $(S_1,S_2)$ are to be transmitted over a MAC with channel inputs $(X_1,X_2)$, then the CPM encoded channel codewords are generated according to ${ \prod_{k=1}^{n}{p(x_{1,k}|s_{1,k})}}$. 
	The main benefit of the CPM technique is enlarging the set of possible joint input distributions, thereby improving the performance compared to separately constructing the source code and the channel code. 
	For an illustrative example we refer the reader to the example presented in \cite[pg. 649]{Cover:80}, which demonstrates the sub-optimality of separate source-channel coding, compared to the CPM technique, for the transmission of correlated sources over a DM MAC.

\vspace{-0.2cm}
\section{Previous Schemes and Motivation for a New Scheme} \label{sec:prevSchemes}

Before introducing the new coding scheme we motivate our work by briefly reviewing the two sets of sufficient conditions for reliable transmission of correlated sources over DM MARCs derived in \cite{Murin:ISIT12} and in \cite{Murin:IT11}.

\vspace{-0.2cm}
\subsection{Previously Derived Joint Source-Channel Coding Schemes for DM MARCs} \label{subsec:ExistingSchemes}

\vspace{-0.15cm}
In \cite{Murin:ISIT12} two JSCC schemes for reliable transmission of correlated sources over DM MARCs were derived. The corresponding sufficient conditions are as follows:
\begin{theorem}
    \thmlabel{thm:jointCond}
    (\cite[Thm. 1]{Murin:ISIT12}) A source pair $(S_1,S_2)$ can be reliably transmitted over a DM MARC with relay and receiver side information as defined in Section \ref{subsec:model} if,
    \vspace{-0.2cm}
    \begin{subequations} \label{bnd:Joint}
    \begin{eqnarray}
        H(S_1|S_2,W_3) &<& I(X_1;Y_3|S_2, V_1, X_2, X_3, W_3) \label{bnd:Joint_rly_S1} \\
        H(S_2|S_1,W_3) &<& I(X_2;Y_3|S_1, V_2, X_1, X_3, W_3) \label{bnd:Joint_rly_S2} \\
        H(S_1,S_2|W_3) &<& I(X_1,X_2;Y_3|V_1, V_2, X_3, W_3) \label{bnd:Joint_rly_S1S2} \\
        H(S_1|S_2,W) &<& I(X_1,X_3;Y|S_1, V_2, X_2) \label{bnd:Joint_dst_S1} \\
        H(S_2|S_1,W) &<& I(X_2,X_3;Y|S_2, V_1, X_1) \label{bnd:Joint_dst_S2} \\
        H(S_1,S_2|W) &<& I(X_1,X_2,X_3;Y|S_1,S_2), \label{bnd:Joint_dst_S1S2}
    \end{eqnarray}
    \end{subequations}
		
		\vspace{-0.15cm}
    \noindent are satisfied for some joint distribution that factorizes as:
    \vspace{-0.2cm}
    \begin{align}
        & p(s_1,s_2,w_3,w) p(v_1) p(x_1|s_1,v_1) p(v_2) p(x_2|s_2,v_2) p(x_3|v_1,v_2) p(y_3,y|x_1,x_2,x_3).
    \label{eq:JntJointDist}
		\vspace{-0.2cm}
    \end{align}
\end{theorem}

\begin{theorem}
    \thmlabel{thm:jointCondFlip}
    (\cite[Thm. 2]{Murin:ISIT12}) A source pair $(S_1,S_2)$ can be reliably transmitted over a DM MARC with relay and receiver side information as defined in Section \ref{subsec:model} if,
    \vspace{-0.2cm}
    \begin{subequations} \label{bnd:JointFlip}
    \begin{eqnarray}
        H(S_1|S_2,W_3) &<& I(X_1;Y_3|S_1, X_2, X_3) \label{bnd:JointFlip_rly_S1} \\
        H(S_2|S_1,W_3) &<& I(X_2;Y_3|S_2, X_1, X_3) \label{bnd:JointFlip_rly_S2} \\
        H(S_1,S_2|W_3) &<& I(X_1,X_2;Y_3|S_1, S_2, X_3 ) \label{bnd:JointFlip_rly_S1S2} \\
        H(S_1|S_2,W) &<& I(X_1,X_3;Y|S_2, X_2, W) \label{bnd:JointFlip_dst_S1} \\
        H(S_2|S_1,W) &<& I(X_2,X_3;Y|S_1, X_1, W) \label{bnd:JointFlip_dst_S2} \\
        H(S_1,S_2|W) &<& I(X_1,X_2,X_3;Y| W), \label{bnd:JointFlip_dst_S1S2}
    \end{eqnarray}
    \end{subequations}

		\vspace{-0.15cm}
    \noindent are satisfied for some joint distribution that factorizes as:
    \vspace{-0.2cm}
    \begin{align}
        & p(s_1,s_2,w_3,w) p(x_1|s_1)p(x_2|s_2) p(x_3|s_1,s_2) p(y_3,y|x_1,x_2,x_3).
    \label{eq:JntFlipJointDist}
    \vspace{-0.2cm}
		\end{align}
		\vspace{-0.3cm}
\end{theorem} 

\vspace{-0.5cm}
\begin{remark} \label{rem:oldSchemesComp}

	\Thmref{thm:jointCond} and \Thmref{thm:jointCondFlip} differ in both the decoding constraints and the admissible joint distribution chains, i.e., \eqref{eq:JntJointDist} and \eqref{eq:JntFlipJointDist}. 
	The main difference between \Thmref{thm:jointCond} and \Thmref{thm:jointCondFlip} is the target nodes for CPM and SW coding: 
	In \Thmref{thm:jointCond}, CPM is used for encoding information from the transmitters to the relay and SW coding is used for encoding information cooperatively from the transmitters and the relay to the destination. 
	Thus, in \Thmref{thm:jointCond} the cooperation between the relay and the transmitters is based on the binning information.
	The RVs $V_1$ and $V_2$ in \Thmref{thm:jointCond} carry the bin indices of the SW source code.
	In \Thmref{thm:jointCondFlip}, SW coding is used for encoding information from the transmitters to the relay and CPM is used for cooperatively encoding information to the destination. 
	Thus, in \Thmref{thm:jointCondFlip} the cooperation between the transmitters and the relay is based on the sources $S_1$ and $S_2$.
			
	Recall that in \cite{Cover:80} it was shown that separate source and channel coding is generally suboptimal for transmitting correlated sources over MACs. Thus, it follows that the relay decoding constraints of \Thmref{thm:jointCond} are generally looser compared to the relay decoding constraints of \Thmref{thm:jointCondFlip}. Using similar reasoning we conclude that the destination decoding constraints of \Thmref{thm:jointCondFlip} are looser compared to the destination decoding constraints of \Thmref{thm:jointCond} (as long as coordination is possible, see \cite[Remark 18]{Murin:IT11}).
	
\end{remark}

\begin{remark}
	The work \cite{SalehkalaibarAref:ISIT11} considered JSCC for the relay channel, in which one of the sources is available at the transmitter while the other is known at the relay. 
	The authors presented a transmission scheme similar to \Thmref{thm:jointCondFlip}, where CPM is utilized to transmit the sources from the transmitters to the destination while the relay applies binning for cooperation.
\end{remark}

\begin{remark} \label{rem:MABRC}
		In the multiple-access broadcast relay channel (MABRC) \cite{Murin:IT11}, the relay also wants to reconstruct the sources in a lossless fashion. This channel model is depicted in Figure \ref{fig:MABRCsideInfo}. As both \Thmref{thm:jointCond} and \Thmref{thm:jointCondFlip} use the DF protocol, the conditions of \Thmref{thm:jointCond} and \Thmref{thm:jointCondFlip} are also sufficient conditions for reliable transmission over the MABRC. 
		\vspace{-0.15cm}
		\begin{figure}[ht]
    \centering
		\captionsetup{font=small}
    \scalebox{0.49}{\includegraphics{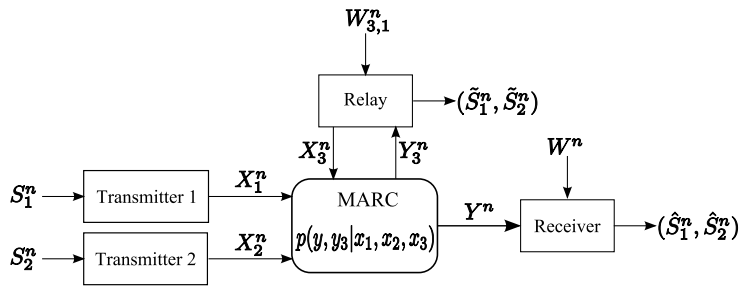}}
    \vspace{-0.05cm}
    \caption{The multiple-access broadcast relay channel with correlated side information.
    $(\tilde{S}^n_{1}, \tilde{S}^n_{2})$ are the reconstructions at the relay, and $(\hat{S}^n_{1}, \hat{S}^n_{2})$ are the reconstructions at the destination.}
    \label{fig:MABRCsideInfo}
    \vspace{-0.6cm}
		\end{figure}
    
\end{remark}

\vspace{-0.55cm}
\subsection{The Motivation for a New JSCC Scheme} \label{subsec:motivation}

\vspace{-0.1cm}
	{\bf Motivating observation 1:}
	As stated in Remark \ref{rem:oldSchemesComp}, the achievability schemes of \Thmref{thm:jointCond} and \Thmref{thm:jointCondFlip} use different combinations of the CPM technique with a SW source code paired with a channel code. 
	The achievability scheme of \Thmref{thm:jointCond} uses SW source coding for cooperatively encoding information from the transmitters and the relay to the destination while CPM is used for encoding information from the transmitters to the relay. In \Thmref{thm:jointCondFlip}, CPM is used for cooperatively encoding information from the transmitters and the relay to the destination while SW source coding is used for encoding information from the transmitters to the relay. 
	Since CPM can generally support the transmission of sources with higher entropies compared to separate source-channel coding, a natural question that arises is {\em whether the CPM technique can be used for simultaneously encoding information to both the relay and the destination}.

	{\bf Motivating observation 2:}
	It was observed in \cite{CoverG:79} that for the relay channel, when decoding at the relay does not constrain the rate, DF as implemented in \cite[Thm. 1]{CoverG:79} is capacity achieving . It follows that cooperation based on binning is optimal in this case.\footnote{We note that in the channel coding problem for the relay channel, other schemes, e.g. the regular encoding schemes of \cite{Carleial:82}, \cite{Willems:82}, achieve the DF-rate without binning, but these schemes are not directly applicable for this scenario, see also \cite{Murin:IT11}.} This raises the question {\em whether it is possible to construct a scheme that combines CPM from the sources to the destination with binning from the relay to the destination, and how does such a scheme compare with \Thmref{thm:jointCond} and \Thmref{thm:jointCondFlip}}.

\begin{figure}[h]
    \vspace{-0.35cm}
    \centering
		\captionsetup{font=small}
    \scalebox{0.49}{\includegraphics{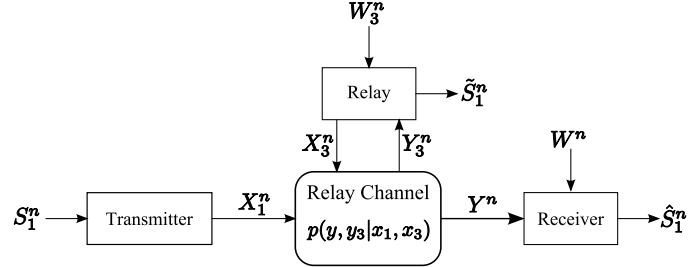}}
    \vspace{-0.1cm}
    \caption{The cooperative relay broadcast channel. $\tilde{S}_1^n$ and $\hat{S}_1^n$ are the reconstructions of the source sequence, $S_1^n$, at the relay and the destination, respectively.}
    \label{fig:CBRCsideInfo}
    \vspace{-0.5cm}
\end{figure}

	{\bf Motivating observation 3:} The cooperative relay-broadcast channel (CRBC) model is a special case of the MABRC obtained by setting $\mS_2 \mspace{-4mu} = \mspace{-4mu} \mX_2 \mspace{-4mu} = \mspace{-4mu} \phi$, such that there is a single transmitter \cite{ErkipGunduz:07}. 
Figure \ref{fig:CBRCsideInfo} depicts the CRBC model.

\noindent For this channel model \cite{ErkipGunduz:07} presented the following necessary and sufficient conditions:
\begin{MyProposition} (\cite[Thm. 3.1]{ErkipGunduz:07})
   A source $S_1$ can be reliably transmitted over a DM CRBC with relay and receiver side information if:
    \vspace{-0.25cm}
		\begin{subequations} \label{eq:relayJoint}
        \begin{eqnarray}
            H(S_1|W_3) &<& I(X_1;Y_3|X_3) \label{eq:RelayJnt_cond} \\
            H(S_1|W) &<& 	 I(X_1,X_3;Y), \label{eq:DestJnt_cond}
        \end{eqnarray}
    \end{subequations}
   
	\vspace{-0.15cm}
\noindent	for some input distribution $p(s_1, w_3, w)p(x_1, x_3)$. Conversely, if a source $S_1$ can be reliably transmitted over the CRBC then the conditions in \eqref{eq:RelayJnt_cond} and \eqref{eq:DestJnt_cond} are satisfied with $<$ replaced by $\leq$ for some input distribution $p(s_1, w_3, w)p(x_1, x_3)$.
\end{MyProposition}

	In \cite[Remark 6]{Murin:ISIT12} it is shown that for a CRBC, the conditions of  \Thmref{thm:jointCond} can be specialized to the conditions of \cite[Thm. 3.1]{ErkipGunduz:07}, while the conditions obtained from \Thmref{thm:jointCondFlip} are generally more restrictive. The reason is that when specializing \Thmref{thm:jointCondFlip} to the case of a single transmitter, the set of joint distributions of the source and relay channel inputs which satisfy \eqref{eq:JntFlipJointDist} does not exhaust the entire space of joint distributions, and in particular, does not include the optimal distribution according to \cite[Thm. 3.1]{ErkipGunduz:07}. We conclude that the downside of using CPM for encoding information to the destination, as implemented in \Thmref{thm:jointCondFlip}, is that it restricts the set of admissible joint distributions; thereby constrains the achievable coordination between the sources and the relay when cooperating to send information to the destination. This leads to the question {\em whether it is possible to construct a scheme in which CPM is used for encoding information to the destination, while the constraints on the source-relay coordination imposed by the distribution chain \eqref{eq:JntFlipJointDist} are relaxed or entirely removed}.
	
\noindent In the next section a new JSCC scheme is derived which gives affirmative answers to the above three questions.

\vspace{-0.1cm}
\section{A New Joint Source-Channel Coding Scheme} \label{sec:MixedJointAchiev}

	\vspace{-0.1cm}
	We now present a new set of sufficient conditions for reliable transmission of correlated sources over DM MARCs with side information. The achievability scheme (\Thmref{thm:jointCond_NewSimult}) is based on DF at the relay, and uses CPM for encoding information to {\em both} the relay and the destination and successive decoding at the relay. Cooperation in the new scheme is based on {\em binning implemented via SW source coding.}
    The decoding method applied at the destination in the new scheme is simultaneous backward decoding of the cooperation information and the transmitted source sequences.
		By combining cooperation based on binning with CPM for encoding information to the destination, the constraints on the distribution chain imposed by the scheme of \Thmref{thm:jointCondFlip} are removed.
		    %
	
\begin{figure}[ht]
\vspace{-1.0cm}
\begin{center}
\captionsetup{font=small}
\subfloat[]{\scalebox{0.31}{\includegraphics{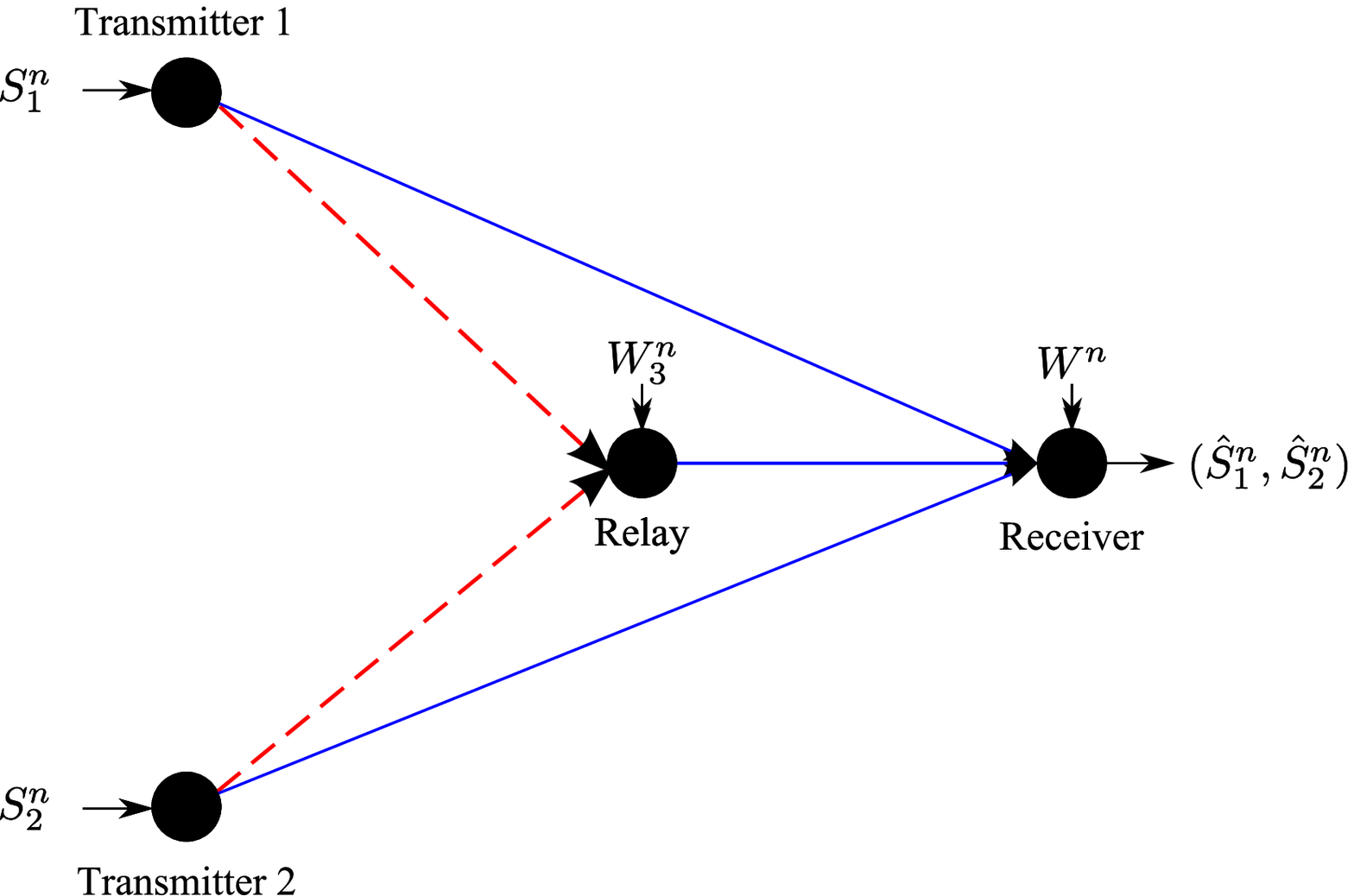}} \label{fig:cpmRelay}}
\subfloat[]{\scalebox{0.31}{\includegraphics{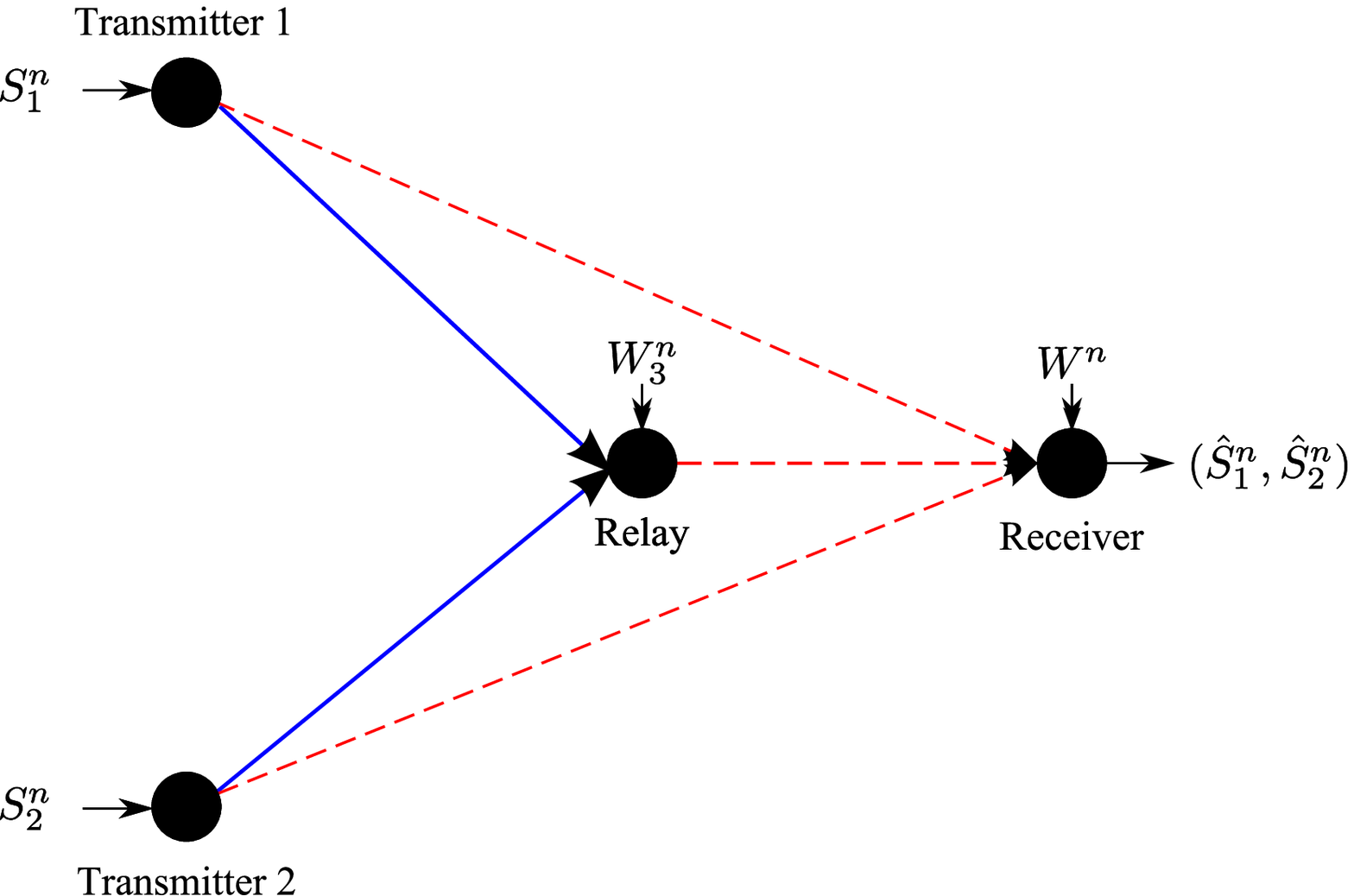}} \label{fig:cpmDest}}
\subfloat[]{\scalebox{0.31}{\includegraphics{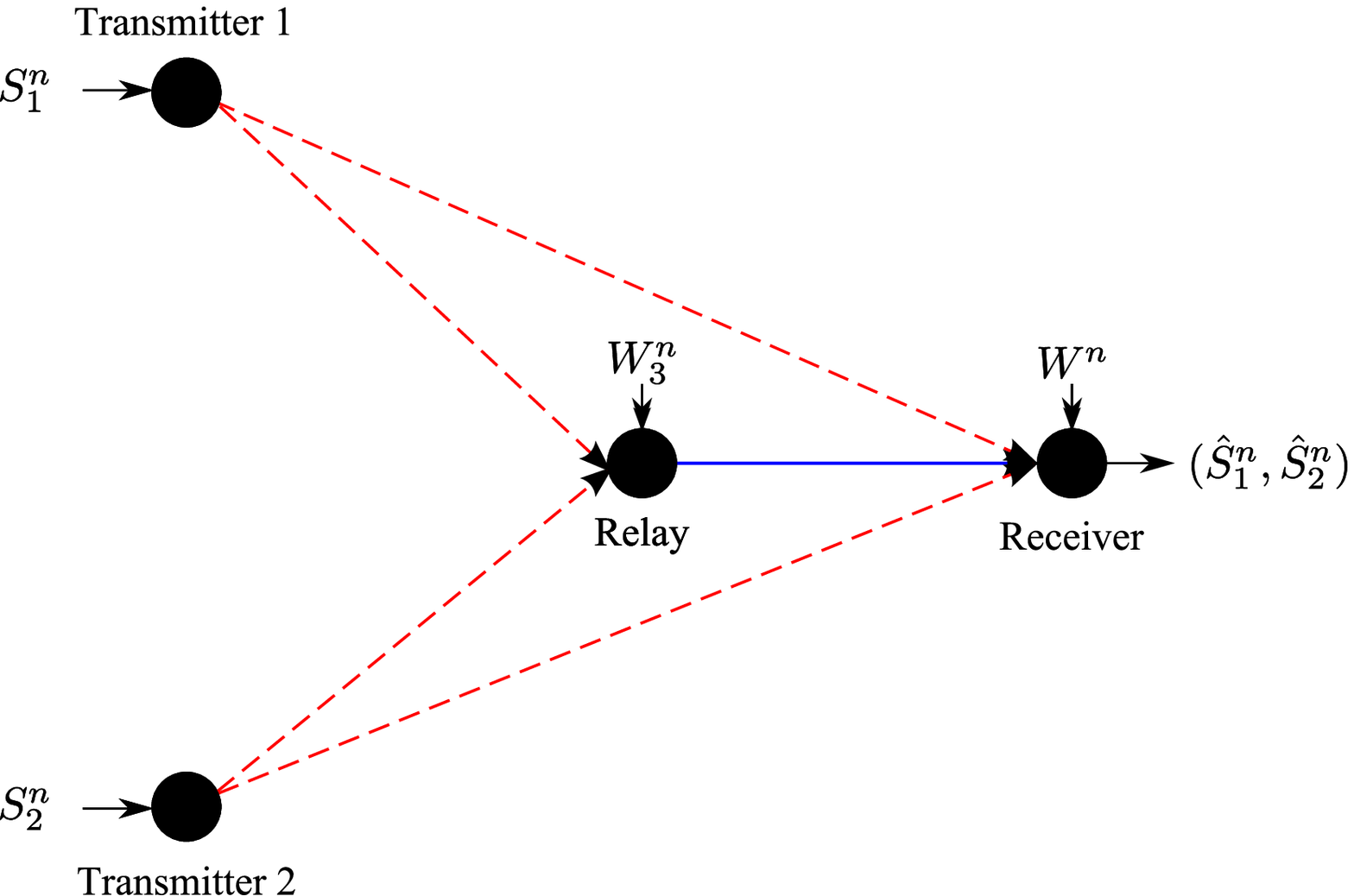}} \label{fig:cpmNew}}
\vspace{-0.15cm}
\caption{Types of information sent to the destination in the schemes of (a) \Thmref{thm:jointCond}; (b) \Thmref{thm:jointCondFlip}; and (c) the new proposed scheme of \Thmref{thm:jointCond_NewSimult}. Solid arrows indicate bin indices, while dashed arrows indicate source-channel codewords.}
\vspace{-1.2cm}
\label{fig:cpm_binning_combinations}
\end{center}
\end{figure}

	Note that in the schemes implemented in \Thmref{thm:jointCond} and in \Thmref{thm:jointCondFlip} the same type of information is sent to the destination from both the relay and from the sources, while in the new scheme implemented in \Thmref{thm:jointCond_NewSimult} {\em different types} of information are sent to the destination from the relay and from the sources. This is illustrated in Figure \ref{fig:cpm_binning_combinations}. It can be observed that in \Thmref{thm:jointCond} (Figure \ref{fig:cpmRelay}) both the relay and the sources send bin indices to the destination, while in \Thmref{thm:jointCondFlip} (Figure \ref{fig:cpmDest}) both the relay and the sources send source-channel codewords. However, this is not the case in \Thmref{thm:jointCond_NewSimult} (Figure \ref{fig:cpmNew}), in which the relay sends bin indices while the sources send source-channel codewords.

	
\vspace{-0.25cm}
\subsection{Sufficient Conditions for Simultaneous Backward Decoding at the Destination} \label{subsec:newScheme}

\vspace{-0.1cm}
Using simultaneous backward decoding the following sufficient conditions are obtained:

\begin{theorem}
    \thmlabel{thm:jointCond_NewSimult}
    A source pair $(S_1,S_2)$ can be reliably transmitted over a DM MARC with relay and receiver side information as defined in Section \ref{subsec:model} if the conditions
    \vspace{-0.2cm}
		\begin{subequations} \label{bnd:JointNewSimult}
    \begin{align}
        H(S_1|S_2,W_3) &< I(X_1;Y_3|S_2, V_1, X_2, X_3, W_3) \label{bnd:JointNewSimult_rly_S1} \\
        H(S_2|S_1,W_3) &< I(X_2;Y_3|S_1, V_2, X_1, X_3, W_3) \label{bnd:JointNewSimult_rly_S2} \\
        H(S_1,S_2|W_3) &< I(X_1,X_2;Y_3| V_1, V_2, X_3, W_3) \label{bnd:JointNewSimult_rly_S1S2} \\
				H(S_1|S_2,W) &< \min \Big\{I(X_1,X_3;Y|S_2,V_2,X_2,W), \nonumber \\
				& \qquad \qquad I(X_1,X_3;Y|S_1,V_2,X_2) + I(X_1;Y|S_2,V_1,X_2,X_3,W) \Big\} \label{bnd:JointNewSimult_dst_S1}
		\end{align}
		\begin{align}
				H(S_2|S_1,W) &< \min \Big\{I(X_2,X_3;Y|S_1,V_1,X_1,W), \nonumber \\
				& \qquad \qquad I(X_2,X_3;Y|S_2,V_1,X_1) + I(X_2;Y|S_1,V_2,X_1,X_3,W) \Big\} \label{bnd:JointNewSimult_dst_S2} \\
				H(S_1,S_2|W) &< I(X_1,X_2,X_3;Y|W), \label{bnd:JointNewSimult_dst_S1S2} 
		\end{align}
    \end{subequations}

		\vspace{-0.2cm}
    \noindent are satisfied for some joint distribution that factorizes as"
    \vspace{-0.25cm}
		\begin{align}
        & p(s_1,s_2,w_3,w)p(v_1)p(x_1|s_1,v_1) p(v_2)p(x_2|s_2,v_2)p(x_3|v_1,v_2)p(y_3,y|x_1,x_2,x_3). 
    \label{eq:JntNewJointDistSimult}
		    \end{align}
\end{theorem}

\vspace{-0.2cm}
\begin{proof}
    The proof is given in Appendix \ref{annex:jointNewSimultProof}.
\end{proof}  

\vspace{-0.3cm}
\subsection{Discussion} \label{subsec:jointDiscussion}

\vspace{-0.15cm}
\begin{remark} \label{rem:SameDist}
    The achievability schemes of \Thmref{thm:jointCond} and \Thmref{thm:jointCond_NewSimult} require the same joint distribution (cf. equations \eqref{eq:JntJointDist} and \eqref{eq:JntNewJointDistSimult}). 

\end{remark}

\vspace{-0.15cm}
\begin{remark} \label{rem:decConstraints}
    Conditions \eqref{bnd:JointNewSimult_rly_S1}--\eqref{bnd:JointNewSimult_rly_S1S2} in \Thmref{thm:jointCond_NewSimult} are constraints due to decoding at the relay, while conditions \eqref{bnd:JointNewSimult_dst_S1}--\eqref{bnd:JointNewSimult_dst_S1S2} are decoding constraints at the destination. 
    Note that the decoding constraints at the relay in \Thmref{thm:jointCond_NewSimult} are identical to \eqref{bnd:Joint_rly_S1}--\eqref{bnd:Joint_rly_S1S2} in \Thmref{thm:jointCond}.
\end{remark}

\begin{remark}
		Note that as \Thmref{thm:jointCond_NewSimult} uses the DF scheme, the conditions of \Thmref{thm:jointCond_NewSimult} are also sufficient conditions for reliable transmission over the MABRC.
\end{remark}

    %

\begin{remark} \label{rem:ExpressionsMeaning}
    In \Thmref{thm:jointCond_NewSimult}, $V_1^n$ and $V_2^n$ represent the binning information for $S_1^n$ and $S_2^n$, respectively.
    Consider \Thmref{thm:jointCond_NewSimult} which uses simultaneous backward decoding: condition \eqref{bnd:JointNewSimult_dst_S1} can be written as follows:
    \vspace{-0.15cm}
		\begin{align}
        H(S_1|S_2,W) &< I(X_1;Y|S_2,V_1,X_2,X_3,W) + \nonumber \\
        & \qquad \qquad \min \big\{ I(V_1, X_3;Y|S_2,V_2,X_2,W),I(X_1,X_3;Y|S_1,V_2,X_2)  \big\}.
    \label{eq:SimultIndivInterp}
    \end{align}

\vspace{-0.15cm}
\noindent On the right-hand side (RHS) of \eqref{eq:SimultIndivInterp}, the mutual information expression $I(X_1;Y|S_2,V_1,X_2,X_3,W)$ represents the available rate for encoding information on the {\em source sequence} $S_1^n$, in excess of the bin index conveyed by the sequence $V_1^n$. This is because $S_2$, $V_1$, $X_2$, $X_3$ and $W$ are known. 
The expression $I(V_1, X_3;Y|S_2,V_2,X_2,W)$ represents the rate of binning information on $S_1$ that can be utilized at the destination.
Also the expression $I(X_1,X_3;Y|S_1,V_2,X_2)$, as $S_1$ and $V_2$ are known, represents the rate for sending the bin index of the source sequence $S_1$, cooperatively from Transmitter 1 and the relay to the destination.
The reason for the two possible binning rates is that $I(V_1, X_3;Y|S_2,V_2,X_2,W)$ represents the maximal rate increase that can be achieved due to the binning information available on the current message in the backward decoding scheme, while $I(X_1,X_3;Y|S_1,V_2,X_2)$ represents the maximal rate for decoding the binning information for the next step in the backward decoding scheme. 
Therefore, decoding via simultaneous backward decoding results in two constraints on the binning rate.
\end{remark}

\begin{remark} \label{rem:MAC}
  \Thmref{thm:jointCond_NewSimult} can be specialized to the MAC with correlated sources by letting $\mV_1= \mV_2= \mX_3= \mW=\phi$. For this setting the conditions \eqref{bnd:JointNewSimult_dst_S1}--\eqref{bnd:JointNewSimult_dst_S1S2} specialize to the ones in \cite[Eqn. (12)]{Cover:80} with $Y$ as the destination.
  Similarly, the MABRC, under $\mV_1= \mV_2= \mX_3= \mW_3= \mW=\phi$, specializes to the compound MAC \cite[Section VI]{GunduzErkip:09}, and \Thmref{thm:jointCond_NewSimult} specializes to \cite[Thm. 6.1]{GunduzErkip:09}.
  We conclude that \Thmref{thm:jointCond_NewSimult} implements a {\em CPM encoding for both the relay and the destination}. This is in contrast to the previous results of \Thmref{thm:jointCond} and \Thmref{thm:jointCondFlip} in which CPM is used for encoding information {\em either} to the relay {\em or} to the destination.
\end{remark}

\begin{remark} \label{rem:CRBC}
    The CRBC model with correlated relay and destination side information can be obtained as a special case of the MABRC model by letting $\mX_2=\mS_2=\phi$. The sufficient conditions for the CRBC given in \cite[Thm. 3.1]{ErkipGunduz:07} can also be obtained from \Thmref{thm:jointCond_NewSimult} by letting $V_1=X_3$, $\mathcal{S}_2 =\mathcal{X}_2 =\mathcal{V}_2=\phi$, and considering an input distribution independent of the sources.
    This is in contrast to \Thmref{thm:jointCondFlip} which specializes to more restrictive conditions (see Subsection \ref{subsec:motivation}).
    We conclude that \Thmref{thm:jointCond_NewSimult} allows more flexibility in the achievable coordination between the sources and the relay compared to \Thmref{thm:jointCondFlip}.
 \end{remark}

\begin{remark}
	Using successive backward decoding at the destination the following sufficient conditions are obtained:

\vspace{-0.15cm}	
\begin{MyProposition} \label{prop:jointCond_New}
    A source pair $(S_1,S_2)$ can be transmitted reliably over a DM MARC with relay and receiver side information as defined in Section \ref{subsec:model} if,
    \vspace{-0.15cm}
		\begin{subequations} \label{bnd:JointNew}
    \begin{align}
        H(S_1|S_2,W_3) &< I(X_1;Y_3|S_2, V_1, X_2, X_3, W_3) \label{bnd:JointNew_rly_S1} \\
        H(S_2|S_1,W_3) &< I(X_2;Y_3|S_1, V_2, X_1, X_3, W_3) \label{bnd:JointNew_rly_S2} \\
        H(S_1,S_2|W_3) &< I(X_1,X_2;Y_3| V_1, V_2, X_3, W_3) \label{bnd:JointNew_rly_S1S2} \\
        H(S_1|S_2,W) & < I(X_1;Y|S_2,V_1,X_2,X_3, W) + I(V_1,X_3;Y|V_2,W) \label{bnd:JointNew_dst_S1} \\
        H(S_2|S_1,W) & < I(X_2;Y|S_1,V_2,X_1,X_3,W) + I(V_2,X_3;Y|V_1,W) \label{bnd:JointNew_dst_S2} \\
        H(S_1,S_2|W) & < I(X_1,X_2;Y|V_1, V_2, X_3,W) + I(V_1,V_2,X_3;Y|W), \label{bnd:JointNew_dst_S1S2}
    \end{align}
    \end{subequations}

		\vspace{-0.2cm}
    \noindent are satisfied for some joint distribution that factorizes as:
    \vspace{-0.25cm}
		\begin{align}
        & p(s_1,s_2,w_3,w)p(v_1)p(x_1|s_1,v_1) p(v_2)p(x_2|s_2,v_2)p(x_3|v_1,v_2)p(y_3,y|x_1,x_2,x_3).
    \label{eq:JntNewJointDist}
		\end{align}
\end{MyProposition}

\vspace{-0.2cm}
\begin{proof}
   The proof is given in Appendix \ref{annex:jointNewProof}.
\end{proof}

\end{remark}

\begin{remark}
		As the scheme of \Thmref{thm:jointCond_NewSimult} applies simultaneous backward decoding at the destination, then the source vectors and the binning information are {\em jointly} decoded (see Appendix \ref{subsec:jointNewProofDecoding}). On the other hand, the scheme of Prop. \ref{prop:jointCond_New} applies successive backward decoding at the destination, thus, first the binning information is decoded, and then, the source vectors are decoded (see Appendix \ref{annex:jointNewProof_decoding}). 
		Since in the latter scheme decoding the binning information uses only part of the available information, the sufficient conditions obtained for the scheme of Prop. \ref{prop:jointCond_New} are more restrictive than those obtained for the scheme of \Thmref{thm:jointCond_NewSimult} This is rigorously shown in the following section.
		
\end{remark}

\vspace{-0.3cm}
\section{Comparison of the Different Achievability Schemes} \label{sec:Comparison}

	\vspace{-0.1cm}
	We now present a detailed comparison of the sufficient conditions established by \Thmref{thm:jointCond_NewSimult}, \Thmref{thm:jointCond}, \Thmref{thm:jointCondFlip} and Prop. \ref{prop:jointCond_New}. 
	Specifically, we show the following:
	\begin{itemize}
		\item 
			In Subsection \ref{subsec:CompareCorrSources} we show that for correlated sources and side information the scheme of \Thmref{thm:jointCond_NewSimult} outperforms the schemes of \Thmref{thm:jointCond} and Prop. \ref{prop:jointCond_New}. 
			%
		
		\item In Subsection \ref{subsec:ExampMixedOutperform} we show that there are scenarios for which the scheme of \Thmref{thm:jointCond_NewSimult} strictly outperforms the schemes of \Thmref{thm:jointCond} and \Thmref{thm:jointCondFlip}.
	\end{itemize}

\vspace{-0.4cm}	
\subsection{Correlated Sources and Side Information} \label{subsec:CompareCorrSources}

   
	\vspace{-0.15cm}
  We now compare \Thmref{thm:jointCond}, \Thmref{thm:jointCond_NewSimult} and Prop. \ref{prop:jointCond_New} for the general input distributions \eqref{eq:JntJointDist}, \eqref{eq:JntNewJointDistSimult} and \eqref{eq:JntNewJointDist}. As stated in Remark \ref{rem:decConstraints}, the decoding constraints at the relay in \Thmref{thm:jointCond_NewSimult} are identical to the decoding constraints at the relay in \Thmref{thm:jointCond} and Prop. \ref{prop:jointCond_New}. 
	Therefore, in the following we compare only the decoding constraints at the destination. The conclusion is summarized in the following proposition:


\begin{MyProposition} \label{prop:CorrSources}
		The scheme of \Thmref{thm:jointCond_NewSimult} is at least as good as the schemes of \Thmref{thm:jointCond} and Prop. \ref{prop:jointCond_New}.
	\end{MyProposition}
	
	\vspace{-0.2cm}
	\begin{proof}	
		The proof is given in Appendix \ref{annex:ProofPropCorrSources}.		
	\end{proof}
	
	\begin{remark}
		We emphasize that Prop. \ref{prop:CorrSources} implies that the superiority of the scheme of \Thmref{thm:jointCond_NewSimult} over the scheme of \Thmref{thm:jointCond} and the scheme of Prop. \ref{prop:jointCond_New} holds in general.
	\end{remark}
	
	Proposition \ref{prop:CorrSources} implies that 
	for JSCC for MARCs, simultaneous backward decoding outperforms sequential backward decoding.
	For the case of separate source and channel codes, \cite[Thm. 1]{Murin:IT11} presented a separation-based achievability scheme subject to the input distribution:
  \vspace{-0.25cm}
  \begin{eqnarray}
    p(s_1,s_2,w_3,w,v_1,v_2,x_1,x_2,x_3) = p(s_1,s_2,w_3,w)p(v_1)p(x_1|v_1)p(v_2)p(x_2|v_2)p(x_3|v_1,v_2).
    \label{eq:sepDist}
  \end{eqnarray}
	
	In this case, we have $p(x_i|s_i,v_i)=p(x_i|v_i), i=1,2$, the joint distributions in \eqref{eq:JntJointDist} and \eqref{eq:JntNewJointDistSimult} specialize to the one in \eqref{eq:sepDist}, and the sufficient conditions of \Thmref{thm:jointCond} and \Thmref{thm:jointCond_NewSimult} specialize to the conditions of \cite[Thm. 1]{Murin:IT11}.

\begin{remark}
When the source and side information sequences are independent, that is $p(s_1,s_2,w_3,w)=$ $p(s_1)p(s_2)$ $p(w_3)p(w)$, the joint distributions in \eqref{eq:JntNewJointDistSimult} and \eqref{eq:JntNewJointDist} specialize to $p(s_1)p(s_2)p(w_3)p(w)p(v_1)p(x_1|v_1)p(v_2)$ $p(x_2|v_2)$ $p(x_3|v_1,v_2)$.
    %
\noindent In this case, the conditions of Prop. \ref{prop:jointCond_New} specialize to the conditions obtained for sending independent messages over the MARC using sliding-window decoding at the destination \cite[Section III.B]{Sankar:07}, while the conditions of \Thmref{thm:jointCond_NewSimult} specialize to the conditions obtained for sending independent messages over the MARC using backward decoding at the destination \cite[Section III.A]{Sankar:07}.\footnote{The same observation holds when the side information is not present. This follows since when the side information is independent of the sources then it cannot help in decoding the sources. Thus, we can set $\mW = \mW_3 = \phi$.}
\end{remark}
\vspace{-0.35cm}
\subsection{Mixed JSCC Can Strictly Outperform the Schemes of \Thmref{thm:jointCond} and \Thmref{thm:jointCondFlip}} \label{subsec:ExampMixedOutperform}

\vspace{-0.1cm}
Recall Remark \ref{rem:SameDist}, which states that the underlying input distributions of \Thmref{thm:jointCond_NewSimult} and \Thmref{thm:jointCond} are identical, while the underlying input distribution for \Thmref{thm:jointCondFlip} is different. Here, we present a comparison of all three schemes for a special case in which the two input distribution chains are the same.
In this example the sources can be reliably transmitted by using the scheme of \Thmref{thm:jointCond_NewSimult}, while reliable transmission is not possible via the schemes of \Thmref{thm:jointCond} and \Thmref{thm:jointCondFlip}.
%
%
	Consider a PSOMARC, defined by $\mX_1 = \mX_2 = \{0,1\}, \mY_3 = \{0,1,2\}, \mY_S = \{0,1\}$.
%
\noindent Let $C_3 = 1$, and consider the deterministic channel mapping $(X_1,X_2) \mapsto (Y_3,Y_S)$ specified in Table \ref{tab:PrimitiveSOMARC}.

	\vspace{-0.25cm}
	\begin{table}[h]
  \begin{center}
  \begin{tabular}[t]{|c|c|c|c|c|}
  	\hline
  	$(X_1, X_2)$ & $(0,0)$ & $(0,1)$ & $(1,0)$ & $(1,1)$ \\
  	\hline 
  	$ Y_3$ & 0 & 1 & 1 & 2 \\
  	\hline
  	$ Y_S$ & 0 & 0 & 1 & 1 \\
  	\hline
   \end{tabular}
  \vspace{-0.1cm}
	\captionsetup{font=small}
  \caption{A deterministic channel mapping $(X_1,X_2) \mapsto (Y_3,Y_S)$ for the PSOMARC. \label{tab:PrimitiveSOMARC}}
  \vspace{-0.8cm}
  \end{center}
  \end{table}

\vspace{-0.1cm}
\noindent The sources $(S_1,S_2)$ are defined over the sets $\mS_1=\mS_2=\{0,1\}$ with the joint distribution specified in Table \ref{tab:SourceDist}. 
	\begin{table}[h!]
  \begin{center}
	\begin{tabular}[t]{|c|c|c|}
	\hline
	$S_1$ \textbackslash $S_2$ & 0 & 1 \\
	\hline
	0 & 1/3 & 1/3 \\
	\hline
	1 & 0 & 1/3 \\
	\hline
	\end{tabular}
	\vspace{-0.1cm}
	\captionsetup{font=small}
	\caption{The joint distribution of $(S_1,S_2)$. The entry in the $j^{\text{th}}$ row and $m^{\text{th}}$ column, $j,m = 0,1$, corresponds to $\Pr \left((S_1,S_2) = (j,m) \right)$. \label{tab:SourceDist}}
	\vspace{-0.8cm}
  \end{center}
  \end{table}

\vspace{-0.2cm}
These sources can be reliably transmitted by letting $X_1 = S_1$ and $X_2 = S_2$. The probability of decoding error at the relay is zero since there is a one-to-one mapping between the channel inputs from the sources and the channel output at the relay. The probability of decoding error at the destination
can be made arbitrarily small by using the fact that each channel output at the destination corresponds only to two possible pairs of channel inputs. This ambiguity can be resolved using the relay-destination link whose capacity is 1 bit per channel use.
	Next, consider the transmission 
	via the schemes of \Thmref{thm:jointCond}, \Thmref{thm:jointCondFlip} and \Thmref{thm:jointCond_NewSimult}. 
	%
	%
	For transmission via the schemes of \Thmref{thm:jointCond} and \Thmref{thm:jointCondFlip} we have the following proposition:
		\begin{MyProposition} \label{prop:NotFeasible}
   The sources defined in Table \ref{tab:SourceDist} {\em cannot} be reliably transmitted over the PSOMARC defined in Table \ref{tab:PrimitiveSOMARC}, by using the schemes of \Thmref{thm:jointCond} and \Thmref{thm:jointCondFlip}.
\end{MyProposition}

\begin{IEEEproof}
	First we make the following claim:
	\begin{claim} \label{clm:lowerPerr}
		If an inequality sign in the conditions of \Thmref{thm:jointCond} and \Thmref{thm:jointCondFlip} is {\em reversed}, then reliable transmission is not possible with the corresponding schemes.
	\end{claim}
	
	\noindent {\em Proof sketch:} The average probability of error for decoding the sources transmitted via the scheme of \Thmref{thm:jointCond} can be {\em lower bounded} by using the properties of jointly typical sequences, \cite[Ch. 6.3]{YeungBook}. This can be done by following arguments similar to those used in \cite[Appendix B.D]{Murin:IT11}, but instead of upper bounding the different quantities in the calculation of the probability of error, we apply lower bounds, see the left-hand side (LHS) of \cite[Eqns. (6.106)--(6.108)]{YeungBook}.
	In particular it follows that if conditions \eqref{bnd:Joint} hold with {\em opposite strict inequality}, e.g., $H(S_1|S_2,W_3) > I(X_1;Y_3|S_2, V_1, X_2, X_3, W_3)$, see \eqref{bnd:Joint_rly_S1}, then reliable transmission is not possible {\em via the scheme of \Thmref{thm:jointCond}}.
	These arguments also apply to \Thmref{thm:jointCondFlip}, that is, if conditions \eqref{bnd:JointFlip} hold with {\em opposite strict inequality}, e.g., $H(S_1|S_2,W_3) > I(X_1;Y_3|S_1, X_2, X_3)$ , see \eqref{bnd:JointFlip_rly_S1}, then reliable transmission is not possible {\em via the scheme of \Thmref{thm:jointCondFlip}}.
	
		
		In Appendix \ref{annex:proofNotFeasible} we show that indeed evaluating both \Thmref{thm:jointCond} and \Thmref{thm:jointCondFlip} for the example in this section, some conditions in \Thmref{thm:jointCond} and \Thmref{thm:jointCondFlip} hold with opposite strict inequality to what is required by the theorems.
		This shows that reliable transmission of the sources is not possible via the schemes of \Thmref{thm:jointCond} and \Thmref{thm:jointCondFlip}.
\end{IEEEproof}

	%

	%
	
	In contrast to \Thmref{thm:jointCond} and \Thmref{thm:jointCondFlip}, we have the following proposition for \Thmref{thm:jointCond_NewSimult}:	
\begin{MyProposition} \label{prop:feasible}
    The sources defined in Table \ref{tab:SourceDist} can be reliably transmitted over the PSOMARC specified in Table \ref{tab:PrimitiveSOMARC}, by using the scheme of \Thmref{thm:jointCond_NewSimult}.
\end{MyProposition}

\begin{IEEEproof}
	Conditions \eqref{bnd:JointNewSimult} can be specialized to the PSOMARC
	by letting $\mV_1=\mV_2=\mW_3=\mW=\phi$ and $I(X_3;Y_R)=C_3$. In particular, a specialization of the conditions of \Thmref{thm:jointCond_NewSimult} which involve $H(S_1,S_2)$, i.e. \eqref{bnd:JointNewSimult_rly_S1S2} and \eqref{bnd:JointNewSimult_dst_S1S2}, gives the following condition:
	\vspace{-0.2cm}
	\begin{align}
     H(S_1,S_2) &< \min \{I(X_1, X_2;Y_3), I(X_1, X_2;Y_S) + C_3 \},  \label{bnd:PrimitiveSOMARC_JointNew_Sum}
  \end{align}
	
	\vspace{-0.2cm}
	\noindent where the joint distribution \eqref{eq:JntNewJointDistSimult} specializes to $p(s_1,s_2)p(x_1|s_1)p(x_2|s_2)p(y_3,y_S|x_1,x_2)$. Next, note that for the sources defined in Table \ref{tab:SourceDist} we have $H(S_1,S_2) = \log_2 3$. Moreover, as $|\mY_3|=3,|\mY_S|=2$ and $C_3=1$, the RHS of \eqref{bnd:PrimitiveSOMARC_JointNew_Sum} is upper bounded by $\log_2 3$, thus, the LHS of \eqref{bnd:PrimitiveSOMARC_JointNew_Sum} equals to the RHS of \eqref{bnd:PrimitiveSOMARC_JointNew_Sum}. However, as condition \eqref{bnd:PrimitiveSOMARC_JointNew_Sum} requires strict inequality, {\em the conditions provided in the statement of \Thmref{thm:jointCond_NewSimult} do not imply that reliable transmission is possible} in the present example. 
	Note that this case is different than the case of Prop. \ref{prop:NotFeasible}, see Remark \ref{rem:possibleVsimpossible} below. 
In Appendix \ref{annex:proofFeasible} we specify an explicit p.m.f $p(x_i|s_i),i=1,2$, for which we show, through an explicit calculation of the probability of decoding error, that reliable transmission 
is possible via the scheme of \Thmref{thm:jointCond_NewSimult}.
\end{IEEEproof}

\begin{remark} \label{rem:possibleVsimpossible}
	The case of Prop. \ref{prop:feasible} is different than the case of Prop. \ref{prop:NotFeasible}. 
	In the case of Prop. \ref{prop:feasible} we have an equality between the LHS and RHS,\footnote{Conditions \eqref{bnd:JointNewSimult}, specialized to the PSOMARC, evaluated by setting $p(x_i|s_i), i=1,2$, to be the deterministic distribution $p(x_i|s_i)=\delta(x_i-s_i)$, where $\delta(x)$ is the Kronecker Delta function, hold with an {\em equality}.}
	while for Prop. \ref{prop:NotFeasible}, evaluating the conditions of \Thmref{thm:jointCond} and \Thmref{thm:jointCondFlip} we show that the inequality sign is reversed compared to what is required by the theorems. Then, in the proof of Prop. \ref{prop:NotFeasible} we show that such reversal implies that reliable transmission is impossible (see Appendix \ref{annex:proofNotFeasible}). Since in the case of Prop. \ref{prop:feasible} we have an equality between the LHS and the RHS quantities, we examine the situation in more detail in Appendix~\ref{annex:proofFeasible}.
\end{remark}





\vspace{-0.1cm}
\section{Necessary Conditions for Reliable Transmission of Correlated Sources Over DM MARCs} \label{sec:NecessaryConditions}	


In this section three sets of necessary conditions for reliable transmission of correlated sources over DM MARCs with side information are derived. 
These new converse results are based on the fact that only certain joint input distributions $p(x_1,x_2)$ can be achieved. 
%
	Observe that from Def. \ref{def:MABRCcodeDef} it follows that valid channel input distributions must obey the Markov chain:
	\vspace{-0.15cm}
	\begin{equation}
		X_1 \leftrightarrow S_1^n \leftrightarrow S_2^n \leftrightarrow X_2.
	\label{eq:MarkovChain}
	\end{equation}

\vspace{-0.15cm}	
\noindent In the following we use the technique introduced by Kang and Ulukus in \cite{Kang:2011} to constrain the achievable joint input distributions to take into account \eqref{eq:MarkovChain}. We start by reviewing some basic definitions and results from \cite{Kang:2011} and \cite{Witsenhausen:75}.
%

\vspace{-0.3cm}
\subsection{Definitions and Known Results}

\begin{MyDefinition} \label{def:maxCorr}
	({\em Maximal correlation}, \cite[Sec. 2]{Witsenhausen:75}) 
	The maximal correlation between the RVs $X$ and $Y$ is defined as $\rho^{\ast}_{XY} \triangleq \sup \E \left\{ f(X) g(Y) \right\}$, where the supremum is taken over $f: \mX \mapsto \setR, g: \mY \mapsto \setR$, s.t $\E \left\{ f(X) \right\} = \E \left\{ g(Y) \right\} = 0$, $\E \left\{ f^2 (X) \right\} = \E \left\{ g^2(Y) \right\} = 1$, and with the convention that the supremum over the empty set equals to 0. The conditional maximal correlation $\rho^{\ast}_{XY|z}$ is defined similarly.
	
\end{MyDefinition}

\begin{MyDefinition} \label{def:KangDefs}
	({\em Matrix notation for probability distributions}, \cite[Eqn. (6)]{Kang:2011})
	%
	%
		%
	Let $X \in \mX$, and $Y \in \mY$, be two discrete random variables with finite cardinalities. The joint probability distribution matrix $\dsP_{XY}$ is defined as $\dsP_{XY}(i,j) \triangleq \Pr \left( X\mspace{-2mu}=\mspace{-2mu}x_i, Y\mspace{-2mu}=\mspace{-2mu}y_j \right), i=1,2,\dots,|\mX|, j=1,2,\dots,|\mY|$.
	The marginal distribution matrix of an RV $X$ is defined as the diagonal matrix $\dsP_X$ such that $\dsP_{X}(i,i) = \Pr \left( X=x_i \right), x_i \in \mX$; $\quad \dsP_{X}(i,j)=0, \quad i \neq j$. This marginal distribution can also be represented in a vector form denoted by $\pvec_X$. The $i$'th element of $\pvec_X$ is $\pvec_X(i) \triangleq \Pr \left( X=x_i \right)$.
The conditional joint probability distribution matrix $\dsP_{XY|z}$ is defined similarly.
\end{MyDefinition}

\begin{MyDefinition} \label{def:spectralRep}
	({\em Spectral representation}, \cite[Eqns. (12)--(13)]{Kang:2011})
	We define the matrix $\tilde{\dsP}_{XY}$ as $\tilde{\dsP}_{XY} \triangleq \dsP^{- \frac{1}{2}}_X \dsP_{XY} \dsP^{- \frac{1}{2}}_Y$,
	\noindent and the vector $\tilde{\pvec}_X$ as $\tilde{\pvec}_X = \pvec_X^{\frac{1}{2}}$,
	\noindent where $\pvec_X^{\frac{1}{2}}$ stands for an element-wise square root of $\pvec_X$. The conditional distributions $\tilde{\dsP}_{XY|z}$ and $\tilde{\pvec}_{X|y}$ are defined similarly.
\end{MyDefinition}

	Note that not every matrix $\tilde{\dsP}_{XY}$ can correspond to a given joint distribution matrix $\dsP_{XY}$. This is because a valid joint distribution matrix $\dsP_{XY}$ must have all its elements to be nonnegative and add to 1. \cite[Thm. 1]{Kang:2011} gives a necessary and sufficient condition for $\tilde{\dsP}_{XY}$ to correspond to a joint distribution matrix $\dsP_{XY}$:
	
\begin{theoremA}
    (\cite[Thm. 1]{Kang:2011}) Let $\dsP_X$ and $\dsP_Y$ be a pair of marginal distributions. A nonnegative matrix $\dsP_{XY}$ is a joint distribution matrix with marginal distributions $\dsP_X$ and $\dsP_Y$ if and only if the singular value decomposition (SVD) of the corresponding nonnegative matrix $\tilde{\dsP}_{XY}$
    %
    \noindent satisfies:
    \vspace{-0.15cm}
		\begin{equation}
			\tilde{\dsP}_{XY} = \dsM \dsSig \dsN^{T} = \pvec^{\frac{1}{2}}_X \left( \pvec^{\frac{1}{2}}_Y \right)^T + \sum_{i=2}^{l}{\sigma_i \boldsymbol{\mu}_i \boldsymbol{\nu}^T_i},
		\label{eq:KangThm1}
		\end{equation}
		
		\vspace{-0.15cm}
		\noindent where $l = \min \{ |\mX|, |\mY| \}$, $\dsM \triangleq [\boldsymbol{\mu}_1, \boldsymbol{\mu}_2, \dots \boldsymbol{\mu}_l]$ and $\dsN \triangleq [\boldsymbol{\nu}_1, \boldsymbol{\nu}_2, \dots \boldsymbol{\nu}_l]$ are two matrices such that $\dsM^T \dsM = \dsI$ and $\dsN^T \dsN=\dsI$, and $\dsSig \triangleq \text{diag}[\sigma_1, \sigma_2, \dots, \sigma_l]$\footnote{We use $\dsSig = \text{diag}[\avec]$ to denote a rectangular matrix $\dsSig$ s.t $\dsSig_{i,i} = a_i, \dsSig_{i,j} = 0, \forall i \ne j$.
		}; $\boldsymbol{\mu}_1 = \pvec^{\frac{1}{2}}_X, \boldsymbol{\nu}_1 = \pvec^{\frac{1}{2}}_Y$, and $\sigma_1 = 1 \ge \sigma_2 \ge \dots \ge \sigma_l \ge 0$. That is, all the singular values of $\tilde{\dsP}_{XY}$ are non-negative and smaller than or equal to 1. 
		We sometime denote $\sigma_i = \sigma_i(\tilde{\dsP}_{X Y})$ to explicitly indicate the matrix for which the singular value is computed.
		The largest singular value of $\tilde{\dsP}_{XY}$ is 1, and its corresponding left and right singular vectors are $\pvec^{\frac{1}{2}}_X$ and $\pvec^{\frac{1}{2}}_Y$.	
\end{theoremA}	

	
	Next, we define the set of all possible conditional distributions  $p(x_1,\mspace{-1mu} x_2|s_{1,1}, \mspace{-1mu} s_{2,1} \mspace{-1mu} )$ satisfying the Markov chain~\eqref{eq:MarkovChain}:
	\begin{align}
		\mB_{X_1 X_2|S_1 S_2} \triangleq \begin{Bmatrix*}[l] & p_{X_1,X_2|S_1,S_2}(x_1,x_2|s_{1,1}, s_{2,1}): \nonumber \\
		& \exists n \in \setN^{+}, p_{X_1|S_1^n}(x_1|s_1^n), p_{X_2|S_2^n}(x_2|s_2^n) \nonumber \\
		& \text{s.t. } \forall (x_1,x_2,s_{1,1},s_{2,1}) \in \mX_1 \times \mX_2 \times \mS_1 \times \mS_2, \nonumber \\
		& p_{X_1,X_2|S_1,S_2}(x_1,x_2|s_{1,1}, s_{2,1}) = \nonumber \\
		& \quad \frac{1}{p_{S_1,S_2}(s_{1,1}, s_{2,1})} {\dst \sum_{\substack{s_{1,2}^n \in \mS_1^{n-1} \\ s_{2,2}^n \in \mS_2^{n-1}}}{p_{X_1|S_1^n}(x_1|s_1^n)p_{X_2|S_2^n}(x_2|s_2^n)p_{S_1^n,S_2^n}(s_1^n, s_2^n)}} \end{Bmatrix*},
	\label{eq:B_Set_Def}
	\end{align}
	
	\noindent where ${p_{S_1^n,S_2^n}(s_1^n, s_2^n) = \prod_{k=1}^{n}{p_{S_1,S_2}(s_{1,k}, s_{2,k})}}$. 
	Note that as $n$ can be arbitrarily large, the set of all conditional distributions $p_{X_1|S_1^n}(x_1|s_1^n)$ and $p_{X_2|S_2^n}(x_2|s_2^n)$, for all positive integers $n$, is countably infinite. Therefore, we are interested in a characterization of the $n$-letter Markov chain \eqref{eq:MarkovChain} via a set which has a {\em bounded and finite cardinality}.
	
	
	In order to achieve this, we first note that as $p_{S_1,S_2}(s_{1,1},s_{2,1})$ is given, $p_{X_1,X_2}(x_1,x_2), p_{X_1,X_2|S_1}(x_1,x_2|s_{1,1})$ and $p_{X_1,X_2|S_2}(x_1,x_2|s_{2,1})$ are all uniquely determined by $p_{X_1,X_2|S_1,S_2}(x_1,x_2|s_{1,1}, s_{2,1})$. 
	Furthermore, in \cite[Sec. 4]{Witsenhausen:75} it is shown that $\sigma_2(\tilde{\dsP}_{X_1 X_2}) = \rho^{\ast}_{X_1 X_2}$.
	Therefore, $\rho^{\ast}_{X_1 X_2}$, $\rho^{\ast}_{X_1 X_2| s_{1,1}}, \rho^{\ast}_{X_1 X_2| s_{2,1}}$ and $\rho^{\ast}_{X_1 X_2| s_{1,1}, s_{2,1}}$
	are all functions of $p_{X_1,X_2|S_1,S_2}(x_1,x_2|s_{1,1}, s_{2,1})$ for a given $p_{S_1,S_2}(s_{1,1}, s_{2,1})$. 
	The following theorem characterizes constraints on these maximal correlations, and thereby gives a necessary condition for the $n$-letter Markov chain~\eqref{eq:MarkovChain}:\footnote{Here we present a simplified version of \cite[Thm. 4]{Kang:2011}.}
	\begin{theoremA} 
    (\cite[Thm. 4]{Kang:2011}) Let $(S_1^n,S_2^n)$ be a pair of length-$n$ independent and identically distributed (i.i.d.) sequences such that $p_{S_{1,k},S_{2,k}}(a,b) = p_{S_{1},S_{2}}(a,b), \forall (a,b) \in \mS_1 \times \mS_2, \forall k \in \{1,2,\dots,n \}$, and let the variables $X_1$ and $X_2$ satisfy the Markov chain \eqref{eq:MarkovChain}. Let $S_{1,k}$ and $S_{2,j}$ be arbitrary elements of $\Svec_{1,1}^n$ and $\Svec_{2,1}^n$, respectively, that is, $k,j \in \left\{1,2,\dots,n \right\}$, then
		\vspace{-0.15cm}
		\begin{equation}
			\rho^{\ast}_{X_1 X_2| s_{1,k}, s_{2,k}} \le \rho^{\ast}_{S_1 S_2}.
		\end{equation}
	
	\end{theoremA}
	
	\vspace{-0.15cm}
	Now, we define the set $\mB_{X_1 X_2|S_1 S_2}'$ as follows:
	\begin{align} 
		\mB_{X_1 X_2|S_1 S_2}' \triangleq \begin{Bmatrix*}[l] & p_{X_1,X_2|S_1,S_2}(x_1,x_2|s_{1,1}, s_{2,1}): \nonumber \\
		& \forall (s_{1,1}, s_{2,1}) \in \mS_1 \times \mS_2 \\
		& \rho^{\ast}_{X_1 X_2} \le \rho^{\ast}_{S_1 S_2}, \nonumber \\
		& \rho^{\ast}_{X_1 X_2| s_{1,1}} \le \rho^{\ast}_{S_1 S_2}, \nonumber \\
		& \rho^{\ast}_{X_1 X_2| s_{2,1}} \le \rho^{\ast}_{S_1 S_2}, \nonumber \\ 
		& \rho^{\ast}_{X_1 X_2| s_{1,1}, s_{2,1}} \le \rho^{\ast}_{S_1 S_2} \end{Bmatrix*}. 
	\end{align}
	
	\noindent Note that by \cite[Thm. 4]{Kang:2011}	the set $\mB_{X_1 X_2|S_1 S_2}'$ is invariant to the symbol index, that is, $s_{1,1}$ and $s_{2,1}$ can be replaced by $s_{1,k}$ and $s_{2,k}$ for any $k \in \{2,3,\dots,n \}$.
	\noindent Since \cite[Thm. 4]{Kang:2011} gives necessary conditions for the $n$-letter Markov chain \eqref{eq:MarkovChain}, it follows that $\mB_{X_1 X_2|S_1 S_2} \subseteq \mB_{X_1 X_2|S_1 S_2}'$.
	Furthermore, the set $\mB_{X_1 X_2|S_1 S_2}'$ is characterized by the singular values\footnote{Recall that $\sigma_2(\tilde{\dsP}_{X_1 X_2}) = \rho^{\ast}_{X_1 X_2}$.} of the matrices $\tilde{\dsP}_{X_1 X_2}, \tilde{\dsP}_{X_1 X_2| s_{1,1}}, \tilde{\dsP}_{X_1 X_2| s_{2,1}}$ and $\tilde{\dsP}_{X_1 X_2| s_{1,1},s_{2,1}}$. Therefore, while the set $\mB_{X_1 X_2|S_1 S_2}$ has countably infinite dimensions, the set $\mB_{X_1 X_2|S_1 S_2}'$ has finite and bounded dimensions.	
	
\vspace{-0.25cm}
\subsection{A MAC Bound}

	Next, we derive a new set of necessary conditions which is a reminiscent of the so-called ``MAC bound'' for the relay channel, \cite[Ch. 16]{KimElGamal:12}, that takes into account \eqref{eq:MarkovChain}.

	\begin{theorem}
    \thmlabel{thm:OuterMarkov}
			Any source pair $(S_1,S_2)$ that can be reliably transmitted over the DM MARC with receiver side information $W$, as defined in Section \ref{subsec:model}, must satisfy the constraints:
\vspace{-0.15cm}
\begin{subequations} \label{bnd:outr_markov_dst}
\begin{eqnarray}
        H(S_1|S_2,W) &\leq& I(X_1,X_3;Y|S_2,X_2,W,Q) \label{bnd:outr_markov_dst_S1} \\
        H(S_2|S_1,W) &\leq& I(X_2,X_3;Y|S_1,X_1,W,Q) \label{bnd:outr_markov_dst_S2} \\
        H(S_1,S_2|W) &\leq& I(X_1,X_2,X_3;Y|W,Q), \label{bnd:outr_markov_dst_S1S2}
\end{eqnarray}
\end{subequations}

	\vspace{-0.15cm}
  \noindent for a joint distribution that factorizes as: 
	\vspace{-0.15cm}
	\begin{align}    
    p(q,s_1,s_2,w,x_1,x_2,x_3,y) = p(q)p(s_1,s_2,w)p(x_1,x_2|s_1,s_2,q)p(x_3|x_1,x_2,s_1,s_2,q)p(y|x_1,x_2,x_3),
    \label{eq:MarkovBound_dist}
	\end{align}

\vspace{-0.15cm}	
\noindent	with $\left| \mQ \right| \leq 4$, and for every $q \in \mQ$, it follows that:
	\vspace{-0.15cm}
	\begin{align}
		p(x_1,x_2|s_1,s_2,Q=q) & \in \mB_{X_1 X_2|S_1 S_2} \subseteq \mB_{X_1 X_2|S_1 S_2}'.
	\label{eq:MarkovBound_dist_Bset}
	\end{align}
       
\end{theorem}

\vspace{-0.2cm}
\begin{IEEEproof}
	The proof is given in Appendix \ref{subsec:MarkovBound_proof}.
\end{IEEEproof}

\begin{remark}
	This bound does not include $W_3$ because decoding is done based only on the information available at the destination, while the relay channel input is allowed to depend on $X_1,X_2,S_1$ and $S_2$. Therefore, $W_3$ does not add any useful information for generating the relay channel input.
\end{remark}

\vspace{-0.4cm}
\subsection{Broadcast Bounds} \label{sec:MAC_BC_bound}

	The next two new sets of necessary conditions are a reminiscent of the so-called ``broadcast bound'' for the relay channel, \cite[Ch. 16]{KimElGamal:12}.
%
%

\begin{MyProposition}
    \label{prop:VGeneral}
    Any source pair $(S_1,S_2)$ that can be reliably transmitted over the DM MARC with relay side information $W_3$ and receiver side information $W$, as defined in Section \ref{subsec:model}, must satisfy the constraints:
\vspace{-0.2cm}
\begin{subequations} \label{bnd:V_general_dst}
\begin{align}
        H(S_1|S_2,W,W_3) & \leq I(X_1;Y,Y_3|S_2,X_2,W,V) \label{bnd:outr_V_dst_S1} \\
        H(S_2|S_1,W,W_3) & \leq I(X_2;Y,Y_3|S_1,X_1,W,V) \label{bnd:outr_V_dst_S2} \\
        H(S_1,S_2|W,W_3) & \leq I(X_1,X_2;Y,Y_3|W,V), \label{bnd:outr_V_dst_S1S2}
\end{align}
\end{subequations}

	\vspace{-0.15cm}
  \noindent for some joint distribution of the form: 
	\vspace{-0.15cm}
	\begin{align}  
    & p(v,s_1,s_2,w,w_3,x_1,x_2,x_3,y,y_3) = p(v,s_1,s_2,w,w_3)p(x_1,x_2|s_1,s_2,v)p(x_3|v)p(y,y_3|x_1,x_2,x_3), \label{eq:BCbound_dist}
	\end{align}    
	
	\vspace{-0.15cm}
	\noindent with $\left| \mV \right| \leq 4$.
       
\end{MyProposition}

\vspace{-0.2cm}
\begin{IEEEproof}
	The proof is given in Appendix \ref{annex:ProofOuterV}.
\end{IEEEproof}	

\begin{remark} \label{rem:outerVImprove}
	In Prop. \ref{prop:VGeneral} we did not place restrictions on $p(x_1,x_2|s_1,s_2)$ as in \Thmref{thm:OuterMarkov}. This is because \cite[Thm. 4]{Kang:2011} requires $(S_1^n, S_2^n)$ to be a pair of {\em i.i.d sequences} of length $n$. However, in the proof of Prop. \ref{prop:VGeneral} $V^n$ is {\em not} an {\em i.i.d sequence}, and therefore $(S_1^n, S_2^n, V^n)$  is {\em not} a triplet of {\em i.i.d sequences}. Hence, it is not possible to use the approach of \cite{Kang:2011} to tighten Prop. \ref{prop:VGeneral}. It is possible, however, to establish a different set of ``broadcast-type" necessary conditions which benefits from the results of \cite{Kang:2011}. This is stated in \Thmref{thm:BCMarkov}.
\end{remark}

	
		\begin{theorem}
    \thmlabel{thm:BCMarkov}
			Any source pair $(S_1,S_2)$ that can be reliably transmitted over the DM MARC with relay side information $W_3$ and receiver side information $W$, as defined in Section \ref{subsec:model}, must satisfy the constraints:
\vspace{-0.15cm}
\begin{subequations} \label{bnd:BC_markov_dst}
\begin{eqnarray}
        H(S_1|S_2,W,W_3) &\leq& I(X_1;Y,Y_3|S_2,X_2,X_3,W,Q) \label{bnd:BC_markov_dst_S1} \\
        H(S_2|S_1,W,W_3) &\leq& I(X_2;Y,Y_3|S_1,X_1,X_3,W,Q) \label{bnd:BC_markov_dst_S2} \\
        H(S_1,S_2|W,W_3) &\leq& I(X_1,X_2;Y, Y_3|X_3,W,Q), \label{bnd:BC_markov_dst_S1S2}
\end{eqnarray}
\end{subequations}

\vspace{-0.15cm}
\noindent for a joint distribution that factorizes as: 
	\vspace{-0.2cm}
	\begin{align}    
    & p(q,s_1,s_2,w,w_3,x_1,x_2,x_3,y,y_3) = \nonumber \\ 
		& \qquad p(q)p(s_1,s_2,w,w_3)p(x_1,x_2|s_1,s_2,q)p(x_3|x_1,x_2,w_3,q)p(y,y_3|x_1,x_2,x_3),
    \label{eq:BCMarkovBound_dist}
	\end{align}
	
	\vspace{-0.2cm}
\noindent	with $\left| \mQ \right| \leq 4$, and for every $q \in \mQ$, it follows that:
	\vspace{-0.2cm}
	\begin{align}
%
		p(x_1,x_2|s_1,s_2,Q=q) & \in \mB_{X_1 X_2|S_1 S_2} \subseteq \mB_{X_1 X_2|S_1 S_2}',
\label{eq:BCMarkovBound_dist_Bset}
\end{align}
	
       
\end{theorem}	

\vspace{-0.2cm}
\begin{IEEEproof}
	The proof follows similar arguments to the proofs of \Thmref{thm:OuterMarkov} and Prop. \ref{prop:VGeneral}, thus, it is omitted here.
\end{IEEEproof}

\vspace{-0.4cm}
\subsection{Discussion}

\begin{remark}
	Note that the side information may affect the corresponding chain, see e.g., \Thmref{thm:BCMarkov}.
\end{remark}

\begin{remark}
	For independent sources ($p(s_1,s_2)=p(s_1)p(s_2)$) and $\mW = \mW_3 = \phi$, a combination of \Thmref{thm:OuterMarkov} and \Thmref{thm:BCMarkov} specializes to the cut-set bound for the MARC derived in \cite[Thm. 1]{KramerMandayam:04}. To see this, note that in this case the RHSs of \eqref{bnd:BC_markov_dst} are identical to the first term in the RHS of \cite[Eqn. (7)]{KramerMandayam:04}, while the RHSs of \eqref{bnd:outr_markov_dst} are identical to the second term in the RHS of \cite[Eqn. (7)]{KramerMandayam:04}, for $G=\{1\}, \{2\}, \{1,2\}$, respectively. Furthermore, we have that \eqref{eq:MarkovBound_dist} and \eqref{eq:BCMarkovBound_dist} are the same. Next, note that for independent sources, $\rho^{\ast}_{S_1 S_2} = 0$, which implies that $\rho^{\ast}_{X_1 X_2}= \rho^{\ast}_{X_1 X_2| s_{1,1}} = \rho^{\ast}_{X_1 X_2| s_{2,1}} = \rho^{\ast}_{X_1 X_2| s_{1,1}, s_{2,1}} = 0$. Therefore, $X_1$ and $X_2$ are independent and conditions \eqref{eq:MarkovBound_dist_Bset} and \eqref{eq:BCMarkovBound_dist_Bset} are satisfied for any $p_{S_1,S_2}(s_1,s_2)=p_{S_1}(s_1)p_{S_2}(s_2)$. 
	Finally, letting $R_1 \triangleq H(S_1), R_2 \triangleq H(S_2)$ implies that $H(S_1,S_2)=R_1+R_2$, and therefore for independent sources the combination of \Thmref{thm:OuterMarkov} and \Thmref{thm:BCMarkov} coincides with \cite[Eqn. (7)]{KramerMandayam:04}.
\end{remark}	
	
\begin{remark}
	For Gaussian MARCs subject to i.i.d phase fading, and for the channel inputs that maximize the achievable region at the destination obtained via DF, the achievable region at the destination is a subset of the corresponding  achievable region at the relay (i.e., decoding at the relay does not constrain the rate to the destination). In this case, \Thmref{thm:OuterMarkov} specializes to \cite[Prop. 1]{Murin:ISWCS11}.\footnote{In \cite[Thm. 4]{Murin:IT11} we showed that for Gaussian MARCs subject to i.i.d phase fading, when decoding at the relay does not constrain the rate to the destination, then source-channel separation is optimal.} From \cite[Thm. 8]{Kramer:2005} it follows that in this case mutually independent channel inputs simultaneously maximize the RHSs of \cite[Eqns. (3)]{Murin:ISWCS11}. Additionally, note that for mutually independent channel inputs, Eqns. \eqref{bnd:outr_markov_dst} coincide with \cite[Eqns. (3)]{Murin:ISWCS11}. Lastly we observe that the mutual independence of the channel inputs implies that $\rho^{\ast}_{X_1 X_2}= \rho^{\ast}_{X_1 X_2| s_{1,1}} = \rho^{\ast}_{X_1 X_2| s_{2,1}} = \rho^{\ast}_{X_1 X_2| s_{1,1}, s_{2,1}} = 0$, thus \eqref{eq:MarkovBound_dist_Bset} is satisfied for any joint distribution of the sources.
		
\end{remark}
	
\begin{remark}
	When specialized to the MAC with correlated sources \Thmref{thm:OuterMarkov} and \Thmref{thm:BCMarkov} coincide and both are tighter than Prop. \ref{prop:VGeneral}.
	Setting $\mX_3 = \mY_3 = \mW_3 = \phi$, 
	the expressions in \eqref{bnd:outr_markov_dst}, \eqref{bnd:V_general_dst} and \eqref{bnd:BC_markov_dst} become identical. 
	However, note that in \eqref{eq:BCbound_dist} a general joint distribution $p(v,s_1,s_2,w)$ is considered, while in \eqref{eq:MarkovBound_dist} and \eqref{eq:BCMarkovBound_dist} $Q \independent (S_1,S_2,W)$. 
	Moreover, the required Markov chain of \eqref{eq:MarkovChain} is not accounted for by the chain of Prop. \ref{prop:VGeneral}, contrary to \Thmref{thm:OuterMarkov} and \Thmref{thm:BCMarkov}.	
	Therefore, we conclude that when specialized to the MAC scenario, \Thmref{thm:OuterMarkov} and \Thmref{thm:BCMarkov} give the same bound which is tighter then the one in Prop. \ref{prop:VGeneral}.	
	
	Setting $\mX_3 = \mY_3 = \mW_3 = \phi$ as well as $\mW = \phi$, specializes our model to the MAC with no side information at the receiver. For this model, both \Thmref{thm:OuterMarkov} and \Thmref{thm:BCMarkov} specialize to \cite[Thm. 7]{Kang:2011}, which establishes necessary conditions for the MAC with correlated sources.
\end{remark}

\vspace{-0.4cm}
\subsection{Numerical Examples}

	\vspace{-0.1cm}
	We now demonstrate the improvement of \Thmref{thm:OuterMarkov} and \Thmref{thm:BCMarkov} upon the cut-set bound of \cite[Ch. 18.1]{KimElGamal:12}. 
In order to simplify the arguments, we consider a scenario with no side information $\mW=\mW_3=\phi$, and focus on the bound on $H(S_1,S_2)$. 
	In the following, we consider explicit PSOMARC and sources for which we show that the cut-set bound fails to indicate whether reliable transmission of the sources over the channel is possible, while a relaxed version of our outer bounds do indicate that reliable transmission of the sources over the channel is impossible.
	
	Consider the PSOMARC defined by $\mX_1 = \mX_2 = \mY_3 = \mY_S = \{0,1\}$, the channel transition probabilities detailed in Tables \ref{tab:TranProbY3} and \ref{tab:TranProbY}, and let $C_3 = 0.1$.
\begin{table}[h]
  \vspace{-0.25cm}
	\begin{center}
	\begin{tabular}[t]{|c|c|c|c|c|}
	\hline
	$Y_3$ \textbackslash $(X_1,X_2)$ & (0,0) & (0,1) & (1,0) & (1,1) \\
	\hline
	0 & 0.87 & 0.25 & 0.51 & 0.24 \\
	\hline
	1 & 0.13 & 0.75 & 0.49 & 0.76 \\
	\hline
	\end{tabular}
	\captionsetup{font=small}
	\caption{The transition probability $(X_1,X_2) \mapsto Y_3$. \label{tab:TranProbY3}}
	\vspace{-0.6cm}
  \end{center}
  \end{table}	
  
  \begin{table}[h]
  \vspace{-0.5cm}
	\begin{center}
	\begin{tabular}[t]{|c|c|c|c|c|}
	\hline
	$Y$ \textbackslash $(X_1,X_2)$ & (0,0) & (0,1) & (1,0) & (1,1) \\
	\hline
	0 & 0.23 & 0.19 & 0.65 & 0.91 \\
	\hline
	1 & 0.77 & 0.81 & 0.35 & 0.09 \\
	\hline
	\end{tabular}
	\captionsetup{font=small}
	\caption{The transition probability $(X_1,X_2) \mapsto Y$. \label{tab:TranProbY}}
	\vspace{-1cm}
  \end{center}
  \end{table}	
  
	\noindent Next, consider the cut-set bound for the sum-rate of the PSOMARC, \cite[Eqn. (9)]{Tandon:CISS:11}. When evaluated for the PSOMARC defined in Tables \ref{tab:TranProbY3}, \ref{tab:TranProbY} the necessary conditions of \cite[Eqn. (9)]{Tandon:CISS:11} yield:
  \vspace{-0.1cm}
	\begin{equation}
		H(S_1,S_2) \le \text{I}_{\text{cut-set}} \triangleq \max_{p(x_1,x_2)} \Big\{ I(X_1,X_2;Y_S) + \min \big\{ C_3, I(X_1,X_2;Y_3|Y_S) \big\} \Big\} \approx 0.516.\footnote{Note that the cut-set bound in \eqref{eq:Example_cutset} depends only on the channel transition probabilities and {\em not} on the joint distribution of the sources.} 
	\label{eq:Example_cutset}
	\end{equation}
	
	\vspace{-0.1cm}
	\noindent The ~maximum ~in ~\eqref{eq:Example_cutset} ~is ~achieved ~by ~$\Pr \left((X_1,X_2)=(0,0) \right) \approx 0.1$, $\Pr \left((X_1,X_2)=(0,1) \right) \approx 0.39$, $\Pr \left((X_1,X_2)=(1,0) \right) \approx 0$, $\Pr \left((X_1,X_2)=(1,1) \right) \approx 0.51$. This and the following optimizations are done numerically using an exhaustive search over all relevant parameters with a step size of 0.01 in each variable.
	Next, we consider the combination of the relaxed versions of \eqref{bnd:outr_markov_dst_S1S2} and \eqref{bnd:BC_markov_dst_S1S2}, with $\mW=\mW_3=\phi$, specialized to the PSOMARC:
	\vspace{-0.25cm}
	\begin{equation}
		H(S_1,S_2) \le \text{I}_{\text{new}} \triangleq \max_{p(x_1,x_2): \rho^{\ast}_{X_1 X_2} \le \rho^{\ast}_{S_1 S_2}} \Big\{ I(X_1,X_2;Y_S) + \min \big\{ C_3, I(X_1,X_2;Y_3|Y_S) \big\} \Big\}.
	\label{eq:Example_markov}
	\end{equation}

\vspace{-0.05cm}	
\noindent Note that \eqref{eq:Example_markov} is less restrictive than \eqref{bnd:outr_markov_dst_S1S2} and \eqref{bnd:BC_markov_dst_S1S2}, as the maximization in \eqref{eq:Example_markov} includes only the restriction due to $\tilde{\dsP}_{X_1 X_2}$, while the restrictions due to the conditional distributions $\tilde{\dsP}_{X_1 X_2|S_1}, \tilde{\dsP}_{X_1 X_2|S_2}$ and $\tilde{\dsP}_{X_1 X_2|S_1,S_2}$ are ignored.
	Finally, we recall the sum-rate condition of \Thmref{thm:jointCond_NewSimult} stated in \eqref{bnd:PrimitiveSOMARC_JointNew_Sum} obtained by combining \eqref{bnd:JointNewSimult_rly_S1S2} and \eqref{bnd:JointNewSimult_dst_S1S2} and specializing the expressions to the PSOAMRC:
	\vspace{-0.15cm}
	\begin{equation}
		H(S_1,S_2) < \text{I}_{\text{suff}} \triangleq \max_{ p(s_1,s_2)p(x_1|s_1)p(x_2|s_2) } \min \big\{I(X_1,X_2;Y_3), I(X_1,X_2;Y_S) + C_3 \big\}.
	\label{eq:Example_suff}
	\end{equation}
	
	\vspace{-0.15cm}
	Let $(S_1,S_2)$ be a pair of sources such that $\mS_1 = \mS_2 = \{0,1\}$, and their joint distribution is	given in Table \ref{tab:SourceDist_p2}. 
	\begin{table}[h!]
  \vspace{-0.25cm}
	\begin{center}
	\begin{tabular}[t]{|c|c|c|}
	\hline
	$S_1$ \textbackslash $S_2$ & 0 & 1 \\
	\hline
	0 & 0 & 0.04 \\
	\hline
	1 & 0.045 & 0.915  \\
	\hline
	\end{tabular}
	\captionsetup{font=small}
	\caption{
	The joint distribution $p(s_1,s_2)$. 
	\label{tab:SourceDist_p2}}
  \vspace{-1cm}
  \end{center}
  \end{table}
	
	\noindent For this joint distribution we evaluate $H(S_1,S_2) \approx 0.504$, therefore, the cut-set necessary condition \eqref{eq:Example_cutset} does not indicate whether these sources can be transmitted reliably or not.
	Furthermore, for the joint distribution given in Table \ref{tab:SourceDist_p2}, the RHS of \eqref{eq:Example_suff} is evaluated as $\text{I}_{\text{suff}} \approx 0.274$. 
	This value is achieved by $\Pr \left(X_1=0|S_1=0 \right) \approx 0$, $\Pr \left(X_1=0|S_1=1 \right) \approx 1$, $\Pr \left(X_1=1|S_1=0 \right) \approx 0.84$, $\Pr \left(X_1=1|S_1=1 \right) \approx 0.16$, $\Pr \left(X_2=0|S_2=0 \right) \approx 0.98$, $\Pr \left(X_2=0|S_2=1 \right) \approx 0.02$, $\Pr \left(X_2=1|S_2=0 \right) \approx 0.49$, $\Pr \left(X_2=1|S_2=1 \right) \approx 0.51$. Thus, the scheme of \Thmref{thm:jointCond_NewSimult} cannot transmit these sources reliably since condition \eqref{eq:Example_suff} is not satisfied.
	
	In contrast to \eqref{eq:Example_cutset}, which is larger than $H(S_1,S_2)$, for the joint distribution given in Table \ref{tab:SourceDist_p2} we have $\text{I}_{\text{new}} \approx 0.485$. This value is
	achieved by $\Pr \left((X_1,X_2)\mspace{-2mu}=\mspace{-2mu}(0,0) \right) \approx 0.08$, $\Pr \left((X_1,X_2)\mspace{-2mu}=\mspace{-2mu}(0,1) \right) \approx 0.41$, $\Pr \left((X_1,X_2)\mspace{-2mu}=\mspace{-2mu}(1,0) \right) \approx 0.07$, $\Pr \left((X_1,X_2)\mspace{-2mu}=\mspace{-2mu}(1,1) \right) \approx 0.44$.
		Hence, our new necessary condition \eqref{eq:Example_markov}, explicitly indicates that reliable transmission of these sources is impossible.
	
	%
	
		This demonstrates the improvement of \Thmref{thm:OuterMarkov} and \Thmref{thm:BCMarkov} upon the cut-set bound.
		
\begin{remark}
	This numerical example {\em does not follow immediately from the results of Kang and Ulukus for the MAC}, detailed in \cite[Subsection III.C]{Kang:2011}. To see this, consider the PSOMARC and sources as defined in Tables \ref{tab:TranProbY3}, \ref{tab:TranProbY} and \ref{tab:SourceDist_p2}, and let $C_3 = 0.2$ (instead of $0.1$).
	Here, \eqref{eq:Example_cutset} is evaluated as $\text{I}_{\text{cut-set}} \approx 0.600$\footnote{This value was found via an exhaustive search over over all $p(x_1,x_2)$ and can be achieved by $\Pr \left((X_1,X_2)\mspace{-2mu}=\mspace{-2mu}(0,0) \right) \approx 0.26$, $\Pr \left((X_1,X_2)\mspace{-2mu}=\mspace{-2mu}(0,1) \right) \approx 0.24$, $\Pr \left((X_1,X_2)\mspace{-2mu}=\mspace{-2mu}(1,0) \right) \approx 0$, $\Pr \left((X_1,X_2)\mspace{-2mu}=\mspace{-2mu}(1,1) \right) \approx 0.5$.}, while \eqref{eq:Example_markov} is evaluated as $\text{I}_{\text{new}} \approx 0.514$\footnote{This value was found via an exhaustive search over over all $p(x_1,x_2)$ s.t  $\rho^{\ast}_{X_1 X_2} \le \rho^{\ast}_{S_1 S_2}$, and can be achieved by $\Pr \left( (X_1,X_2)\mspace{-2mu}=\mspace{-2mu}(0,0) \right) \approx 0.2$, $\Pr \left( (X_1,X_2)\mspace{-2mu}=\mspace{-2mu}(0,1) \right) \approx 0.36$, $\Pr \left( (X_1,X_2)\mspace{-2mu}=\mspace{-2mu}(1,0) \right) \approx 0.14$, $\Pr \left( (X_1,X_2)\mspace{-2mu}=\mspace{-2mu}(1,1) \right) \approx 0.3$.}. Moreover, recall that $H(S_1,S_2) \approx 0.504$. Hence, for $C_3 = 0.2$, \eqref{eq:Example_markov} does not indicate whether reliable transmission of the sources is possible, while for $C_3 = 0.1$, \eqref{eq:Example_markov} explicitly indicates that reliable transmission is impossible. Observe that the necessary conditions are affected by the presence of the relay. Also note that the cut-set conditions \eqref{eq:Example_cutset} does not indicate whether reliable transmission is possible or not, for either value of $C_3$.
\end{remark}

\begin{remark}
	In the above numerical example we assume that side information is not present. To see the effect of side information at the relay on \eqref{eq:Example_markov} consider the PSOMARC and sources as defined in Tables \ref{tab:TranProbY3}, \ref{tab:TranProbY} and \ref{tab:SourceDist_p2}, and let $C_3 = 0.5$. Here, $I(X_1,X_2;Y_2|Y_S) \approx 0.185, I(X_1,X_2;Y_S) \approx 0.329$ and $\text{I}_{\text{new}} \approx 0.514$\footnote{These value were found via an exhaustive search over over all $p(x_1,x_2)$ s.t  $\rho^{\ast}_{X_1 X_2} \le \rho^{\ast}_{S_1 S_2}$, and can be achieved by $\Pr \left( (X_1,X_2)\mspace{-2mu}=\mspace{-2mu}(0,0) \right) \approx 0.04$, $\Pr \left( (X_1,X_2)\mspace{-2mu}=\mspace{-2mu}(0,1) \right) \approx 0.46$, $\Pr \left( (X_1,X_2)\mspace{-2mu}=\mspace{-2mu}(1,0) \right) \approx 0.03$, $\Pr \left( (X_1,X_2)\mspace{-2mu}=\mspace{-2mu}(1,1) \right) \approx 0.47$.}. Therefore, in this case $I(X_1,X_2;Y_2|Y_S)$ is the dominant term in the minimization on the RHS of \eqref{eq:Example_markov}. Now, let $W_3 = (S_1,S_2)$, which makes \eqref{bnd:BC_markov_dst_S1S2} redundant.\footnote{When $W_3 = (S_1,S_2)$ the chains \eqref{eq:MarkovBound_dist} and \eqref{eq:BCMarkovBound_dist} are the same, and $H(S_1,S_2|W,W_3)=0$.} In this case, the RHS of \eqref{eq:Example_markov} becomes ${\dst \mspace{-15mu} \max_{\mspace{15mu} p(x_1,x_2): \rho^{\ast}_{X_1 X_2} \le \rho^{\ast}_{S_1 S_2}} \mspace{-15mu} I(X_1,X_2;Y_S) + C_3}$,~and we have $\text{I}_{\text{new}} \approx 0.919$\footnote{This value is found via an exhaustive search over over all $p(x_1,x_2)$ s.t  $\rho^{\ast}_{X_1 X_2} \le \rho^{\ast}_{S_1 S_2}$, and can be achieved by $\Pr \left( (X_1,X_2)\mspace{-2mu}=\mspace{-2mu}(0,0) \right) \approx 0.01$, $\Pr \left( (X_1,X_2)\mspace{-2mu}=\mspace{-2mu}(0,1) \right) \approx 0.47$, $\Pr \left( (X_1,X_2)\mspace{-2mu}=\mspace{-2mu}(1,0) \right) \approx 0.01$, $\Pr \left( (X_1,X_2)\mspace{-2mu}=\mspace{-2mu}(1,1) \right) \approx 0.51$.}. To conclude, in this case, the presence of side information at the relay significantly enlarges~$\text{I}_{\text{new}}$.
\end{remark}

\begin{remark}
			We note that the necessary conditions presented in \Thmref{thm:OuterMarkov} and \Thmref{thm:BCMarkov} are not tight in general. For instance, consider the PSOMARC specified in Table \ref{tab:PrimitiveSOMARC} with $C_3 = 1$, and the pair of sources defined in Table \ref{tab:SourceDist}. Prop. \ref{prop:feasible} implies that the sources defined in Table \ref{tab:SourceDist} can be reliably transmitted over this PSOMARC by using the scheme of \Thmref{thm:jointCond_NewSimult}. Here, the maximal sum-rate sufficient condition which is evaluated using \eqref{eq:Example_suff} is $\text{I}_{\text{suff}} = \log_2 3$.
			For this combination of sources and channel, the sum-rate necessary condition due to the cut-set bound is evaluated via \eqref{eq:Example_cutset} as $\text{I}_{\text{cut-set}} = 2$, which is achieved by setting $\Pr \left((X_1,X_2)=(0,0) \right) = \Pr \left((X_1,X_2)=(0,1) \right) = \Pr \left((X_1,X_2)=(1,0) \right) = \Pr \left((X_1,X_2)=(1,1) \right) = 0.25$. Furthermore, using the same $p_{X_1,X_2}(x_1,x_2)$ we also evaluate the newly derived sum-rate necessary condition (from either \Thmref{thm:OuterMarkov} or \Thmref{thm:BCMarkov}) via \eqref{eq:Example_markov} as $\text{I}_{\text{new}} = 2$. Thus, for this combination of channel and sources			
			the RHSs of \eqref{eq:Example_cutset} and \eqref{eq:Example_markov} are strictly larger than the RHS of \eqref{eq:Example_suff}. 
			
			On the other hand, there are sources and channels for which $\text{I}_{\text{cut-set}} = \text{I}_{\text{new}} = \text{I}_{\text{suff}}$. As an example, consider a PSOMARC, defined by $\mX_1 = \mX_2 = \{0,1,2\}, \mY_3 = \{0,1,2,3,4,5\}$ and $\mY_S = \{0,1,2\}$.
			Let $C_3 = 1$, and consider the deterministic channel mapping $(X_1,X_2) \mapsto (Y_3,Y_S)$ specified in Table~\ref{tab:PrimitiveSOMARC_Ex2}.
			\vspace{-0.1cm}
			\begin{table}[h]
			\begin{center}
			\begin{tabular}[t]{|c|c|c|c|c|c|c|}
				\hline
				$(X_1, X_2)$ & $(0,0)$ & $(1,1)$ & $(1,2)$ & $(2,0)$ & $(2,2)$ & \mbox{Otherwise} \\
				\hline 
				$ Y_3$ & 0 & 2 & 3 & 4 & 5 & 1 \\
				\hline
				$ Y_S$ & 0 & 2 & 1 & 2 & 0 & 1 \\
				\hline
			 \end{tabular}
			\captionsetup{font=small}
			\caption{A deterministic channel mapping $(X_1,X_2) \mapsto (Y_3,Y_S)$ for the PSOMARC. \label{tab:PrimitiveSOMARC_Ex2}}
			\vspace{-0.75cm}
			\end{center}
			\end{table}
			
			\noindent The sources $(S_1,S_2)$ are defined over the sets $\mS_1 \mspace{-3mu} = \mspace{-3mu} \mS_2 \mspace{-3mu} = \mspace{-3mu} \{0,1,2\}$ with the joint distribution specified in Table \ref{tab:SourceDist_Ex2}. 
			\vspace{-0.15cm}
			\begin{table}[h]
				\begin{center}
				\begin{tabular}[t]{|c|c|c|c|}
				\hline
				$S_1$ \textbackslash $S_2$ & 0 & 1 & 2 \\
				\hline
				0 & 1/6 & 1/6 & 0 \\
				\hline
				1 & 0 & 1/6 & 1/6 \\
				\hline
				2 & 1/6 & 0 & 1/6 \\
				\hline
				\end{tabular}
				\captionsetup{font=small}
				\caption{The joint distribution of $(S_1,S_2)$. The entry in the $j^{\text{th}}$ row and $m^{\text{th}}$ column, $j,m = 0,1,2$, corresponds to $\Pr \left((S_1,S_2) = (j,m) \right)$. \label{tab:SourceDist_Ex2}}
				\vspace{-0.75cm}
				\end{center}
				\end{table}
			
			\noindent Following the arguments presented in Appendix E, it can be shown that, using the scheme of \Thmref{thm:jointCond_NewSimult} the sources defined in Table \ref{tab:SourceDist_Ex2} can be reliably transmitted over the PSOMARC defined in Table \ref{tab:PrimitiveSOMARC_Ex2}, with $C_3 = 1$. In particular, we have $H(S_1,S_2) = \text{I}_{\text{suff}} = \log_2 6$ (note that since $|\mY_3| = 6$, it follows from \eqref{eq:Example_suff} that $\text{I}_{\text{suff}} \le \log_2 6$).
			For the channel mapping specified in Table \ref{tab:PrimitiveSOMARC_Ex2}, we also have $\text{I}_{\text{new}} \le \log_2 6$ and $\text{I}_{\text{cut-set}} \le \log_2 6$. This follows from the fact that $|\mY_S| = 3$ and from the fact that $C_3=1$. In fact, $\text{I}_{\text{cut-set}} = \text{I}_{\text{new}} = \log_2 6$ is obtained by setting $p(x_1,x_2) = p(s_1,s_2)$. 
			Hence, for this combination of channel and sources the RHSs of \eqref{eq:Example_cutset}, \eqref{eq:Example_markov} and \eqref{eq:Example_suff} coincide and tightness in sum-rate is achieved.
			Furthermore, for every $C_3 \ge 1$ we obtain $\text{I}_{\text{new}} = \text{I}_{\text{suff}}$. 
			To understand this equality, first recall from the above discussion that $\text{I}_{\text{suff}} \le \log_2 6$ with equality obtained with the assignment $p(x_1,x_2) = p(s_1,s_2)$. For evaluating $\text{I}_{\text{new}}$, we recall the expression for $\text{I}_{\text{new}}$ given by \eqref{eq:Example_markov}, repeated here for ease of reference:
			\vspace{-0.15cm}
			\begin{equation*}
				\text{I}_{\text{new}} = \max_{p(x_1,x_2): \rho^{\ast}_{X_1 X_2} \le \rho^{\ast}_{S_1 S_2}} \Big\{ I(X_1,X_2;Y_S) + \min \big\{ C_3, I(X_1,X_2;Y_3|Y_S) \big\} \Big\}.
			\end{equation*}
			
			Now, since $|\mY_S| = 3$ we have that $I(X_1,X_2;Y_S) \le \log_2 3$, and this is achieved with equality by the assignment $p(x_1,x_2) = p(s_1,s_2)$. For $I(X_1,X_2;Y_3|Y_S)$ we write:
			\vspace{-0.15cm}
			\begin{equation*}
				I(X_1,X_2;Y_3|Y_S) \stackrel{(a)}{=} H(Y_3|Y_S) \stackrel{(b)}{\le} 1,
			\end{equation*}
			
			\vspace{-0.15cm}
			\noindent where (a) follows from the the fact that in the considered PSOMARC the mapping from $(X_1,X_2)$ to $Y_3$ is deterministic, and (b) follows from the fact that for every possible value of $Y_S$ there are only two possible values of $Y_3$. An equality in (b) is achieved with the assignment $p(x_1,x_2) = p(s_1,s_2)$. Hence, for $C_3 \ge 1$ the active term in the minimization on the RHS of \eqref{eq:Example_markov} is $I(X_1,X_2;Y_3|Y_S)$, and we have $\text{I}_{\text{new}} = \text{I}_{\text{suff}}$, both maximized with the assignment $p(x_1,x_2) = p(s_1,s_2)$.
			Finally, note that if $C_3 < 1$ then the necessary conditions \eqref{eq:Example_cutset} and \eqref{eq:Example_markov} are not satisfied.
			
\end{remark}

\vspace{-0.3cm}
\section{Conclusions} \label{sec:conclusions}
\vspace{-0.1cm}
In this work we studied JSCC for lossless transmission of correlated sources over DM MARCs.
We derived a new DF-based JSCC scheme which uses the CPM technique for encoding the correlated source sequences for transmission to both the relay and the destination, while SW source coding is used for cooperation between the sources and the relay. This combination allows removing the constraints on the distribution chain required by a previously derived scheme which used CPM to the destination \cite[Thm. 2]{Murin:ISIT12} (quoted as \Thmref{thm:jointCondFlip} in this manuscript).
The new scheme of \Thmref{thm:jointCond_NewSimult} applies simultaneous backward decoding at the destination to simultaneously decode both source sequences and the cooperation information. As the scheme implements CPM-based encoding of the source sequences at the transmitters, both the relay and the destination benefit from the joint source-channel encoding. This is in contrast to the JSCC schemes derived in \cite{Murin:ISIT12} (quoted as \Thmref{thm:jointCond} and \Thmref{thm:jointCondFlip} in this manuscript), in which either the relay or the destination benefits from the CPM encoding, but not both simultaneously. 

We then provided a detailed comparison of the new scheme of \Thmref{thm:jointCond_NewSimult} with the two JSCC schemes of \cite{Murin:ISIT12} and with the scheme of Prop. \ref{prop:jointCond_New} which apply sequential decoding of the source sequences and the cooperation information at the destination. 
		We showed that the scheme of \Thmref{thm:jointCond_NewSimult} is better than the scheme derived in \cite[Thm. 1]{Murin:ISIT12} and the scheme of Prop. \ref{prop:jointCond_New}. 
		We also showed that there are cases in which the scheme of \Thmref{thm:jointCond_NewSimult} strictly outperforms the schemes of \Thmref{thm:jointCond} and \Thmref{thm:jointCondFlip}.
		However, we cannot show that the new scheme of \Thmref{thm:jointCond_NewSimult} is universally better than the scheme of \cite[Thm. 2]{Murin:ISIT12}. This follows from the different admissible joint distributions (see Remarks \ref{rem:oldSchemesComp} and~\ref{rem:SameDist}). 
		 

	Finally, we derived three different sets of necessary conditions for reliable transmission of correlated sources over DM MARCs. We also showed that the newly derived sets are at least as tight as previously known results.	
	One of the new sets is in the spirit of the ``MAC bound" for the classic relay channel, while the other two sets are in the spirit of the ``broadcast bound" for the relay channel. Two of the new sets use the Markov relationship between the sources and the channel inputs to restrict the set of feasible distributions. 
			
\appendices
\numberwithin{equation}{section}
\numberwithin{MyProposition}{section}

\vspace{-0.2cm}
\section{Proof of Theorem \thmref{thm:jointCond_NewSimult}} \label{annex:jointNewSimultProof}
	
	\vspace{-0.2cm}
\subsection{Codebook Construction} \label{subsecsec:Thm3_codeconst}
	
\vspace{-0.1cm}
\begin{itemize}
	\item 
		
		For each $i=1,2$, consider a set of $2^{nR_i}$ bins and let $\msgCal_i \triangleq \{1,2,\dots,2^{nR_i}\}, i=1,2$, be the corresponding set of bin indices. For $i=1,2$, assign every $\svec_i \in \mS_i^n$ to one of the $2^{nR_i}$ bins independently according to a uniform distribution over the bin indices. Denote this assignment by $f_i: \mS_i^n \mapsto \msgCal_i, i=1,2$.

	\item
		For $i=1,2$, generate $2^{nR_i}$ codewords $\vvec_i(\msg_i), \msg_i \in \msgCal_i $, by choosing the letters $v_{i,k}(\msg_i), k = 1,2,\dots,n$,
independently according to the p.m.f $p_{V_i}(v_{i,k}(\msg_i))$.
For each pair $(\svec_i, \msg_i) \in \mS_i^n \times \msgCal_i, i=1,2$, generate one codeword $\xvec_i(\svec_i, \msg_i)$ by choosing the letters $x_{i,k}(\svec_i, \msg_i)$ independently according to the p.m.f $p_{X_i|S_i,V_i}(x_{i,k}|s_{i,k},v_{i,k}(\msg_i))$, $k=1,2,\dots, n$. Finally, generate one relay codeword $\xvec_3(\msg_1,\msg_2)$ for each pair $(\msg_1, \msg_2) \in \msgCal_1 \times \msgCal_2$, by choosing the letters $x_{3,k}(\msg_1,\msg_2)$ independently according to the p.m.f $p_{X_3|V_1,V_2}(x_{3,k}|v_{1,k}(\msg_1),v_{2,k}(\msg_2))$, $k=1,2,\dots, n$.
\end{itemize}

\vspace{-0.25cm}
\subsection{Encoding} \label{subsecsec:Thm3_enc} 

\vspace{-0.1cm}
    Consider two source sequences each of length $Bn$, $s^{Bn}_{i,1} \in \mS^{Bn}_i, i=1,2$. Partition each sequence into $B$ length-$n$ subsequences, $\svec_{i,b} \in \mS_i^n, b=1,2,\dots,B$. Similarly partition the side information sequences $w_{3,1}^{Bn}$ and $w^{Bn}$ into $B$ length-$n$ subsequences $\wvec_{3,b} \in \mW_3^n, \wvec_{b} \in \mW^n, b=1,2,\dots,B$, respectively. A total of $Bn$ source samples is transmitted over $B+1$ blocks, such that at each block $n$ channel symbols are transmitted.

At block $1$, transmitter $i, i=1,2$, transmits the channel codeword $\xvec_i(\svec_{i,1}, 1)$.
At block $b, b=2,3,\dots,B$, transmitter $i$ transmits the channel codeword $\xvec_i(\svec_{i,b}, \msg_{i,b-1})$, where $\msg_{i,b-1} = f_i(\svec_{i,b-1}) \in \msgCal_i$ is the bin index of source vector $\svec_{i,b-1}$.
Let $(\avec_1, \avec_2) \in \mS_1^n \times \mS_2^n$ be two sequences generated according to $p(\avec_1,\avec_2) = \prod_{k=1}^{n}{p_{S_1,S_2}(a_{1,k}, a_{2,k})}$. These sequences are known to all nodes. At block $B+1$, transmitter $i, i=1,2$, transmits $\xvec_i(\avec_i,\msg_{i,B})$.

At block $b=1$, the relay transmits $\xvec_3(1,1)$.
Assume that at block $b, b=2,3,\dots,B,B+1$, the relay has the estimates $(\tilde{\svec}_{1,b-1}, \tilde{\svec}_{2,b-1})$ of $(\svec_{1,b-1}, \svec_{2,b-1})$. It then finds the corresponding bin indices $\tilde{\msg}_{i,b-1} = f_i(\tilde{\svec}_{i,b-1}) \in \msgCal_i, i=1,2$, and transmits the channel codeword $\xvec_3(\tilde{\msg}_{1,b-1},\tilde{\msg}_{2,b-1})$ at time $b$.

\vspace{-0.4cm}
\subsection{Decoding} \label{subsec:jointNewProofDecoding}

\vspace{-0.1cm}
The relay decodes the source sequences sequentially. At the end of channel block $b$ the relay decodes $\svec_{i,b}, i=1,2$, as follows: 
Using the estimates $(\tmsg_{1,b-1},\tmsg_{2,b-1})$, the received signal $\yvec_{3,b}$ and the side information $\wvec_{3,b}$, the relay decodes $(\svec_{1,b}, \svec_{2,b})$ by looking for a unique pair $(\tsvec_{1}, \tsvec_{2}) \in \mS_1^n \times \mS_2^n$ such that:
\vspace{-0.2cm}
\begin{align}
    & \big(\tsvec_{1}, \tsvec_{2}, \vvec_1(\tmsg_{1,b-1}), \vvec_2(\tmsg_{2,b-1}), \xvec_1(\tsvec_{1}, \tmsg_{1,b-1}), \xvec_2(\tsvec_{2}, \tmsg_{2,b-1}), \xvec_3(\tmsg_{1,b-1}, \tmsg_{2,b-1}), \wvec_{3,b}, \yvec_{3,b}\big) \in \styp.
\label{eq:RelayJntNewDecType}
\end{align}

	\vspace{-0.2cm}
	Decoding at the destination is done via simultaneous backward decoding. 
Let $\bs{\alpha} \in \mW^n$ be an i.i.d sequence such that each letter $\alpha_k$ is selected independently according to $p_{W|S_1,S_2}(\alpha_k | a_{1,k}, a_{2,k}), k=1,2,\dots,n$.	
The destination node waits until the end of channel block $B+1$. It first tries to decode $(\msg_{1,B},\msg_{2,B})$ using the received signal at channel block $B+1$, $\yvec_{b+1}$, and using $\avec_1, \avec_2$, and $\bs{\alpha}$. 
	Going backwards from the last channel block to the first, we assume that at block $b$ the destination has estimates $(\hmsg_{1,b},\hmsg_{2,b})$ of $(\msg_{1,b},\msg_{2,b})$.
The destination simultaneously decodes $(\svec_{1,b},\svec_{2,b}, \msg_{1,b-1},\msg_{2,b-1})$ based on the received signal $\yvec_{b}$, and the side information $\wvec_{b}$, by looking for a unique combination $(\hsvec_{1},\hsvec_{2}, \hmsg_{1},\hmsg_{2})\in \mS_1^n \times \mS_2^n \times \msgCal_1 \times \msgCal_2$ such that:
\vspace{-0.25cm}
\begin{align}
   & \big( \hsvec_{1}, \hsvec_{2}, \vvec_1(\hmsg_{1}), \vvec_2(\hmsg_{2}), \xvec_1(\hsvec_{1}, \hmsg_{1}), \xvec_2(\hsvec_{2}, \hmsg_{2}), \xvec_3(\hmsg_{1}, \hmsg_{2}), \wvec_{b}, \yvec_{b} \big) \in \styp,
\label{eq:DestJntNewSimultDecTypeList}
\end{align}

\vspace{-0.25cm}
\noindent and $f_1(\hsvec_{1})=\hmsg_{1,b}, f_2(\hsvec_{2,b})=\hmsg_{2}$. Denote the decoded variables by $(\hsvec_{1,b},\hsvec_{2,b}, \hmsg_{1,b-1},\hmsg_{2,b-1})$.

\vspace{-0.4cm}
\subsection{Error Probability Analysis} \label{subsec:jointNewProofErrorAnalysis}
	\vspace{-0.1cm}
	\textbf{Relay error probability:} The relay error probability analysis follows the same arguments as the relay error probability analysis detailed in \cite[Appendix B]{Murin:IT11}.
	
	\textbf{Destination error probability:} The average probability of error in decoding at the destination at block $b$, $\bar{P}_{\mbox{\scriptsize dest},b}^{(n)}$, is defined by:
\vspace{-0.45cm}
\begin{align}
    \bar{P}_{\mbox{\scriptsize dest},b}^{(n)} & \triangleq \Pr\big((\hSvec_{1,b},\hSvec_{2,b}) \ne (\Svec_{1,b},\Svec_{2_b})\big). \nonumber
\end{align}

\vspace{-0.2cm}
Due to backward decoding, the pair of source sequences sent at time $b$ is decoded after the pair at time $b+1$ is decoded. Let $ \mF_b \triangleq \big\{\big(\hSvec_{1,b}, \hSvec_{2,b}, \hU_{1,b-1}, \hU_{2,b-1}\big) \ne \big(\Svec_{1,b}, \Svec_{2,b}, U_{1,b-1}, U_{2,b-1} \big)\big\}$. Then, as in \cite[Eqn. (40)]{CoverG:79}, we write:
\vspace{-0.2cm}
\begin{equation}
    \bar{P}_{\mbox{\scriptsize dest}}^{(n)}\le \sum_{b=1}^B\Pr\big(\mF_b\cap\mF_{b+1}^c\big).\footnote{As stated in Subsection \ref{subsecsec:Thm3_enc}, at block $B+1$, source terminal $i$ transmits $\xvec_i(\avec_i,\msg_{i,B})$, where $\avec_i \in \mS_i^n, i=1,2$, is known to all nodes. Therefore, at block $B+1$ we define $\mF_{B+1} \triangleq \big\{\big( \hU_{1,B}, \hU_{2,B}\big) \ne \big(U_{1,B}, U_{2,B} \big)\big\}$.}
\end{equation}

\vspace{-0.15cm}
\noindent Let $\eps_0, \eps_1, \dots, \eps_8$ be positive numbers such that $\eps_0 \ge \eps_1 > \eps, \eps_m > \eps$ and $\eps_m \rightarrow 0$ as $\eps \rightarrow 0$, for $ m=0,1,\dots,8$. 

\noindent Now, define two error events at block $b$:
\vspace{-0.15cm}
\begin{itemize}
    \item Joint-typicality fails:
\vspace{-0.3cm}
\begin{align*}
	\mE_{1,b} & \triangleq \Big\{\big(\Svec_{1,b}, \Svec_{2,b}, \Vvec_1(U_{1,b-1}), \Vvec_2(U_{2,b-1}),\Xvec_1(\Svec_{1,b},U_{1,b-1}), \\
& \qquad \qquad \Xvec_2(\Svec_{2,b},U_{2,b-1}), \Xvec_3(U_{1,b-1},U_{2,b-1}), \Wvec_b, \Yvec_b\big)\notin\styp\Big\}.
\end{align*}

	\vspace{-0.25cm}
  \item Simultaneous decoding of the bin indices (for the next step) and the source sequences fails:
    \vspace{-0.3cm}
		\begin{align*}
    \mE_{2,b} & \triangleq \Big\{\exists \big(\hsvec_1, \hsvec_2, \hu_1, \hu_2\big)\in\mS_1^n\times\mS_2^n\times\mU_1\times\mU_2, \\
    &  \qquad \qquad \big(\hsvec_1, \hsvec_2, \hu_1, \hu_2\big) \ne \big(\Svec_{1,b}, \Svec_{2,b}, U_{1,b-1}, U_{2,b-1} \big), f_1(\tsvec_1) = \hU_{1,b}, f_2(\tsvec_2) = \hU_{2,b},\\
    &  \qquad \qquad  \qquad     \big(\hsvec_{1}, \hsvec_{2}, \hVvec_1(\hu_{1}), \hVvec_2(\hu_{2}), \hXvec_1(\hsvec_{1},\hu_{1}), \hXvec_2(\hsvec_{2},\hu_{2}), \hXvec_3(\hu_{1},\hu_{2}), \Wvec_b,\Yvec_b\big) \in \styp\Big\}.
    \end{align*}
\end{itemize}

\vspace{-0.3cm}
\noindent Then, $\mF_b = \mE_{1,b}\cup\mE_{2,b}$, and we bound:
\vspace{-0.3cm}
\[
      \Pr\big(\mF_b\cap\mF_{b+1}^c\big) \le \Pr\big(\mE_{1,b}\cup\mE_{2,b}\big|\mF_{b+1}^c\big) = \Pr\big(\mE_{1,b}\big|\mF_{b+1}^c\big) + \Pr\big(\mE_{2,b}\big|\mE_{1,b}^c\cap \mF_{b+1}^c\big).
\]

\vspace{-0.25cm}
\noindent By ~applying ~the ~properties ~of ~strong ~typicality, ~\cite[Theorem 6.9]{YeungBook} ~we ~have ~that ~for ~$n$ ~sufficiently ~large, $\Pr\Big(\mE_{1,b}\Big|\mF_{b+1}^c\Big)\le \eps$.
%
%
For bounding $\Pr\big(\mE_{2,b}\big|\mE_{1,b}^c\cap \mF_{b+1}^c\big)$ we consider the following error events:
\begin{flalign*}
        & \mE_{2,b}^{(1)} \triangleq \Big\{ \exists \hu_1 \in \mU_1, \hu_1 \ne U_{1,b-1}, \big(\Svec_{1,b}, \Svec_{2,b}, \hVvec_1(\hu_1), \Vvec_2(U_{2,b-1}), & \\
				& \mspace{80mu} \hXvec_1(\Svec_{1,b},\hu_1),   \Xvec_2(\Svec_{2,b},U_{2,b-1}), \tXvec_3(\hu_1,U_{2,b-1}), \Wvec_b, \Yvec_b \big)\in\styp\Big\}. & \\
        & \mE_{2,b}^{(2)}  \triangleq \Big\{ \exists \hu_2 \in \mU_2, \hu_2 \ne U_{2,b-1}, \big(\Svec_{1,b}, \Svec_{2,b}, \Vvec_1(U_{1,b-1}), \hVvec_2(\hu_2), & \\
				& \mspace{80mu} \Xvec_1(\Svec_{1,b},U_{1,b-1}), \hXvec_2(\Svec_{2,b},\hu_2), \hXvec_3(U_{1,b-1},\hu_{2}), \Wvec_b, \Yvec_b \big)\in\styp\Big\}. & \\
				&\mE_{2,b}^{(3)} \triangleq \Big\{ \exists \hu_1 \in \mU_1, \hu_1 \ne U_{1,b-1}, \exists \hu_2 \in \mU_2, \hu_2 \ne U_{2,b-1},& \\
        & \mspace{80mu} \big(\Svec_{1,b}, \Svec_{2,b}, \hVvec_1(\hu_1), \hVvec_2(\hu_2), \hXvec_1(\Svec_{1,b},\hu_1),    \hXvec_2(\Svec_{2,b},\hu_2), \hXvec_3(\hu_1,\hu_{2}), \Wvec_b, \Yvec_b \big)\in\styp\Big\}.& \\
				&\mE_{2,b}^{(4)}  \triangleq \Big\{ \exists \hsvec_1 \in \mS_1^n, \hsvec_1 \ne \Svec_{1,b}, f_1(\hsvec_1) = \hU_{1,b}, \big(\hsvec_1,\Svec_{2,b}, \Vvec_1(U_{1,b-1}), \Vvec_2(U_{2,b-1}), & \\
        & \mspace{80mu} \hXvec_1(\hsvec_1,U_{1,b-1}),  \Xvec_2(\Svec_{2,b},U_{2,b-1}), \Xvec_3(U_{1,b-1}, U_{2,b-1}), \Wvec_b, \Yvec_b\big) \in \styp\Big\}. & \\
				&\mE_{2,b}^{(5)} \triangleq \Big\{ \exists \hsvec_2 \in \mS_2^n, \hsvec_2 \ne \Svec_{2,b}, f_2(\hsvec_2) = \hU_{2,b}, \big(\Svec_{1,b}, \hsvec_{2}, \Vvec_1(U_{1,b-1}), \Vvec_2(U_{2,b-1}), & \\
			& \mspace{80mu} \Xvec_1(\Svec_{1,b},U_{1,b-1}), \hXvec_2(\tsvec_{2},U_{2,b-1}), \Xvec_3(U_{1,b-1}, U_{2,b-1}), \Wvec_b, \Yvec_b\big) \in \styp\Big\}. & \\
				&\mE_{2,b}^{(6)} \triangleq \Big\{ \exists (\hsvec_1, \hsvec_2) \in \mS_1^n\times\mS_2^n, \hsvec_1 \ne \Svec_{1,b}, \hsvec_2 \ne \Svec_{2,b}, f_1(\hsvec_1) = \hU_{1,b}, f_2(\hsvec_2) = \hU_{2,b}, \big(\hsvec_{1}, \hsvec_{2}, \Vvec_1(U_{1,b-1}), & \\
				& \mspace{80mu} \Vvec_2(U_{2,b-1}), \hXvec_1(\hsvec_{1},U_{1,b-1}), \hXvec_2(\hsvec_{2},U_{2,b-1}), \Xvec_3(U_{1,b-1}, U_{2,b-1}), \Wvec_b, \Yvec_b\big) \in \styp\Big\}. & 
\end{flalign*}
\begin{flalign*}				
				&\mE_{2,b}^{(7)} \triangleq \Big\{ \exists \hsvec_1 \in \mS_1^n, \hsvec_1 \ne \Svec_{1,b},  f_1(\hsvec_1) = \hU_{1,b}, \exists \hu_1\in\mU_1, \hu_1 \ne U_{1,b-1}, \big(\hsvec_{1}, \Svec_{2,b}, \hVvec_1(\hu_{1}), & \\
				& \mspace{80mu} \Vvec_2(U_{2,b-1}), \hXvec_1(\hsvec_{1},\hu_{1}),   \Xvec_2(\Svec_{2,b},U_{2,b-1}), \hXvec_3(\hu_{1}, U_{2,b-1}), \Wvec_b, \Yvec_b\big) \in \styp\Big\}. & \\
				&\mE_{2,b}^{(8)} \triangleq \Big\{ \exists \hsvec_2 \in \mS_2^n, \hsvec_2 \ne \Svec_{2,b},  f_2(\hsvec_2) = \hU_{2,b}, \exists \hu_2\in\mU_2, \hu_2 \ne U_{2,b-1}, \big(\Svec_{1,b}, \hsvec_{2}, \Vvec_1(U_{1,b-1}), & \\
		& \mspace{80mu}	\hVvec_2(\hu_{2}), \Xvec_1(\Svec_{1,b},U_{1,b-1}),   \hXvec_2(\hsvec_{2},\hu_{2}), \hXvec_3(U_{1,b-1}, \hu_{2}), \Wvec_b, \Yvec_b\big) \in \styp\Big\}. & \\
			  &\mE_{2,b}^{(9)} \triangleq \Big\{ \exists \hsvec_1 \in \mS_1^n, \hsvec_1 \ne \Svec_{1,b},  f_1(\hsvec_1) = \hU_{1,b}, \exists \hu_2\in\mU_2, \hu_2\ne U_{2,b-1}, \big(\hsvec_{1}, \Svec_{2,b}, \Vvec_1(U_{1,b-1}), & \\
				& \mspace{80mu} \hVvec_2(\hu_{2}), \hXvec_1(\hsvec_{1},U_{1,b-1}),  \hXvec_2(\Svec_{2,b},\hu_{2}), \hXvec_3(U_{1,b-1}, \hu_{2}), \Wvec_b, \Yvec_b\big) \in \styp\Big\}. & \\
%
				&\mE_{2,b}^{(10)} \triangleq \Big\{ \exists \hsvec_2 \in \mS_2^n, \hsvec_2 \ne \Svec_{2,b},  f_2(\hsvec_2) = \hU_{2,b}, \exists \hu_1\in\mU_1, \hu_1\ne U_{1,b-1}, \big(\Svec_{1,b}, \hsvec_{2}, \hVvec_1(\hu_{1}), & \\
			& \mspace{80mu} \Vvec_2(U_{2,b-1}), \hXvec_1(\Svec_{1,b},\hu_{1}),   \hXvec_2(\hsvec_{2},U_{2,b-1}), \hXvec_3(\hu_{1}, U_{2,b-1}), \Wvec_b, \Yvec_b\big) \in \styp\Big\}. & \\
			  &\mE_{2,b}^{(11)} \triangleq \Big\{ \exists \hsvec_1 \in \mS_1^n, \hsvec_1 \ne \Svec_{1,b},  f_1(\hsvec_1) = \hU_{1,b}, \exists \hu_1\in\mU_1, \hu_1\ne U_{1,b-1}, \exists \hu_2\in\mU_2, \hu_2\ne U_{2,b-1}, & \\
        & \mspace{120mu} \big(\hsvec_{1}, \Svec_{2,b}, \hVvec_1(\hu_{1}), \hVvec_2(\hu_{2}), \hXvec_1(\hsvec_{1},\hu_{1}),   \hXvec_2(\Svec_{2,b},\hu_{2}), \hXvec_3(\hu_{1}, \hu_{2}), \Wvec_b, \Yvec_b\big) \in \styp\Big\}. & \\
				&\mE_{2,b}^{(12)} \triangleq \Big\{ \exists \hsvec_2 \in \mS_2^n, \hsvec_2 \ne \Svec_{2,b},  f_2(\hsvec_2) = \hU_{2,b}, \exists \hu_1\in\mU_1, \hu_1\ne U_{1,b-1}, \exists \hu_2\in\mU_2, \hu_2\ne U_{2,b-1}, & \\
        & \mspace{80mu} \big(\Svec_{1,b}, \hsvec_{2}, \hVvec_1(\hu_{1}), \hVvec_2(\hu_{2}), \hXvec_1(\Svec_{1,b},\hu_{1}),   \hXvec_2(\hsvec_{2},\hu_{2}), \hXvec_3(\hu_{1}, \hu_{2}), \Wvec_b, \Yvec_b\big) \in \styp\Big\}. & \\
				&\mE_{2,b}^{(13)} \triangleq \Big\{ \exists \hsvec_1 \in \mS_1^n, \hsvec_1 \ne \Svec_{1,b},  f_1(\hsvec_1) = \hU_{1,b}, \exists \hsvec_2 \in \mS_2^n, \hsvec_2 \ne \Svec_{2,b},  f_2(\hsvec_2) = \hU_{2,b}, \exists \hu_1\in\mU_1, \hu_1\ne U_{1,b-1}, & \\
        & \mspace{80mu} \big(\hsvec_{1}, \hsvec_{2}, \hVvec_1(\hu_{1}), \Vvec_2(U_{2,b-1}), \hXvec_1(\hsvec_{1},\hu_{1}),   \hXvec_2(\hsvec_{2},U_{2,b-1}), \hXvec_3(\hu_{1}, U_{2,b-1}), \Wvec_b, \Yvec_b\big) \in \styp\Big\}. & \\
				&\mE_{2,b}^{(14)} \triangleq \Big\{ \exists \hsvec_1 \in \mS_1^n, \hsvec_1 \ne \Svec_{1,b},  f_1(\hsvec_1) = \hU_{1,b}, \exists \hsvec_2 \in \mS_2^n, \hsvec_2 \ne \Svec_{2,b},  f_2(\hsvec_2) = \hU_{2,b}, \exists \hu_2\in\mU_2, \hu_2\ne U_{2,b-1}, & \\
        & \mspace{80mu} \big(\hsvec_{1}, \hsvec_{2}, \Vvec_1(U_{1,b-1}), \hVvec_2(\hu_{2}), \hXvec_1(\hsvec_{1},U_{1,b-1}),   \hXvec_2(\hsvec_{2},\hu_{2}), \hXvec_3(U_{1,b-1}, \hu_{2}), \Wvec_b, \Yvec_b\big) \in \styp\Big\}. & \\
				&\mE_{2,b}^{(15)} \triangleq \Big\{ \exists \hsvec_1 \in \mS_1^n, \hsvec_1 \ne \Svec_{1,b},  f_1(\hsvec_1) = \hU_{1,b}, \exists \hsvec_2 \in \mS_2^n, \hsvec_2 \ne \Svec_{2,b},  f_2(\hsvec_2) = \hU_{2,b}, & \\
        & \mspace{80mu} \exists \hu_1\in\mU_1, \hu_1\ne U_{1,b-1}, \exists \hu_2\in\mU_2, \hu_2\ne U_{2,b-1}, & \\
        & \mspace{80mu} \big(\hsvec_{1}, \hsvec_{2}, \hVvec_1(\hu_{1}), \hVvec_2(\hu_{2}), \hXvec_1(\hsvec_{1},\hu_{1}),   \hXvec_2(\hsvec_{2},\hu_{2}), \hXvec_3(\hu_{1}, \hu_{2}), \Wvec_b, \Yvec_b\big) \in \styp\Big\}. &
    \end{flalign*}

\ifthenelse{\boolean{SquizFlag}}{}{}

\ifthenelse{\boolean{SquizFlag}}{}{}

\section{Proof of Proposition \ref{prop:jointCond_New}} \label{annex:jointNewProof} 

\vspace{-0.2cm}
\subsection{Codebook Construction and Encoding}
	
	\vspace{-0.1cm}
	The codebook construction and encoding are identical to \Thmref{thm:jointCond_NewSimult}, see Appendix \ref{annex:jointNewSimultProof}.

\vspace{-0.3cm}
\subsection{Decoding} \label{annex:jointNewProof_decoding}

	\vspace{-0.1cm}
	Decoding at the relay is identical to \Thmref{thm:jointCond_NewSimult}, see Appendix \ref{annex:jointNewSimultProof}. Decoding at the destination is done using successive backward decoding.
Let $\bs{\alpha} \in \mW^n$ be an i.i.d sequence such that each letter $\alpha_k$ is selected independently according to $p_{W|S_1,S_2}(\alpha_k | a_{1,k}, a_{2,k}), k=1,2,\dots,n$.
The destination node waits until the end of channel block $B+1$. It first tries to decode $(\msg_{1,B},\msg_{2,B})$ using the received signal at channel block $B+1$, $\yvec_{B+1}$, and $\bs{\alpha}$.
Going backwards from the last channel block to the first, the destination has the estimates $(\hmsg_{1,b},\hmsg_{2,b})$ of $(\msg_{1,b},\msg_{2,b})$ when decoding at block $b$.
Now, for decoding at block $b$ the destination first recovers the bin indices $\hmsg_{i,b-1}, i=1,2$, corresponding to $\svec_{i,b-1}$, based on its received signal $\yvec_{b}$ and the side information $\wvec_{b}$. This is done by looking for a unique pair $(\hmsg_{1}, \hmsg_{2}) \in \msgCal_1 \times \msgCal_2$ such that:
\vspace{-0.2cm}
\begin{align}
    & \big(\vvec_1(\hmsg_{1}), \vvec_2(\hmsg_{2}), \xvec_3(\hmsg_{1}, \hmsg_{2}), \wvec_{b}, \yvec_{b}\big) \in \styp.
\label{eq:DestJntNewDecTypeIndices}
\end{align}

\vspace{-0.2cm}
\noindent Denote the decoded indices by $(\hmsg_{1,b-1},\hmsg_{2,b-1})$. Next, the destination decodes $\left(\svec_{1,b}, \svec_{2,b}\right)$ by looking for a unique pair $\left(\hsvec_{1}, \hsvec_{2}\right)$ such that:
\vspace{-0.2cm}
\begin{align}
   & \big( \hsvec_{1}, \hsvec_{2}, \vvec_1(\hmsg_{1,b-1}), \vvec_2(\hmsg_{2,b-1}), \xvec_1(\hsvec_{1}, \hmsg_{1,b-1}), \xvec_2(\hsvec_{2}, \hmsg_{2,b-1}), \xvec_3(\hmsg_{1,b-1}, \hmsg_{2,b-1}), \wvec_{b}, \yvec_{b} \big) \in \styp,
\label{eq:DestJntNewDecTypeList}
\end{align}

\vspace{-0.2cm}
\noindent and $f_1(\hsvec_{1})=\hmsg_{1,b}, f_2(\hsvec_{2})=\hmsg_{2,b}$. Denote the decoded sequences with $\left(\hsvec_{1,b}, \hsvec_{2,b}\right)$.

\vspace{-0.3cm}
\subsection{Error Probability Analysis}

	\vspace{-0.1cm}
	Following arguments similar to those in Appendix \ref{subsec:jointNewProofErrorAnalysis} it can be shown that decoding the source sequences at the relay can be done reliably as long as \eqref{bnd:JointNew_rly_S1}--\eqref{bnd:JointNew_rly_S1S2} hold, and decoding the source sequences at the destination can be done reliably as long as \eqref{bnd:JointNew_dst_S1}--\eqref{bnd:JointNew_dst_S1S2} hold.

\vspace{-0.3cm}
\section{Proof of Proposition \ref{prop:CorrSources}}  \label{annex:ProofPropCorrSources}

\vspace{-0.2cm}
\subsection{\Thmref{thm:jointCond_NewSimult} Vs. \Thmref{thm:jointCond}}

\vspace{-0.2cm}
First we compare \eqref{bnd:JointNewSimult_dst_S1} and \eqref{bnd:Joint_dst_S1}. The first term on the RHS of \eqref{bnd:JointNewSimult_dst_S1} can be written as:
\vspace{-0.2cm}
\begin{align}
	I(X_1,X_3;Y|S_2,V_2,X_2,W) & \stackrel{(a)}{=} I(S_1;Y|S_2,V_2,X_2,W) + I(X_1,X_3;Y|S_1,V_2,X_2) \nonumber \\
	&  \ge I(X_1,X_3;Y|S_1,V_2,X_2),
\end{align}

\vspace{-0.2cm}
\noindent where (a) follows from the Markov chains $S_1 \leftrightarrow\ (S_2,V_2,X_1,X_2,X_3,W) \leftrightarrow Y$, $(S_2,W) \leftrightarrow (S_1,V_2,X_2) \leftrightarrow Y$ and from the chain rule for mutual information. From the non-negativity of mutual information it follows that the second term on the RHS of \eqref{bnd:JointNewSimult_dst_S1}, $I(X_1,X_3;Y|S_1,V_2,X_2) + I(X_1;Y|S_2,V_1,X_2,X_3,W)$ is greater than or equal to $I(X_1,X_3;Y|S_1,V_2,X_2)$. As the LHSs of \eqref{bnd:JointNewSimult_dst_S1} and \eqref{bnd:Joint_dst_S1} are the same, we conclude that \eqref{bnd:JointNewSimult_dst_S1} is less restrictive than \eqref{bnd:Joint_dst_S1}. Using similar arguments it also follows that \eqref{bnd:JointNewSimult_dst_S2} is less restrictive than \eqref{bnd:Joint_dst_S2}.
	Next, compare \eqref{bnd:JointNewSimult_dst_S1S2} and~\eqref{bnd:Joint_dst_S1S2}:
	\vspace{-0.2cm}
	\begin{align}
		I(X_1,X_2,X_3;Y|W) & \ge I(X_1,X_2,X_3;Y|S_1,S_2), \label{eq:compare_Simult_Thm1_f}
	\end{align}

\vspace{-0.2cm}	
\noindent where \eqref{eq:compare_Simult_Thm1_f} follows from the Markov chain $(S_1,S_2) \leftrightarrow (X_1,X_2,X_3,W) \leftrightarrow Y$, and from the non-negativity of mutual information. As the LHSs of \eqref{bnd:JointNewSimult_dst_S1S2} and \eqref{bnd:Joint_dst_S1S2} are the same, we conclude that \eqref{bnd:JointNewSimult_dst_S1S2} is less restrictive than \eqref{bnd:Joint_dst_S1S2}.
%
	In conclusion: \Thmref{thm:jointCond_NewSimult} is at least as good as \Thmref{thm:jointCond}.

\vspace{-0.4cm}
\subsection{\Thmref{thm:jointCond_NewSimult} Vs. Prop. \ref{prop:jointCond_New}}
\vspace{-0.2cm}
First consider \eqref{bnd:JointNewSimult_dst_S1} and \eqref{bnd:JointNew_dst_S1}. We begin with the first term on the RHS of \eqref{bnd:JointNewSimult_dst_S1}:
\vspace{-0.2cm}
\begin{align}
	& I(X_1,X_3;Y|S_2,V_2,X_2,W) - I(X_1;Y|S_2,V_1,X_2,X_3,W) - I(V_1,X_3;Y|W, V_2) \nonumber \\
	& \qquad \stackrel{(a)}{=} I(V_1,X_3;Y|S_2,V_2,X_2,W) - I(V_1,X_3;Y|W, V_2) \nonumber \\
%
%
%
	& \qquad \stackrel{(b)}{=} I(V_1,X_3; S_2,X_2|V_2,W,Y) \ge 0,
\end{align}

\vspace{-0.2cm}
\noindent where (a) follows from the chain rule for mutual information; and (b) follows from the Markov relationship $(S_2,X_2) \leftrightarrow (V_2,W) \leftrightarrow (V_1,X_3)$. Next, consider the second term on the RHS of \eqref{bnd:JointNewSimult_dst_S1}:
\vspace{-0.2cm}
\begin{align}
	& I(X_1,X_3;Y|S_1,V_2,X_2) + I(X_1;Y|S_2,V_1,X_2,X_3,W) -	I(X_1;Y|S_2,V_1,X_2,X_3,W) - I(V_1,X_3;Y|V_2,W) \nonumber \\
%
%
	& \qquad \stackrel{(a)}{=} I(X_1;Y|S_1,V_2,X_1,X_2,W) + I(V_1,X_3;Y|S_1,V_2,X_2,W) - I(V_1,X_3;Y|W,V_2) \nonumber \\
	& \qquad \stackrel{(b)}{=} I(X_1;Y|S_1,V_2,X_1,X_2,W) + I(V_1,X_3;S_1,X_2|V_2,W,Y) \ge 0,
\end{align}

\vspace{-0.2cm}
\noindent where (a) follows from the chain rule for mutual information; and (b) follows from the Markov relationship $(S_1,X_2) \leftrightarrow (V_2,W) \leftrightarrow (V_1,X_3)$. As the LHS of \eqref{bnd:JointNewSimult_dst_S1} and \eqref{bnd:JointNew_dst_S1} is the same, we conclude that \eqref{bnd:JointNewSimult_dst_S1} is less restrictive than \eqref{bnd:JointNew_dst_S1}. Using similar arguments it follows that \eqref{bnd:JointNewSimult_dst_S2} is less restrictive than \eqref{bnd:JointNew_dst_S2}. For the expressions involving $H(S_1,S_2|W)$, note that the RHS of \eqref{bnd:JointNewSimult_dst_S1S2} equals to the RHS of \eqref{bnd:JointNew_dst_S1S2}.
Therefore, we conclude that \Thmref{thm:jointCond_NewSimult} is at least as good as Prop. \ref{prop:jointCond_New}.

\vspace{-0.45cm}
\section{Proof of Proposition \ref{prop:NotFeasible}} 	\label{annex:proofNotFeasible}

\vspace{-0.2cm}
	It is enough to show that if at least one of the conditions in \eqref{bnd:Joint} holds with opposite strict inequality, then reliable transmission is not possible via the scheme of \Thmref{thm:jointCond}. The same statement holds for \eqref{bnd:JointFlip} and \Thmref{thm:jointCondFlip}. Furthermore, note that for the deterministic PSOMARC specified in Table \ref{tab:PrimitiveSOMARC}, and for the pair of correlated sources specified in Table \ref{tab:SourceDist}, reliable transmission to the destination requires assistance from the relay. To see this note that $H(S_1,S_2)= \log_2 3$, while $|\mY_S|=2$, which implies that the sources cannot be decoded at the destination without the help of the relay. 
	In Appendix \ref{annex:proofNotFeasible_thm1} we show that when the scheme of \Thmref{thm:jointCond} is used, if the sources can be decoded at the relay then they {\em cannot} be decoded at the destination, i.e., condition \eqref{bnd:Joint_dst_S1S2} holds with strict inequality. In Appendix \ref{annex:proofNotFeasible_thm2} we show that when the scheme of \Thmref{thm:jointCondFlip} is used, then the sources cannot be decoded at the relay, i.e., condition \eqref{bnd:JointFlip_rly_S1S2} holds with strict inequality.
	

	
\vspace{-0.45cm}		
\subsection{Transmission Using the Scheme of Theorem \thmref{thm:jointCond}} \label{annex:proofNotFeasible_thm1}

\vspace{-0.15cm}		 

	We begin with specializing the conditions of \Thmref{thm:jointCond} in \eqref{bnd:Joint_rly_S1}--\eqref{bnd:Joint_dst_S1S2} to the PSOMARC by letting $\mW_3=\mW=\phi$ and $I(X_3;Y_R)=C_3$. From the orthogonality of the relay-destination link it follows that the scheme of \Thmref{thm:jointCond} is optimized by letting $\mV_1=\mV_2=\phi$.
		This fact and the resulting sufficient conditions are stated in the following proposition:
		
		\begin{MyProposition}
			The sufficient conditions of \Thmref{thm:jointCond} in \eqref{bnd:Joint_rly_S1}--\eqref{bnd:Joint_dst_S1S2}, specialized to the PSOMARC, are optimized by letting $\mV_1=\mV_2=\phi$. The resulting conditions are:
			\vspace{-0.2cm}
			\begin{subequations} \label{bnd:Thm7}
			\begin{align}
        H(S_1|S_2) &< \min \{I(X_1;Y_3|S_2, X_2), I(X_1;Y_S|S_1,X_2) + C_3 \} \label{bnd:Thm7_rly_S1} \\
        H(S_2|S_1) &< \min \{I(X_2;Y_3|S_1, X_1), I(X_2;Y_S|S_2,X_1) + C_3 \} \label{bnd:Thm7_rly_S2} \\
        H(S_1,S_2) &< \min \{I(X_1, X_2;Y_3), I(X_1, X_2;Y_S|S_1,S_2) + C_3 \}, \label{bnd:Thm7_rly_S1S2} 
			\end{align}
			\end{subequations}

			\vspace{-0.15cm}
			\noindent subject to a joint distribution that factorizes as
			\vspace{-0.2cm}
			\begin{align}\label{eq:PrimSOMARC_JointDist}
				p(s_1,s_2)p(x_1|s_1)p(x_2|s_2)p(y_3,y_S|x_1,x_2).
			\end{align}
		\end{MyProposition}
		
		\vspace{-0.15cm}
		\begin{IEEEproof}
			We begin with the constraints due to decoding at the relay given by \eqref{bnd:Joint_rly_S1}--\eqref{bnd:Joint_rly_S1S2}. For the RHS of condition \eqref{bnd:Joint_rly_S1} (with $\mW_3=\phi$) we write:	
			\begin{subequations} \label{eq:ExtraThm3RlySum}
			\vspace{-0.2cm}
			\begin{align}
				I(X_1;Y_3|S_2, V_1,X_2,X_3) & \stackrel{(a)}{=} H(Y_3|S_2, V_1,X_2, X_3) - H(Y_3|S_2, X_1,X_2) \nonumber \\
				& \stackrel{(b)}{\le} H(Y_3|S_2,X_2) - H(Y_3|S_2,X_1,X_2) \nonumber \\
				& = I(X_1;Y_3|S_2,X_2), \label{eq:ExtraThm3RlySingle}
			\end{align}
			
			\vspace{-0.15cm}
			\noindent where (a) follows from the definition of the PSOMARC which implies that the 
			Markov chain $(V_1, X_3) \leftrightarrow (S_2, X_1,X_2) \leftrightarrow Y_3$ holds; and (b) follows from the fact the conditioning reduces entropy. Similarly, for the RHS of conditions \eqref{bnd:Joint_rly_S2}--\eqref{bnd:Joint_rly_S1S2} we have:
			\vspace{-0.2cm}
			\begin{align}
				I(X_2;Y_3|S_1, V_2,X_1, X_3) & \le I(X_2;Y_3|S_1,X_1) \\
				I(X_1,X_2;Y_3|V_1,V_2,X_3)	& \le I(X_1,X_2;Y_3). 
			\end{align}
			\end{subequations}
			
			\vspace{-0.15cm}
			Next, consider the constraints due to decoding at the destination given by \eqref{bnd:Joint_dst_S1}--\eqref{bnd:Joint_dst_S1S2}, and recall that for the PSOMARC the channel output at the destination, $Y$, is replaced by the pair of channel outputs $(Y_R,Y_S)$. For the RHS of \eqref{bnd:Joint_dst_S1} we write:
			\vspace{-0.25cm}
			\begin{subequations} \label{eq:ExtraThm4DstSingle_2_sum}
			\begin{align}
				& I(X_1,X_3;Y_R,Y_S|S_1,V_2,X_2) \nonumber \\
				& \qquad = I(X_1;Y_R,Y_S|S_1,V_2,X_2) + I(X_3;Y_R|S_1,V_2,X_1,X_2) + I(X_3;Y_S|S_1,V_2,X_1,X_2,Y_R) \nonumber \\
				& \qquad \stackrel{(a)}{=} I(X_1;Y_S|S_1,V_2,X_2)	+ I(X_1;Y_R|S_1,V_2,X_2,Y_S) + I(X_3;Y_R|S_1,V_2,X_1,X_2) \nonumber \\
				& \qquad = I(X_1;Y_S|S_1,V_2,X_2) + H(Y_R|S_1,V_2,X_2,Y_S) - H(Y_R|S_1,V_2,X_1,X_2,Y_S) \nonumber \\
				& \qquad \qquad + H(Y_R|S_1,V_2,X_1,X_2) - H(Y_R|S_1,V_2,X_1,X_2,X_3) \nonumber \\
				& \qquad  \stackrel{(b)}{=} I(X_1;Y_S|S_1,V_2,X_2) + H(Y_R|S_1,V_2,X_2,Y_S) - H(Y_R|X_3)	\nonumber \\
				& \qquad \stackrel{(c)}{\le} I(X_1;Y_S|S_1,X_2) + I(X_3;Y_R), \label{eq:ExtraThm4DstSingle_1}
			\end{align}
		
			\vspace{-0.2cm}
			\noindent where (a) follows from the fact that $Y_S$ is uniquely determined by $X_1$ and $X_2$, and therefore it follows that $I(X_3;Y_S|S_1,V_2,X_1,X_2,Y_R)=0$; (b) follows from the Markov chain $Y_S \leftrightarrow (S_1,V_2,X_1,X_2) \leftrightarrow Y_R$ (which directly follows from the definition of the conditional distribution function of the SOMARC: $p(y_R,y_S,y_3|x_1,x_2,x_3)=p(y_R|x_3)p(y_S,y_3|x_1,x_2)$), and from the Markov chain $(S_1,V_2,X_1,X_2) \leftrightarrow X_3 \leftrightarrow Y_R$; and (c) follows from the arguments leading to \eqref{eq:ExtraThm3RlySingle} and from the fact that conditioning reduces entropy. Similarly, for the RHS of conditions \eqref{bnd:Joint_dst_S2}--\eqref{bnd:Joint_dst_S1S2} we have:
			\vspace{-0.25cm}
			\begin{align}
				I(X_2,X_3;Y_R,Y_S|S_2,V_1,X_1) & \le I(X_2;Y_S|S_2,X_1) + I(X_3;Y_R) \\
				I(X_1,X_2,X_3;Y_R,Y_S|S_1,S_2) & \le I(X_1,X_2;Y_S|S_1,S_2) + I(X_3;Y_R). 
			\end{align}
			\end{subequations}
			
			\vspace{-0.15cm}
			\noindent Finally, substituting $I(X_3;Y_R)=C_3$ in \eqref{eq:ExtraThm4DstSingle_2_sum} and combining with \eqref{eq:ExtraThm3RlySum}, we obtain the RHSs of conditions \eqref{bnd:Thm7}. 
			Note that conditions \eqref{bnd:Thm7} are subject to the chain: 
			\vspace{-0.2cm}
			\begin{equation*}
				p(s_1,s_2,v_1,v_2,x_1,x_2,y_3,y_s) = p(s_1,s_2)p(v_1)p(x_1|s_1,v_1)p(v_2)p(x_2|s_2,v_2)p(y_3,y_S|x_1,x_2).
			\end{equation*}
			
			\vspace{-0.2cm}
			\noindent Furthermore, as \eqref{bnd:Thm7} is independent of $(V_1,V_2)$ then the resulting chain is:
			\vspace{-0.2cm}
			\begin{equation}
				\sum_{(v_1,v_2) \in \mV_1 \times \mV_2} p(s_1,s_2,v_1,v_2,x_1,x_2,y_3,y_s) \mspace{-3mu} = \mspace{-3mu} p(s_1,s_2)p(x_1|s_1)p(x_2|s_2)p(y_3,y_S|x_1,x_2). \label{eq:chainSum}
			\end{equation}
			
			\vspace{-0.15cm}
			\noindent Lastly, note that the upper bounds \eqref{eq:ExtraThm3RlySum}--\eqref{eq:ExtraThm4DstSingle_2_sum}, subject to the chain \eqref{eq:chainSum}, are obtained by letting $\mV_1=\mV_2=\phi$ in \eqref{bnd:Joint} and \eqref{eq:JntJointDist}. Thus, $\mV_1=\mV_2=\phi$ maximizes the sufficient conditions of \Thmref{thm:jointCond}.
		\end{IEEEproof}

	Next, note that the LHS of condition \eqref{bnd:Thm7_rly_S1S2}, evaluated for the sources defined in Table \ref{tab:SourceDist}, equals $\log_2 3$ bits. Therefore, for successfully transmitting $S_1$ and $S_2$ we must have that the RHS of \eqref{bnd:Thm7_rly_S1S2} is greater than (or equals to) $\log_2 3$.
	Now, consider the RHS of condition \eqref{bnd:Thm7_rly_S1S2} for these sources and the PSOMARC defined in Table \ref{tab:PrimitiveSOMARC}: finding the maximum of $I(X_1, X_2;Y_3)$ over all $p(x_1|s_1)p(x_2|s_2)$ we have:
	\vspace{-0.1cm}
	\begin{equation} 
		\max_{p(x_1|s_1)p(x_2|s_2)} I(X_1,X_2;Y_3) = \max_{p(x_1|s_1)p(x_2|s_2)} H(Y_3),
		\label{eq:Thm1Maximization}
	\end{equation}

	\vspace{-0.1cm}
	\noindent which follows as the channel from $(X_1,X_2)$ to $Y_3$ is deterministic.
	As $|\mY_3|=3$, it follows that ${\dst \max_{p(x_1|s_1)p(x_2|s_2)} \mspace{-12mu} H(Y_3)= }$ ${\log_2 3}$ if and only if $\Pr \{Y_3 = j\}=1/3, j=0,1,2$. This requires that $\Pr \{(X_1,X_2) = (0,0) \} = \Pr \{(X_1,X_2) = (1,1) \} = 1/3$ and $\Pr \{((X_1,X_2) = (0,1)) \cup ((X_1,X_2) = (1,0)) \} = 1/3$.
	Since the sources distribution is given, $\Pr \{(X_1,X_2) = (i,j) \}$ depends only on $p(x_1|s_1)p(x_2|s_2)$, which consists of four unknowns. This corresponds to an algebraic equations system with three equations, four unknowns, and the constraint that all the variables are in the range $[0,1]$. The two possible solutions of this system, solved using {\tt Mathematica}\footnote{Let $p_{i,j} \triangleq \Pr\{X_i = j | S_i = 0 \}, i,j=0,1$. The following algebraic equations system is solved:
	\vspace{-0.15cm}
	\begin{align*}
		\mathtt{Solve[} & \mathtt{p_{00} \cdot p_{10} + p_{00} \cdot p_{11} + p_{01} \cdot p_{11} == 1 \&\& (1 - p_{00}) \cdot (1 - p_{10}) + (1 - p_{00}) \cdot (1 - p_{11}) + (1 - p_{01}) \cdot (1 - p_{11}) == 1 \&\&} \\
  & \mathtt{p_{00}  \cdot (1 - p_{10}) + p_{00} \cdot (1 - p_{11}) + p_{01} \cdot (1 - p_{11}) + (1 - p_{00}) \cdot p_{10} + (1 - p_{00}) \cdot p_{11} + (1 - p_{01}) \cdot p_{11} == 1 \&\&} \\
	& \mathtt{0 <= p_{00} <= 1 \&\& 0 <= p_{01} <= 1 \&\& 0 <= p_{10} <= 1 \&\& 0 <= p_{11} <= 1, \{p_{00}, p_{01}, p_{10}, p_{11}\}]},
	\end{align*}
	
	\vspace{-0.2cm}
	
	to obtain $\mathtt{\{\{p_{00} = 0, p_{01} = 1, p_{10} = 0, p_{11} = 1\}, \{p_{00} = 1, p_{01} = 0, p_{10} = 1, p_{11} = 0\}\}}$.
	}, are deterministic mappings from $s_i$ to $x_i$.\footnote{This is also validated via an exhaustive search.}
	The expression $I(X_1, X_2;Y_S|S_1,S_2) + C_3$, evaluated using each of these conditional distributions, equals $1$ bit. 
	Therefore, the RHS of condition \eqref{bnd:Thm7_rly_S1S2}, when evaluated using 
	these conditional distributions, is strictly smaller than $\log_2 3$.
	This implies that for these sources and PSOMARC, condition \eqref{bnd:Thm7_rly_S1S2} holds with opposite strict inequality, and we conclude that reliable transmission via the scheme of \Thmref{thm:jointCond} is impossible.

	\vspace{-0.25cm}
	\subsection{Transmission Using the Scheme of Theorem \thmref{thm:jointCondFlip}} \label{annex:proofNotFeasible_thm2}
	
	\vspace{-0.10cm}
	Specializing the conditions of \Thmref{thm:jointCondFlip} in \eqref{bnd:JointFlip_rly_S1}--\eqref{bnd:JointFlip_dst_S1S2} to the PSOMARC by letting $\mW_3=\mW=\phi$  and $I(X_3;Y_R)=C_3$, results in the following sufficient conditions:
\ifthenelse{\boolean{SquizFlag}}{}{}
	\vspace{-0.25cm}
	\begin{subequations} \label{bnd:Thm8}
    \begin{align}
        H(S_1|S_2) &< \min \{I(X_1;Y_3|S_1, X_2), I(X_1;Y_S|S_2,X_2) + C_3 \} \label{bnd:Thm8_rly_S1} \\
        H(S_2|S_1) &< \min \{I(X_2;Y_3|S_2, X_1), I(X_2;Y_S|S_1,X_1) + C_3 \} \label{bnd:Thm8_rly_S2} \\
        H(S_1,S_2) &< \min \{I(X_1, X_2;Y_3|S_1,S_2), I(X_1, X_2;Y_S) + C_3 \}, \label{bnd:Thm8_rly_S1S2}
    \end{align}
\end{subequations}

	\vspace{-0.25cm}
	\noindent subject to the input distribution \eqref{eq:PrimSOMARC_JointDist}.

	\noindent Consider maximizing the  mutual information expression $I(X_1, X_2;Y_3|S_1,S_2)$ on the RHS of condition \eqref{bnd:Thm8_rly_S1S2} for the considered sources and PSOMARC, over all $p(x_1|s_1)p(x_2|s_2)$:
	\vspace{-0.2cm}
	\begin{align} 
		 &\mspace{-15mu} \max_{p(x_1|s_1)p(x_2|s_2)} I(X_1,X_2;Y_3|S_1,S_2) \nonumber \\
		& \stackrel{(a)}{=} \max_{p(x_1|s_1)p(x_2|s_2)} H(Y_3|S_1,S_2) \nonumber \\
%
%
		& \stackrel{(b)}{=} \max_{p(x_1|s_1)p(x_2|s_2)} \sum_{\substack{(\tilde{s}_1,\tilde{s}_2) \in \mS_1 \times \mS_2, \\ p(\tilde{s}_1,\tilde{s}_2) \ne 0}}{ p(\tilde{s}_1,\tilde{s}_2) \cdot H\big(Y_3|(S_1,S_2)=(\tilde{s}_1,\tilde{s}_2)\big) } \nonumber \\
%
%
%
		& \stackrel{(c)}{\le} \frac{1}{6} \cdot \sum_{\substack{(\tilde{s}_1,\tilde{s}_2) \in \mS_1 \times \mS_2, \\ p(\tilde{s}_1,\tilde{s}_2) \ne 0}}{ \max_{p(x_1|\tilde{s}_1)p(x_2|\tilde{s}_2)} \left\{ - \sum_{y_3 \in \mY_3}{p(y_3|\tilde{s}_1,\tilde{s}_2) \cdot \log_2 p(y_3|\tilde{s}_1,\tilde{s}_2)} \right\} } \nonumber \\
%
		& = \frac{1}{6} \cdot \sum_{\substack{(\tilde{s}_1,\tilde{s}_2) \in \mS_1 \times \mS_2, \\ p(\tilde{s}_1,\tilde{s}_2) \ne 0}}{ \max_{p(x_1|\tilde{s}_1)p(x_2|\tilde{s}_2)}} \left\{ - \mspace{-10mu} \sum_{y_3 \in \mY_3}{\mspace{-35mu} \sum_{\mspace{50mu} (x_1,x_2) \in \mX_1 \times \mX_2 } \mspace{-50mu} p(y_3,x_1,x_2|\tilde{s}_1,\tilde{s}_2) \cdot \log_2 \left(  \mspace{-35mu} \sum_{\mspace{50mu} (x_1,x_2) \in \mX_1 \times \mX_2 } \mspace{-50mu} p(y_3,x_1,x_2|\tilde{s}_1,\tilde{s}_2) \right) }  \right\}  \nonumber
	\end{align}
	\begin{align}			
%
%
		& \stackrel{(d)}{=} \frac{1}{6} \cdot \sum_{\substack{(\tilde{s}_1,\tilde{s}_2) \in \mS_1 \times \mS_2, \\ p(\tilde{s}_1,\tilde{s}_2) \ne 0}}{ \max_{p(x_1|\tilde{s}_1)p(x_2|\tilde{s}_2)}} \nonumber \\
		& \mspace{50mu}  \left\{ - \mspace{-10mu} \sum_{y_3 \in \mY_3}{\mspace{-40mu} \sum_{ \mspace{50mu} (x_1,x_2) \in \mX_1 \times \mX_2 } \mspace{-50mu} p(x_1|\tilde{s}_1)p(x_2|\tilde{s}_2) p(y_3|x_1,x_2) \cdot \log_2 \left( \mspace{-40mu}  \sum_{\mspace{50mu} (x_1,x_2) \in \mX_1 \times \mX_2 } \mspace{-50mu} p(x_1|\tilde{s}_1)p(x_2|\tilde{s}_2) p(y_3|x_1,x_2) \right) }  \right\} \nonumber \\ 
%
		& \stackrel{(e)}{=} \frac{1}{6} \cdot \mspace{-50mu} \sum_{\mspace{50mu} \substack{(\tilde{s}_1,\tilde{s}_2) \in \mS_1 \times \mS_2, \\ p(\tilde{s}_1,\tilde{s}_2) \ne 0}}{\mspace{-7mu} \max_{p(x_1)p(x_2)}} \left\{ - \mspace{-10mu} \sum_{y_3 \in \mY_3}{\mspace{-40mu} \sum_{\mspace{50mu} (x_1,x_2) \in \mX_1 \times \mX_2 } \mspace{-50mu} p(x_1)p(x_2)p(y_3|x_1,x_2) \cdot \log_2 \left( \mspace{-40mu} \sum_{\mspace{50mu} (x_1,x_2)  \in \mX_1 \times \mX_2 } \mspace{-50mu} p(x_1)p(x_2)p(y_3|x_1,x_2) \right)} \right\} \nonumber \\
%
%
%
%
%
		& = \max_{p(x_1)p(x_2)} H(Y_3) \nonumber \\
%
		& \stackrel{(f)}{=} 1.5, 
	\end{align}
	
	\vspace{-0.25cm}
	\noindent where (a) follows from the fact that $Y_3$ is a deterministic function of $(X_1,X_2)$; (b) follows from the definition of conditional entropy; (c) follows from the joint distribution of the sources in Table \ref{tab:SourceDist} and the fact that the maximum of a sum is less than the sum of the maximum of the summands; (d) follows from the Markov chain $(S_1,S_2)-(X_1,X_2)-Y_3$;	(e) follows from the fact that since $\tilde{s}_1$ and $\tilde{s}_2$ appear only in the conditioning of the conditional distributions $p(x_1|\tilde{s}_1),p(x_2|\tilde{s}_2)$, the maximizing $p(x_1|\tilde{s}_1)p(x_2|\tilde{s}_2)$ is the same for any pair $(\tilde{s}_1,\tilde{s}_2)$. Thus, the maximizing $p(x_1|\tilde{s}_1)p(x_2|\tilde{s}_2)$ is independent of the value of $(\tilde{s}_1,\tilde{s}_2)$; 
	finally, (f) follows from \cite{Cover:80}. 
	
	Recall that $H(S_1,S_2) = \log_2 3$ bits. Thus, $H(S_1,S_2) > \max_{p(x_1|s_1)p(x_2|s_2)} I(X_1,X_2;Y_3|S_1,S_2)$,
	%
	 \noindent and \eqref{bnd:Thm8_rly_S1S2} holds with strict opposite inequality. Therefore we conclude that reliable transmission via the scheme of \Thmref{thm:jointCondFlip} is impossible.
		This concludes the proof of Prop. \ref{prop:NotFeasible}.

\vspace{-0.3cm}
\section{Proof of Proposition \ref{prop:feasible}} 	\label{annex:proofFeasible}

		%
	%
		Here, instead of specializing the conditions of \Thmref{thm:jointCond_NewSimult} to the PSOAMRC, we analyze the decoding rules of \Thmref{thm:jointCond_NewSimult} given in \eqref{eq:RelayJntNewDecType}--\eqref{eq:DestJntNewSimultDecTypeList} for a specific $p(x_i|s_i),i=1,2$.
Let $p(x_i|s_i), i=1,2$, be the deterministic distribution $p(x_i|s_i)=\delta(x_i-s_i)$, 
where $\delta(x)$ is the Kronecker Delta function, 
and set $\mV_1=\mV_2=\phi$. Hence, there is no superposition encoding at the sources, and the cooperation between the sources and the relay is based only on the codeword transmitted by the relay. 

\vspace{-0.4cm}
\subsection{Encoding at the Relay}
Let $\mQ \triangleq \{ 1,2,\dots, 2^n\}$, and let $f_3: (\svec_1,\svec_2) \mapsto \mQ$, be the encoding function at the relay. 
At block $b=1$, the relay transmits the codeword $1$. 
Assume that at block $b, b=2,3,\dots,B,B+1$, the relay has the estimates $(\tilde{\svec}_{1,b-1}, \tilde{\svec}_{2,b-1})$ of $(\svec_{1,b-1}, \svec_{2,b-1})$. Then, at time $b$, the relay transmits the channel codeword $q_{b-1} = f_3(\tsvec_{1,b-1},\tsvec_{2,b-1}), q_{b-1} \in \mQ$.

\vspace{-0.4cm}
\subsection{Decoding at the Relay}

\vspace{-0.2cm}
\subsubsection{Decoding rule} For the mapping defined in Table \ref{tab:PrimitiveSOMARC} and the specified $p(x_i|s_i)$, the relay decoding rule \eqref{eq:RelayJntNewDecType} is specialized to the following decoding rule: {\slshape the relay decodes $(\svec_{1,b}, \svec_{2,b})$ by looking for a unique pair $(\tsvec_{1}, \tsvec_{2}) \in \mS_1^n \times \mS_2^n$ such that $\big(\tsvec_{1}, \tsvec_{2}, \yvec_{3,b}\big) \in \styp$}. Denote the decoded sequences by $(\tsvec_{1,b}, \tsvec_{2,b})$.

\subsubsection{Error probability analysis} Let $\mE_r \triangleq \left\{ \left( \tSvec_{1,b}, \tSvec_{2,b} \right) \ne \left( \Svec_{1,b}, \Svec_{2,b} \right) \right\}$. The average probability of error for decoding at the relay at block $b$, $\bar{P}_{r,b}^{(n)}$, is defined as:
\vspace{-0.2cm}
\begin{align}
    \bar{P}_{r,b}^{(n)} & \triangleq \sum_{(\svec_{1,b},\svec_{2,b}) \in \mS_1^n \times \mS_2^n}{\mspace{-24mu} p(\svec_{1,b},\svec_{2,b})} \Pr \Big( \mE_r | \svec_{1,b},\svec_{2,b} \Big) \nonumber \\
    &\leq  \sum_{(\svec_{1,b},\svec_{2,b}) \notin \styp(S_1,S_2)}{\mspace{-54mu} p(\svec_{1,b},\svec_{2,b})} + \sum_{(\svec_{1,b},\svec_{2,b}) \in \styp(S_1,S_2)}{\mspace{-54mu} p(\svec_{1,b},\svec_{2,b})}  \Pr \Big( \mE_r | (\svec_{1,b},\svec_{2,b})\in \styp \Big).  \label{eq:RelayDecErrProbDef_PrimSOMARC}
\end{align}

\noindent From \cite[Thm. 6.9]{YeungBook} the first sum in \eqref{eq:RelayDecErrProbDef_PrimSOMARC} can be bounded by $\epsilon$. Next, by the union bound we write:
\vspace{-0.2cm}
\begin{align}
	\Pr \Big( \mE_r | (\svec_{1,b},\svec_{2,b})\in \styp \Big) & \leq \Pr \Big( \big(\svec_{1,b}, \svec_{2,b}, \Yvec_{3,b}\big) \notin \styp | (\svec_{1,b},\svec_{2,b})\in \styp \Big) \nonumber \\
	& \quad + \Pr \Big( \exists (\tsvec_{1}, \tsvec_{2}) \neq (\svec_{1,b}, \svec_{2,b}): (\tsvec_{1}, \tsvec_{2}, \Yvec_{3,b}\big) \in \styp | (\svec_{1,b},\svec_{2,b})\in \styp \Big). \label{eq:PrimSOMARC_relayErrorProb}
\end{align}

\vspace{-0.2cm}
\noindent For the specified $p(x_i|s_i), i=1,2$, and the channel mapping defined in Table \ref{tab:PrimitiveSOMARC}, $Y_3$ is a deterministic function of the sources $S_1$ and $S_2$. Moreover, there is one-to-one mapping between the source pairs $(S_1,S_2)$ and $Y_3$. Hence, for each possible source pair $(S_1,S_2)$ there is a unique value of $Y_3$, and we conclude that:
\vspace{-0.2cm}
\begin{equation}
	 \Pr \Big( \big(\svec_{1,b}, \svec_{2,b}, \Yvec_{3,b}\big) \notin \styp | (\svec_{1,b},\svec_{2,b})\in \styp \Big) = 0.
\label{eq:PrimSOMARC_RelayGoodSources}
\end{equation} 

\vspace{-0.2cm}
\noindent From the one-to-one mapping between the source pairs $(S_1,S_2)$ and $Y_3$, and from the definition of strong typicality, \cite[Ch. 6.1]{YeungBook}, it follows that:
\vspace{-0.2cm}
\begin{align}
	 \Pr \Big( \exists (\tsvec_{1}, \tsvec_{2}) \neq (\svec_{1,b}, \svec_{2,b}): (\tsvec_{1}, \tsvec_{2}, \Yvec_{3,b}\big) \in \styp | (\svec_{1,b},\svec_{2,b})\in \styp \Big) = 0.
\label{eq:PrimSOMARC_RelayBadSources}
\end{align}


\noindent Combining \eqref{eq:PrimSOMARC_relayErrorProb}--\eqref{eq:PrimSOMARC_RelayBadSources} yields $\bar{P}_{r,b}^{(n)} \leq \epsilon$ for sufficiently large $n$. We conclude that the sources of Table \ref{tab:SourceDist} can be reliably transmitted over the channel to the relay.


\vspace{-0.4cm}
\subsection{Decoding at the Destination}

\vspace{-0.2cm}
\subsubsection{Decoding rule} 
Recall that $q_{b}$ is available at the destination assuming the relay correctly decoded the source sequences. 
The destination decoding rule of \Thmref{thm:jointCond_NewSimult}, see \eqref{eq:DestJntNewSimultDecTypeList}, is specialized to the following decoding rule:\footnote{This follows from the fact that the relay's information is transmitted via an orthogonal link.} {\slshape the destination decodes $(\svec_{1,b}, \svec_{2,b})$, by looking for a unique pair $(\hsvec_{1}, \hsvec_{2})\in \mS_1^n \times \mS_2^n$ such that $\big(\hsvec_{1}, \hsvec_{2}, \yvec_{S,b}\big) \in \styp$ and $f_3(\hsvec_{1}, \hsvec_{2})=q_{b}$}. Denote the decoded sequences by $(\hsvec_{1,b}, \hsvec_{2,b})$.

\subsubsection{Error probability analysis} Let $\mE_d \triangleq \left\{ \big( \hSvec_{1,b}, \hSvec_{2,b} \big) \ne \left( \Svec_{1,b}, \Svec_{2,b} \right) \right\}$. Following the same arguments that led to \eqref{eq:RelayDecErrProbDef_PrimSOMARC}, the average probability of decoding error at the destination at block $b$, $\bar{P}_{d,b}^{(n)}$ can be upper bounded as:
\vspace{-0.25cm}
\begin{align}
    \bar{P}_{d,b}^{(n)} & \leq \epsilon + \sum_{(\svec_{1,b},\svec_{2,b}) \in \styp}{\mspace{-30mu} p(\svec_{1,b},\svec_{2,b})}  \Pr \Big( \mE_d \big| (\svec_{1,b},\svec_{2,b}) \in \styp  \Big). \label{eq:DestDecErrProbDef_PrimSOMARC}
\end{align}

\vspace{-0.1cm}
\noindent Using the union bound $\Pr \Big( \mE_d \big| (\svec_{1,b},\svec_{2,b}) \in \styp  \Big)$ can be upper bounded by:
\vspace{-0.2cm}
\begin{align}
	& \Pr \Big( \big(\svec_{1,b}, \svec_{2,b}, \Yvec_{S,b}\big) \notin \styp \big| (\svec_{1,b},\svec_{2,b}) \in \styp \Big) + \nonumber \\
	& \qquad \Pr \Big( \exists (\hsvec_{1}, \hsvec_{2}) \neq (\svec_{1,b}, \svec_{2,b}): \big\{(\hsvec_{1}, \hsvec_{2}, \Yvec_{S,b}\big) \in \styp \big\} \cap  \big\{ f_3(\hsvec_{1}, \hsvec_{2})=q_{b}\big\} \big| (\svec_{1,b},\svec_{2,b}) \in \styp \Big). \label{eq:PrimSOMARC_destErrorProb}
\end{align}

\vspace{-0.25cm}
\noindent Since $x_i=s_i, i=1,2$, and $Y_S$ is a deterministic function of $(X_1,X_2)$ then as $(\svec_{1,b}, \svec_{2,b}) \in \styp$ it follows that $(\svec_{1,b}, \svec_{2,b}, \Yvec_{S,b}) \in \styp$, thus
\vspace{-0.25cm}
\begin{equation}
	\Pr \Big( \big(\svec_{1,b}, \svec_{2,b}, \Yvec_{S,b}\big) \notin \styp \big| ((\svec_{1,b},\svec_{2,b}) \in \styp \Big) = 0.
\label{eq:PrimSOMARC_DestGoodSources}
\end{equation}

\vspace{-0.2cm}
\noindent The channel to the destination does not provide a one-to-one mapping between the pair $(S_1,S_2)$ and $Y_S$. 
\noindent Let $\theta(y_S)$ denote the inverse mapping from the channel output $Y_S$ to the sources, e.g., $\theta(0) = \{ (0,0), (0,1)\}$. 
From \cite[Def. 6.6]{YeungBook} it follows that if $\big(\svec_{1}, \svec_{2}, \Yvec_{S}\big) \in \styp$ then: 
\vspace{-0.2cm}
\begin{align}
	\forall y_{S,k}: (s_{1,k}, s_{2,k}) \in \theta(y_{S,k}), \quad k=1,2,\dots,n.
	\label{eq:inverseMapping}
\end{align}

\vspace{-0.2cm}
\noindent Furthermore, $\forall y_S\in\mY_S: \left\| \theta(y_S) \right\| = 2$. Therefore, by mapping the two elements of $\theta(y_S)$ into different symbols transmitted from the relay we can guarantee that the condition $f_3(\hsvec_{1},\hsvec_{2})=q_{b}$ holds only for the transmitted source sequences.\footnote{From the fact that $\forall y\in\mY: \left\| \theta(y) \right\| = 2$ it follows that resolving the ambiguity in $\theta(y)$ requires 1 bit per source pair, and therefore, this information can be transmitted from the relay via the relay-destination link with capacity $C_3=1$ bit. 
}
\noindent Hence, we conclude that the combination of the codeword transmitted by the relay and $\Yvec_S$ uniquely identifies the transmitted source pair. Thus,
\vspace{-0.2cm}
\begin{align}
	& \mspace{-5mu} \Pr \Big( \exists (\hsvec_{1}, \hsvec_{2}) \neq (\svec_{1,b}, \svec_{2,b}): \big\{(\hsvec_{1}, \hsvec_{2}, \Yvec_{S,b}\big) \in \styp\big\} \cap \big\{ f_3(\hsvec_{1}, \hsvec_{2})=q_{b}\big\} \Big| (\svec_{1,b},\svec_{2,b}) \in \styp \Big) = 0. \label{eq:PrimSOMARC_destBadSources}
\end{align}

\vspace{-0.2cm}
\noindent Combining \eqref{eq:DestDecErrProbDef_PrimSOMARC}--\eqref{eq:PrimSOMARC_destBadSources} yields $\bar{P}_{d,b}^{(n)} \leq \epsilon$ for $n$ large enough. We conclude that the sources of Table \ref{tab:SourceDist} can be reliably transmitted over the channel to the destination.


\ifthenelse{\boolean{SquizFlag}}{}{}

\vspace{-0.2cm}
\section{Proofs of Theorem \thmref{thm:OuterMarkov} and Proposition \ref{prop:VGeneral}} \label{annex:MAC_BC_proofs}

\vspace{-0.25cm}
\subsection{Proof of \Thmref{thm:OuterMarkov}} \label{subsec:MarkovBound_proof}
    
		\vspace{-0.15cm}
		Assume a sequence of encoders $f_i^{(n)}, i=1,2,3$, and decoders $g^{(n)}$ is specified such that $P_e^{(n)} \rightarrow 0$ as $n \rightarrow \infty$. Fano's inequality \cite[Ch. 2.8]{YeungBook}, in the context of the current scenario, states that:
    \vspace{-0.25cm}
		\begin{eqnarray}
        H(S_1^n, S_2^n| \hat{S}_{1}^n, \hat{S}_{2}^n) \leq 1 + n P_e^{(n)} \log_2 \left| \mS_1 \times \mS_2 \right|\triangleq n \gamma(P_e^{(n)}),
    \label{eq:FanoDest}
    \end{eqnarray}

		\vspace{-0.2cm}
    \noindent where $\gamma(x)$ is a non-negative function that approaches $\frac{1}{n}$ as $x \rightarrow 0$.
    We also obtain:
    \vspace{-0.2cm}
		\begin{align}
        H(S_1^n, S_2^n| \hat{S}_{1}^n, \hat{S}_{2}^n) & \stackrel{(a)}{\ge} H(S_1^n, S_2^n |W^n, Y^n ) \stackrel{(b)}{\geq} H(S_1^n |S_2^n, W^n, Y^n) \label{eq:generalDestEntropyBasic},
    \end{align}

		\vspace{-0.2cm}
    \noindent where (a) follows from the fact that conditioning reduces entropy, and from the fact that $(\hat{S}_{1}^n, \hat{S}_{2}^n)$ is a deterministic function of $(Y^n,W^n)$; (b) follows from non-negativity of the entropy function for discrete sources.
		Constraint \eqref{bnd:outr_markov_dst_S1} is a consequence of the following chain of inequalities:
		\vspace{-0.15cm}
    \begin{align}
        & \sum_{k=1}^{n}{I(X_{1,k}, X_{3,k};Y_{k}| S_{2,k}, X_{2,k}, W_{k})} \nonumber \\
%
        & \qquad \qquad \stackrel{(a)}{=} \sum_{k=1}^{n}{\Big[H(Y_{k}|S_{2,k}, X_{2,k}, W_{k})} - H\big(Y_{k}|S_1^n, S_2^n, X_{1,1}^{k}, X_{2,1}^{k}, X_{3,1}^{k}, W^n, W_{3,1}^n, Y^{k-1}, Y_{3,1}^{k-1}\big)\Big] \nonumber \\
        & \qquad \qquad \stackrel{(b)}{\geq} \sum_{k=1}^{n}{\Big[H(Y_{k}|S_2^n, X_{2,k}, W^n, Y^{k-1})} - H(Y_{k}|S_1^n, S_2^n, W^n, W_{3,1}^n, Y^{k-1})\Big] \nonumber \\
%
        & \qquad \qquad \stackrel{(c)}{=} I(S_1^n, W_{3,1}^n; Y^n|S_2^n, W^n) \nonumber \\
%
%
        & \qquad \qquad \stackrel{(d)}{\geq} H(S_1^n | S_2^n, W^n) - H(S_1^n |S_2^n, W^n, Y^n) \nonumber \\
        & \qquad \qquad \stackrel{(e)}{\geq} nH(S_1 | S_2, W) - n\gamma(P_e^{(n)}) \label{eq:generalDestSingleChain},
    \end{align}


		\vspace{-0.2cm}
    \noindent where (a) follows from the memoryless channel assumption (see \eqref{eq:MARCchanDist}) and the causal Markov relation $(S_1^n, S_2^n, W^n,$ $W_{3,1}^n) \leftrightarrow (X_{1,1}^{k}, X_{2,1}^{k}, X_{3,1}^{k}, Y^{k-1}, Y_{3,1}^{k-1}) \leftrightarrow Y_{k}$ (see \cite{Massey:90});
(b) follows from the fact that conditioning reduces entropy;
(c) follows from the fact that $X_{2,k}$ is a deterministic function of $S_2^n$;
(d) follows from the non-negativity of the mutual information; 
and (e) follows from the memoryless sources and side information assumption and from \eqref{eq:FanoDest}--\eqref{eq:generalDestEntropyBasic}.

Following arguments similar to those that led to \eqref{eq:generalDestSingleChain} we obtain:
    \vspace{-0.2cm}
		\begin{subequations} \label{eq:generalDestChain}
    \begin{align} 
        H(S_2|S_1, W) &\leq \frac{1}{n} \sum_{k=1}^{n}{I(X_{2,k}, X_{3,k};Y_{k} | S_{1,k}, X_{1,k},  W_{k})} + \gamma(P_e^{(n)}) \label{eq:generalDestSingleChain2}  \\
        H(S_1, S_2| W) & \leq \frac{1}{n} \sum_{k=1}^{n}{I(X_{1,k},X_{2,k}, X_{3,k};Y_{k}| W_{k})} + \gamma(P_e^{(n)}).  \label{eq:generalDestJointChain12}
    \end{align}
    \end{subequations}
   
	\vspace{-0.1cm}
	\noindent Note that the following three expressions, $I(X_{1,k}, X_{3,k};Y_{k}| S_{2,k}, X_{2,k},  W_{k})$, $I(X_{2,k}, X_{3,k};Y_{k} | S_{1,k}, X_{1,k}, W_{k})$, and $I(X_{1,k},X_{2,k}, X_{3,k};Y_{k}| W_{k})$, depend on the marginal conditional distribution:
  \vspace{-0.2cm}
	\begin{equation*}
		p(x_{1,k},x_{2,k},x_{3,k}|s_{1,k},s_{2,k}) = p(x_{1,k},x_{2,k}|s_{1,k},s_{2,k})p(x_{3,k}|s_{1,k},s_{2,k},x_{1,k},x_{2,k}),
	\end{equation*}
	
	\vspace{-0.2cm}
   \noindent and on $p(s_{1,k},s_{2,k}, w_k)$ and $p(y_k|x_{1,k},x_{2,k},x_{2,k})$. Moreover, note that $X_{1,k}$ is a function of $S_1^n$ while $X_{2,k}$ is a function of $S_2^n$, and therefore the Markov chain in \eqref{eq:MarkovChain} holds. Thus, it follows that: 
  \vspace{-0.2cm}
	\begin{align}
%
		p(x_{1,k},x_{2,k}|s_{1,k},s_{2,k}) \in \mB_{X_1 X_2|S_1 S_2} \subseteq \mB_{X_1 X_2|S_1 S_2}'. \label{eq:probLargerSet}  	
  \end{align}  
   
	 \vspace{-0.2cm}
   \noindent Next, we introduce the time-sharing random variable $Q$ uniformly distributed over $\{ 1,2,\dots,n \}$ and independent of all other random variables. We can write the following:
\vspace{-0.15cm}
\begin{align}
%
\frac{1}{n} \sum_{k=1}^{n}{I(X_{1,k}, X_{3,k};Y_{k} | S_{2,k},X_{2,k}, W_{k})} & = I(X_{1,Q}, X_{3,Q};Y_{Q} | S_{2,Q}, X_{2,Q}, W_{Q}, Q) \nonumber \\
	& = I(X_{1}, X_{3};Y | S_{2}, X_{2}, W, Q),
	\label{eq:MarkovBoundConcavity}	
\end{align}

\vspace{-0.2cm}     
\noindent where $X_1 \triangleq X_{1,Q}$, $X_2 \triangleq X_{2,Q}$, $X_3 \triangleq X_{3,Q}$, $Y \triangleq Y_{Q}$, $S_2 \triangleq S_{2,Q}$ and $W \triangleq W_{Q}$. Furthermore, since for all values of $q$ we have $p(x_{1,q},x_{2,q}|s_{1,q},s_{2,q},Q=k) = p(x_{1,k},x_{2,k}|s_{1,k},s_{2,k})$ which satisfies \eqref{eq:probLargerSet}, then we have that for $k=1,2,\dots,n$ it holds that:
  \vspace{-0.3cm}
	\begin{align}
%
		p(x_{1,q},x_{2,q}|s_{1,q},s_{2,q}, Q=k) \in \mB_{X_1 X_2|S_1 S_2}'.
  \end{align}

	\vspace{-0.2cm}
 \noindent Finally, note that for all $k$, the expressions and structural constraints on the distribution chain are identical. Thus, repeating the steps leading to \eqref{eq:MarkovBoundConcavity} for \eqref{eq:generalDestSingleChain2} and \eqref{eq:generalDestJointChain12}, and taking the limit $n \mspace{-4mu} \rightarrow \mspace{-4mu} \infty$, leads to the constraints in~\eqref{bnd:outr_markov_dst}.

\vspace{-0.25cm}	
\subsection{Proof of Proposition \ref{prop:VGeneral}} \label{annex:ProofOuterV}

%
\vspace{-0.1cm}
First, define the auxiliary RV $V_k \triangleq (W_{3,1}^n, Y_{3,1}^{k-1}), k=1,2,\dots,n$. 
%
\noindent Constraint \eqref{bnd:outr_V_dst_S1} is a consequence of the following chain of inequalities:
    \vspace{-0.2cm}
		\begin{align}
        & \sum_{k=1}^{n}{I(X_{1,k};Y_{k}, Y_{3,k}| S_{2,k}, X_{2,k}, W_{k}, V_{k})} \nonumber \\
        & \qquad \stackrel{(a)}{=} \sum_{k=1}^{n}{\Big[H(Y_{k}, Y_{3,k}|S_{2,k}, X_{2,k}, W_{k}, W_{3,1}^n , Y_{3,1}^{k-1})} \nonumber \\
        & \qquad \qquad \qquad - H(Y_{k}, Y_{3,k}|S_{2,k}, X_{1,1}^{k}, X_{2,1}^{k}, X_{3,1}^{k}, W_{k}, W_{3,1}^n, Y^{k-1}, Y_{3,1}^{k-1})\Big] \nonumber \\
%
%
        & \qquad \stackrel{(b)}{\geq} \sum_{k=1}^{n}{\Big[H(Y_{k}, Y_{3,k}|S_2^n, X_{2,k}, Y^{k-1}, W^n, W_{3,1}^n, Y_{3,1}^{k-1})} \nonumber \\
        & \qquad \qquad \qquad - H(Y_{k}, Y_{3,k}|S_1^n, S_2^n, X_{1,1}^{k}, X_{2,1}^{k}, X_{3,1}^{k}, W^n, W_{3,1}^n, Y^{k-1}, Y_{3,1}^{k-1} )\Big] \nonumber 
			\end{align}
			\begin{align}
%
%
        & \qquad \stackrel{(c)}{\geq} \sum_{k=1}^{n}{\Big[H(Y_{k}, Y_{3,k}| S_2^n, W^n,W_{3,1}^n, Y^{k-1}, Y_{3,1}^{k-1})} - H(Y_{k}, Y_{3,k}|S_1^n, S_2^n, W^n, W_{3,1}^n, Y^{k-1},Y_{3,1}^{k-1})\Big] \nonumber \\
%
%
%
        & \qquad \geq H(S_1^n|S_2^n, W^n,W_{3,1}^n ) - H(S_1^n|S_2^n, W^n, W_{3,1}^n, Y^n) \nonumber \\
%
%
        & \qquad \stackrel{(d)}{\geq} nH(S_1 | S_2, W, W_3) - n\gamma(P_e^{(n)}), \label{eq:VDestSingleChain}
    \end{align}

		\vspace{-0.25cm}
    \noindent where (a) follows from the definition of $V_k$, the fact that $X_{3,1}^k$ is a deterministic function of $(W_{3,1}^n,Y_{3,1}^{k-1})$ and from the memoryless channel assumption, see \eqref{eq:MARCchanDist};
(b) follows from the fact that conditioning reduces entropy and, \cite{Massey:90};
(c) follows from the fact that $X_{2,k}$ is a deterministic function of $S_2^n$, and from the property that conditioning reduces entropy; 
(d) follows again from the fact that conditioning reduces entropy, the memoryless sources and side information assumption, and \eqref{eq:FanoDest}--\eqref{eq:generalDestEntropyBasic}.

Following arguments similar to those that led to \eqref{eq:VDestSingleChain} we can also show that:
    \vspace{-0.15cm}
		\begin{subequations} \label{eq:VDestJointChain}
    \begin{align}
        H(S_2 | S_1, W, W_3) & \leq \frac{1}{n} \sum_{k=1}^{n}{I(X_{2,k};Y_{k}, Y_{3,k}| S_{1,k}, X_{1,k}, W_{k}, V_{k})} + \gamma(P_e^{(n)}) \label{eq:VDestJointChain_S2} \\
 				H(S_1,S_2| W, W_3) & \leq \frac{1}{n} \sum_{k=1}^{n}{I(X_{1,k}, X_{2,k};Y_{k}, Y_{3,k}|W_{k}, V_{k})} + \gamma(P_e^{(n)}) \label{eq:VDestJointChain_Sum}.
    \end{align}
    \end{subequations}

\vspace{-0.1cm}
\noindent Next, we define the time-sharing random variable $Q$ uniformly distributed over $\{ 1,2,\dots,n \}$ and independent of all other random variables. We can write the following:
\vspace{-0.15cm}
\begin{align}
	\frac{1}{n} \sum_{k=1}^{n}{I(X_{1,k};Y_{k}, Y_{3,k} | S_{2,k},X_{2,k}, W_{k}, V_{k})} 
%
	& = I(X_{1,Q};Y_{Q}, Y_{3,Q}| S_{2,Q}, X_{2,Q}, W_{Q}, V_{Q}, Q) \nonumber \\
	& = I(X_{1};Y, Y_{3}|S_{2}, X_{2}, W, V),
	\label{eq:VBoundConcavity}	
\end{align}

	\vspace{-0.2cm}
	\noindent where $X_1 \triangleq X_{1,Q}$, $X_2 \triangleq X_{2,Q}$, $Y \triangleq Y_{Q}$, $Y_3 \triangleq Y_{3,Q}$, $S_2 \triangleq S_{2,Q}$, $W \triangleq W_Q$ and $V \triangleq (V_Q, Q)$. Since $(X_{1,k}, X_{2,k})$ and $X_{3,k}$ are independent given $(S_{1,k},S_{2,k},V_k)$, for $\bar{v} = (v,k)$ we have:
    \vspace{-0.2cm}
		\begin{align}
        & \Pr \big( X_1=x_1,X_2=x_2,X_3=x_3|S_1=s_1, S_2=s_2,V=\bar{v} \big) \nonumber \\
%
        & \qquad = \Pr \big( X_1=x_1,X_2=x_2|S_1=s_1, S_2=s_2, V=\bar{v} \big) \Pr \big(X_3=x_3|V=\bar{v} \big).
    \label{eq:VouterDist}
    \end{align}

\vspace{-0.2cm}    
\noindent Hence, the probability distribution is of the form given in \eqref{eq:BCbound_dist}.
Finally, repeating the steps leading to \eqref{eq:VBoundConcavity} for \eqref{eq:VDestJointChain_S2} and \eqref{eq:VDestJointChain_Sum}, and taking the limit $n \rightarrow \infty$, leads to the constraints in \eqref{bnd:V_general_dst}.

%
	
\ifthenelse{\boolean{SquizFlag}}{}{}

\end{document}

%% file: newCommands.tex
\theoremstyle{remark} \newtheorem{theorem}{Theorem}
\theoremstyle{remark} 
\theoremstyle{remark} \newtheorem{claim}{Claim}

\newtheorem*{theoremA}{Theorem}

\theoremstyle{remark} \newtheorem{remark}{Remark}

\newcommand{\eps}{\epsilon}

\newcommand{\styp}{A_{\epsilon}^{*(n)}}

\newcommand{\mE}{\mathcal{E}}

\newcommand{\mX}{\mathcal{X}}

\newcommand{\mY}{\mathcal{Y}}
\newcommand{\mW}{\mathcal{W}}
\newcommand{\mV}{\mathcal{V}}
\newcommand{\mS}{\mathcal{S}}
\newcommand{\mB}{\mathcal{B}}

\newcommand{\mQ}{\mathcal{Q}}

\newcommand{\ts}{\tilde{s}}

\newcommand{\tXvec}{\tilde{\mathbf{X}}}
\newcommand{\txvec}{\tilde{\mathbf{x}}}
\newcommand{\yvec}{\mathbf{y}}
\newcommand{\xvec}{\mathbf{x}}

\newcommand{\zvec}{\mathbf{z}}

\newcommand{\svec}{\mathbf{s}}

\newcommand{\avec}{\mathbf{a}}

\newcommand{\pvec}{\mathbf{p}}

\newcommand{\vvec}{\mathbf{v}}
\newcommand{\wvec}{\mathbf{w}}

\newcommand{\tsvec}{\tilde{\svec}}
\newcommand{\hsvec}{\hat{\svec}}
\newcommand{\hvvec}{\hat{\vvec}}
\newcommand{\hxvec}{\hat{\xvec}}

\newcommand{\msg}{u}

\newcommand{\msgBig}{\MakeUppercase{\msg}}
\newcommand{\msgCal}{\mathcal{\msgBig}}
\newcommand{\tmsg}{\tilde{\msg}}

\newcommand{\hmsg}{\hat{\msg}}
\newcommand{\mU}{\mathcal{U}}
\newcommand{\mF}{\mathcal{F}}
\newcommand{\hU}{\hat{U}}
\newcommand{\hu}{\hat{u}}
\newcommand{\tu}{\tilde{u}}
\newcommand{\Svec}{\mathbf{S}}
\newcommand{\Wvec}{\mathbf{W}}
\newcommand{\hSvec}{\hat{\mathbf{S}}}
\newcommand{\hXvec}{\hat{\mathbf{X}}}
\newcommand{\hVvec}{\hat{\mathbf{V}}}
\newcommand{\tSvec}{\tilde{\mathbf{S}}}
\newcommand{\tVvec}{\tilde{\mathbf{V}}}
\newcommand{\tvvec}{\tilde{\mathbf{v}}}
\newcommand{\tv}{\tilde{v}}
\newcommand{\tx}{\tilde{x}}
\newcommand{\tV}{\tilde{V}}
\newcommand{\tX}{\tilde{X}}

\newcommand{\Xvec}{\mathbf{X}}
\newcommand{\Yvec}{\mathbf{Y}}

\newcommand{\Vvec}{\mathbf{V}}
\newcommand{\E}{\mathds{E}}

\newcommand{\dsP}{\mathds{P}}
\newcommand{\dsM}{\mathds{M}}
\newcommand{\dsN}{\mathds{N}}
\newcommand{\dsI}{\mathds{I}}
\newcommand{\dsSig}{\mathds{D}}

\newcommand{\setR}{\mathfrak{R}}
\newcommand{\setN}{\mathfrak{N}}

\newcommand{\bs}{\boldsymbol}
\newcommand{\dst}{\displaystyle}

\newcommand\independent{\protect\mathpalette{\protect\independenT}{\perp}}
\def\independenT#1#2{\mathrel{\rlap{$#1#2$}\mkern2mu{#1#2}}}
\long\def\symbolfootnote[#1]#2{\begingroup\def\thefootnote{\fnsymbol{footnote}}\footnote[#1]{#2}\endgroup}

\newcommand{\thmlabel}[1]{\label{thm:#1}}
\newcommand{\thmref}[1]{\ref{thm:#1}}
\newcommand{\Thmref}[1]{Thm.~\thmref{#1}}

\newtheorem{MyProposition}{Proposition}

\newtheorem{MyDefinition}{Definition}

%% file: Notations.tex
In this work, we denote random variables (RVs) with upper case letters, e.g. $X$, $Y$, and their realizations with lower case letters , e.g., $x$, $y$. A discrete RV $X$ takes values in a set $\mX$. $|\mX|$ is used to denote the cardinality of a finite, discrete set $\mX$.
We use $p_X(x)$ to denote the probability mass function (p.m.f.) of a discrete RV $X$ on $\mX$; for brevity we may omit the subscript $X$ when it is the uppercase version of the sample symbol $x$.
We denote vectors with boldface letters, e.g. $\xvec$, $\yvec$, 
the $i$'th element of a vector $\xvec$ is denoted by $x_i$, and we use $\xvec_i^j$ where $i<j$ to denote $(x_i, x_{i+1},...,x_{j-1},x_j)$; $x^j$ is a short form notation for $x_1^j$, and unless specified otherwise $\xvec \triangleq x^n$.
Matrices are denoted by doublestroke font, e.g. $\dsP$.
We denote the empty set with $\phi$, and the complement of the set $\mB$ by $\mB^c$.
We use $H(\cdot)$ to denote the entropy of a discrete RV and $I(\cdot;\cdot)$ to denote the mutual information between two RVs, as defined in \cite[Ch. 2.2]{YeungBook}. 
We use $\styp(X)$ to denote the set of $\eps$-strongly typical sequences with respect to (w.r.t.) the p.m.f $p_X(x)$ on $\mX$, as defined in \cite[Ch. 6.1]{YeungBook}. When referring to a typical set we may omit the RVs from the notation when these variables are obvious from the context. 
We use $X \leftrightarrow Y \leftrightarrow Z$ to denote a Markov chain formed by the RVs $X,Y,Z$ as defined in \cite[Ch. 2.1]{YeungBook}.
Finally, we use $X \independent Y$ to denote that $X$ is statistically independent of $Y$, $\setN^{+}$ is used to denote the set of positive integers, $\setR$ is used to denote the set of real numbers and $\E\{ \cdot \}$ is used to denote stochastic expectation.